\documentclass[prb,aps,twocolumn,groupaddress,floatfix]{revtex4-1}
\usepackage{latexsym}
\usepackage{graphicx}
\usepackage{verbatim}
\usepackage{multirow}
\usepackage [english]{babel}
\usepackage [autostyle, english = american]{csquotes}
\MakeOuterQuote{"}
\usepackage{amsmath}
\newcommand{\beq}{\begin{eqnarray}}
\newcommand{\eeq}{\end{eqnarray}}
\usepackage{mathrsfs}
\usepackage{float}
\usepackage{caption}
\usepackage{bm}
\usepackage[usenames, dvipsnames]{color}
\usepackage{tabularx}
\usepackage{subcaption}
\usepackage{mathtools}
\usepackage{mathtools}
\usepackage{slashed}
\usepackage{physics}	
\usepackage{graphicx}   
\usepackage{epstopdf}
\usepackage{hyperref}   
\usepackage{bbold}
\usepackage{wasysym}
\usepackage{amsmath}
\usepackage{bm}

\def\-{\raisebox{.75pt}{-}}

\DeclareMathSizes{10}{10}{7}{5}

\begin{document}

\title{Topological Dipole Conserving Insulators and Multipolar Responses}
\author{Julian May-Mann} 
\affiliation{Department of Physics and Institute of Condensed Matter Theory, University of Illinois at Urbana-Champaign, 1110 West Green Street, Urbana, Illinois 61801-3080, USA}
\author{Taylor L. Hughes} 
\affiliation{Department of Physics and Institute of Condensed Matter Theory, University of Illinois at Urbana-Champaign, 1110 West Green Street, Urbana, Illinois 61801-3080, USA}

\begin{abstract}{Higher order topological insulators (HOTIs) are a novel form of insulating quantum matter, which are characterized by having gapped boundaries that are separated by gapless corner or hinge states. Recently, it has been proposed that the essential features of a large class of HOTIs are captured by topological multipolar response theories. In this work, we show that these multipolar responses can be realized in interacting lattice models, which conserve both charge and dipole. In this work we study several models in both the strongly interacting and mean-field limits. In $2$D we consider a ring-exchange model which exhibits a quadrupole response, and can be tuned to a $C_4$ symmetric higher order topological phase with half-integer quadrupole moment, as well as half-integer corner charges. We then extend this model to develop an analytic description of adiabatic dipole pumping in an interacting lattice model. The quadrupole moment changes during this pumping process, and if the process is periodic, we show the total change in the quadrupole moment is quantized as an integer. We also consider two interacting $3$D lattice models with chiral hinge modes. We show that the chiral hinge modes are heralds of a recently proposed "dipolar Chern-Simons" response, which is related to the quadrupole response by dimensional reduction. Interestingly, we find that in the mean field limit, both the $2$D and $3$D interacting models we consider here are equivalent to known models of non-interacting HOTIs (or boundary obstructed versions). The self-consistent mean-field theory treatment provides insight into the connection between free-fermion (mean-field) theories having vanishing polarization and interacting models where dipole moments are microscopically conserved. }

\end{abstract}
\maketitle
\tableofcontents
\section{Introduction}

Through intense theoretical effort, symmetry protected topological phases (SPTs) have become a well understood area of condensed matter physics\cite{hasan2010, senthil2015, kane2005, bernevig2006, kitaev2009periodic,ryu2010,hasan2011, chen2013, kapustin2015}. The most notable feature of these phases of matter is that they have a gapped bulk and robust surface states. As the name implies, the surface states of an SPT are protected by a symmetry of the system, and in the absence of this symmetry the system can be smoothly deformed into a trivial insulator without closing the bulk gap. For example, in $1$D there exists an SPT protected by inversion symmetry with quantized half-integer surface charges\cite{su1979solitons,jackiw1981solitons}, and in $3$D there exists an SPT protected by time reversal symmetry that hosts a single surface Dirac cone\cite{fu2007topological,moore2007topological}.  Additionally, there are also topological phases, where the surface states remain robust even in the absence of any symmetry. In $2$D such a topological phase is realized in integer quantum hall insulators\cite{klitzing1980}. 

In addition to their gapless surfaces states, the topological nature of SPTs is also manifest through various quantized topological response phenomena. For example, the aforementioned $1$D SPT has a half-integer quantized polarization response, the $2$D topological phase has a quantized Chern-Simons response, and the $3$D SPT has quantized axion electrodynamics \cite{king1993theory,qi2008, qi2013axion,zhang1989effective}. Additionally, certain topological responses are also related to each other through a dimensional hierarchy\cite{qi2008}. For example, the $2$D Chern-Simons/quantum Hall response on a thin torus can be mapped on to the charge polarization response of a $1$D insulator. In this limit, the Laughlin gauge argument for Hall current in $2$D\cite{laughlin1981quantized} maps onto Thouless charge pumping in $1$D\cite{thouless1983}. In SPTs, topological responses such as these are quantized by the interplay of topology and symmetry, and are robust, provided the relevant symmetry is preserved, and the bulk of the system remains gapped. Because of their sharp quantization, topological responses can be used to predict experimentally relevant characteristics of topological materials.

While the study of SPTs initially focused on systems protected by internal symmetries, it has since evolved to show that there are a rich array of topological systems that are protected by spatial symmetries--referred to as topological crystalline insulators\cite{fu2011, hughes2011, hsieh2012, isobe2015, cheng2016, bradlyn2017}. Of particular interest for this work are topological crystalline insulators having \emph{gapped} boundaries, but which exhibit gapless corner, or hinge modes. Systems with these novel surface states are known as higher order topological insulators (HOTIs)\cite{benalcazar2017a, benalcazar2017b, langbehn2017, song2017, serra2018, peterson2018, imhof2018}. The first example was provided in in Ref. \onlinecite{benalcazar2017a}, where it was shown that there exists a non-interacting HOTI in $2$D that can be protected by $C_4$ spatial rotations symmetry (or $M_x$ and $M_y$ mirror symmetries for a boundary obstructed phase\cite{benalcazar2017a,khalaf2019boundary}). This model exhibits a symmetry-quantized quadrupole moment, and when defined a on a lattice with boundary this HOTI has half-integer charges localized at its corners. Similar to how a polarized $1$D system can be related to a $2$D Chern insulator by dimensional reduction, the $2$D quadrupole insulator is related to a $3$D HOTI with chiral hinge modes\cite{benalcazar2017b}. Due to the rich phenomena exhibited by the quadrupole insulator and its cousins, HOTIs have attracted a great deal of theoretical attention. In particular, there have been several works classifying these systems according to their symmetries\cite{song2017b,song2017c,rasmussen2018,thorngren2018}.
Despite these advances, the topological response of HOTIs still remains largely nebulous. This is in striking contrast to SPTs with internal symmetries, and topologically ordered systems, which have well understood topological responses.

Recently, it has been proposed that the topological responses of certain HOTIs can be written in terms of symmetric rank-2 gauge fields\cite{you2019}. Rank-2 gauge fields are symmetric two index gauge fields $A_{ij}$ that, for scalar-charge theories, transform as $A_{ij} \rightarrow A_{ij} + \partial_i \partial_j \Lambda$ under a gauge transformation $\Lambda$. In the last several years, rank-2 gauge fields have attracted attention in condensed matter physics because of their relation to fracton phases\cite{nandkishore2019fractons,pretko2020fracton}. A key feature of these gauge theories is that the Gauss's law for rank-2 gauge fields naturally conserves both charge and dipole \cite{pretko2017a, pretko2017b, pretko2018b}. In particular, this inherent dipole conservation leads to modified dynamics for the excitations in fracton phases\cite{nandkishore2019, haah2011, vijay2016, hsieh2017, pretko2017, ma2018, gromov2019}. Indeed, the exotic conservation laws can force  quasiparticles to be either immobile, or confined to move along sub-dimensional manifolds, such as lines or planes in $3$D. In lattice models, this sub-dimensional confinement can be interpreted as arising from a microscopic dipole conservation law\cite{xu2006novel, xu2010emergent, pretko2017}. Because of this, it has been proposed that the essential physics of fracton systems can be captured by effective rank-2 gauge theories. Generally it has been shown that rank-2 gauge fields can couple to matter theories that conserve dipole\cite{pretko2018}. In these systems, the dipole moment can be treated as the conserved charge of a global $1$-form symmetry\cite{seiberg2019, seiberg2020, dubinkin2020c}. 

Returning to HOTIs, it was shown in Ref. \onlinecite{you2019}, that a rank-2 quadrupole response can describe the quantized corner charges of a $2$D HOTI. This rank-2 quadrupole response can be considered as a rank-2 analog of the (rank-1) charge polarization of a $1$D system\cite{qi2011topological}. Similar to how charge polarization describes the surface charge of a system, the rank-2 quadrupole response describes the boundary polarization and corner charge of a system.  The rank-2 quadrupole response is also related to a topological response in $3$D by dimensional reduction. This $3$D response is a kind of dipolar Chern-Simons response\cite{you2019}, and it can describe a $3$D HOTI with chiral hinge modes. As noted before, rank-2 gauge fields can couple only to matter theories that conserve a dipole moment. In order to conserve dipole, such a theory must not have any single-particle charge dynamics, since single-particle charge transport changes the dipole moment of a system. However, it is possible to have multi-particle charge dynamics in a dipole conserving system, e.g., correlated/pair hopping terms which are represented by quartic interaction terms. Because of this, the multipolar rank-2 topological responses of Ref. \onlinecite{you2019} most naturally describe interacting HOTIs. 

In this work, our goal is to develop tools to further investigate these interacting HOTIs and their topological responses, and then to apply our techniques to several dipole conserving lattice models of interacting fermions. To analyze these systems, we formulate a linear response theory of the rank-2 quadrupole moment of dipole conserving solids. This formulation of the quadrupole moment is largely analogous to the linear response formulation of polarization in charge conserving solids\cite{resta2007}. Similar to how a change in polarization is viewed as a pumping of charge across the system, here the change in the quadrupole moment is viewed as a pumping of dipoles across the system. For a system with boundaries, the change in quadrupole moment can also lead to surface charge currents and a change in corner charges.  Using our linear response formalism, we are able to exactly calculate the quadrupole response of the interacting quadrupole model of Ref. \onlinecite{you2019} and provide an alternative confirmation that it is a HOTI having a quantized quadrupole moment. Interestingly, we are also provide a protocol for tuning the parameters of the model to fully demonstrate a dipole pumping process in this strongly interacting system, i.e., we can see how dipole is pumped across the system as the quadrupole moment of the system is tuned by varying parameters of the model.  Furthermore, we show that analyzing the model within a self-consistent mean field theory approximation, maps this interacting HOTI to the non-interacting quadrupole model of Ref. \onlinecite{benalcazar2017a}. We can use this mean-field approximation to demonstrate a remarkable connection between non-interacting HOTIs and topological rank-2 responses, the latter of which would naively apply only to systems with exact dipole conservation. 

We then move on to consider two interacting $3$D models. The first model we study is related to the dipole conserving $2$D interacting quadrupole model  by dimensional reduction. This model conserves the dipole moment along the $x$ and $y$-directions, and is invariant under $C_4\mathcal{T}$ symmetry (the product of $C_4$ symmetry and time reversal symmetry $\mathcal{T}$).
Since the $3$D dipole Chern-Simons response is related to the $2$D rank-2 quadrupole response via dimensional reduction\cite{you2019}, we expect this $3$D model to exhibit a quantized dipole Chern-Simons response and have chiral hinge modes. 
Indeed, we are able to show that the bulk and surface responses of this $3$D model are identical to those predicted by the dipole Chern-Simons theory. We also verify that this model supports chiral hinge modes that are consistent with $C_4\mathcal{T}$ symmetry. Similar to 2D case, we show that within a self-consistent mean field approximation this interacting $3$D HOTI is equivalent to a known non-interacting chiral hinge insulator in $3$D. The second $3$D model  we consider is a related interacting model that is invariant under $M_x\mathcal{T}$ (the product of $M_x$ mirror symmetry and time reversal symmetry $\mathcal{T}$) and $M_y\mathcal{T}$ (the product of $M_y$ mirror symmetry and time reversal symmetry $\mathcal{T}$) instead of $C_4\mathcal{T}$. This model also has chiral hinge modes, however, in the mean field limit, this model is equivalent to a layered system that harbors dangling Chern insulators on the top and bottom surfaces having Chern number $+1$ and $-1$ respectively. Because of this, the mean field model constitutes a "boundary-obstructed topological insulator" using the terminology of Ref. \onlinecite{khalaf2019boundary}.

In summary, our work on the generic formalism and the explicit $2$D and $3$D models represents a further link between higher order topological phases protected by spatial symmetries and analogous fracton-like systems having stricter multipole conservation laws. Our paper is organized as follows. In Section \ref{sec:PolReview} we motivate our linear response formulation of the quadrupole moment by reviewing the linear response theory of polarization, and considering relevant examples. In Section \ref{sec:DimRed} we formulate a linear response definition of the rank-2 quadrupole response in solids. We use this formalism to calculate the quadrupole response of a dipole conserving $2$D lattice model and show how adiabatic deformations of the model can lead to a change in the quadrupole moment via dipole pumping. We also show that this lattice model can be tuned to a $C_4$ symmetric topological phase, with half-integer quadrupole moment. In Section \ref{sec:ResponseAction} we review the $3$D dipolar Chern-Simons response theory, which is related to the rank-2 quadrupole response in $2$D by dimensional reduction. In Section \ref{sec:LatticeC} we present a $3$D $C_4\mathcal{T}$ symmetric lattice model that is related to the dipole conserving $2$D lattice model we discussed earlier by dimensional reduction. We show that this model realizes the dipole Chern-Simons response and has  protected chiral hinge modes. In Sec. \ref{sec:LatticeM} we present a related $\mathcal{M}_y\mathcal{T}$ and $\mathcal{M}_y\mathcal{T}$ symmetric lattice model in $3$D. We show that this model also has protected hinge modes, and that it is  related to a boundary obstructed phase in the mean-field limit. We conclude our results in Section \ref{sec:conclusion}. We also have several appendices that contain the technical details of our calculations.

\section{Linear Response Theory of Polarization}\label{sec:PolReview}
\begin{figure}
\includegraphics[width=2\textwidth/5]{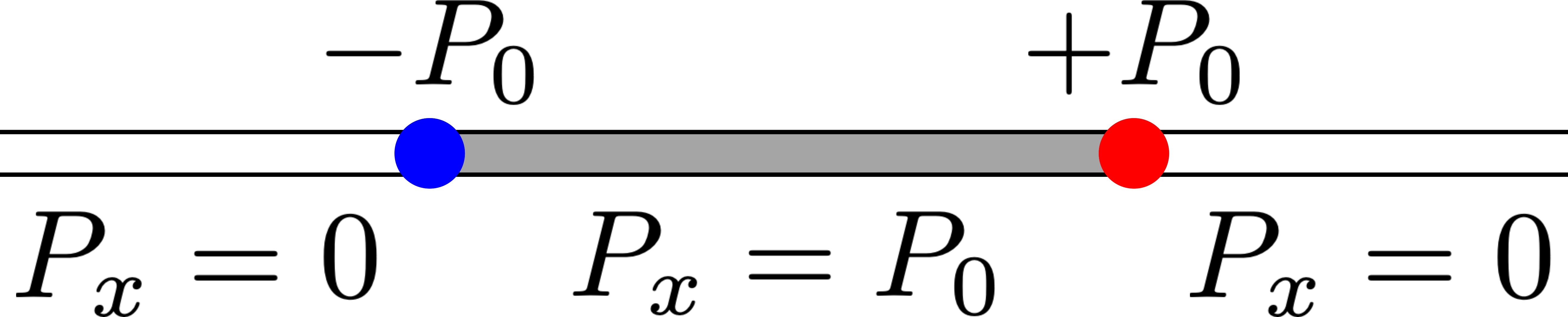}
\caption{The bound charges of a $1$D sample with polarization $P_{x} = P_0$ (gray) embedded in a $1$D vacuum with $P_{x} = 0$ (white).}\label{fig:PolDiagram}
\end{figure}

Before considering the linear response theory of the quadrupole moment in solids, it will be useful to first review the modern linear response theory of polarization\cite{resta2007} (from here on we will use the terms "dipole moment" and polarization interchangeably in this paper). For simplicity we shall focus on systems in $1$D. Our starting point is the effective response action for a system with polarization $P$:
\begin{equation}
S_{P} = \int dt dx P_x (\partial_x A_0 - \partial_t A_x).\label{eq:PolResponse}
\end{equation}
Here, the gauge fields $A_x$ and $A_0$ are background fields that serve as probes for the charge responses of the system. To show that Eq. \ref{eq:PolResponse} does in fact describe the polarization of a $1D$ system, let us consider an infinite line where $P_x = P_0$ for $0\leq x \leq L_x,$ and $P_x = 0$ everywhere else. This corresponds to a finite $1$D system with polarization $P_0$ embedded in an unpolarized vacuum. According to Eq. \ref{eq:PolResponse}, the electric charge is $j_0 = \frac{\delta}{\delta A_0} S_{P} = -\partial_x P_x$.  This means that a charge of $-P_0$ will be localized at the $x=0$ boundary, and a charge of $+P_0$ localized at $x = L_x$ boundary (see Fig. \ref{fig:PolDiagram}). This is exactly the boundary charge distribution we expect for a $1$D system with polarization $P_0$. 

An important feature of Eq. \ref{eq:PolResponse} is that $\partial_x A_t - \partial_t A_x$ is a total derivative. Because of this, there are only non-trivial charge responses when $P_x$ varies in space or time, e.g., when boundaries are present. In other words, only \textit{changes} in polarization are observable. This was a key insight in the development of the modern theory of polarization\cite{resta1992theory, king1993, resta1994macroscopic}. In this framework, the "polarization" of a given system is properly defined as the polarization of a system relative to an (unpolarized) reference state trivial insulator (which is often taken to be a trivial atomic insulator). With this in mind, let us consider the change in polarization during an adiabatic process. Physically, this change in polarization can be attributed to the pumping of charge across the system, i.e., to a charge current. If we parameterize the adiabatic process by $\theta,$ then the change in polarization for a system that is minimally coupled to a flat background gauge field $A_{x}$ is given by
\begin{equation}
\begin{split}
\frac{\partial}{\partial \theta} P = \lim_{\epsilon \rightarrow 0}\frac{i}{\epsilon L}\sum_{n\neq 0}&\Big[\frac{\bra{0} \frac{\partial H}{\partial\theta}\ket{n}\bra{n} \frac{\partial H}{\partial{A_x}} \ket{0}}{\epsilon + E_0 - E_n}\\ &-  \frac{\bra{0} \frac{\partial H}{\partial{A_x}}\ket{n}\bra{n} \frac{\partial H}{\partial\theta} \ket{0}}{\epsilon + E_n - E_0}\Big],\label{eq:KuboSpectralPol}
\end{split}
\end{equation}
where $H$ is the Hamiltonian of the system we are considering, and $\ket{n}$ is an energy eigenstate with energy $E_n$. After some algebra, this expression can be rewritten in terms of a Berry curvature as
\begin{equation}
\frac{\partial}{\partial \theta} P_x =  \frac{i}{ L} \Big[ \frac{\partial}{\partial A_{x}} \bra{0} \frac{\partial}{\partial \theta} \ket{0}-  \frac{\partial}{\partial \theta} \bra{0} \frac{\partial}{\partial A_{x}} \ket{0}\Big],
\label{eq:PolSimp}
\end{equation}
where $\ket{0}$ is the ground state wavefunction of the model being considered, and we are implicitly taking $\ket{0}$ to be a function of both the adiabatic parameter $\theta,$ as well as the background gauge field $A_x$. Eq. \ref{eq:PolSimp} is invariant under shifting $\ket{0}$ by an arbitrary phase that depends on $\theta$ and $A_x$. Because of this, we can choose an overall phase for the ground state wavefunction such that $\frac{\partial}{\partial A_{x}} \bra{0} \frac{\partial}{\partial \theta } \ket{0} = 0$. For this choice, the total change in polarization during an adiabatic process is found to be 
\begin{equation}
\Delta P_x =  \Delta \left[\frac{-i}{ L} \bra{0} \frac{\partial}{\partial A_{x}} \ket{0}\right].
\label{eq:PolSimp2}
\end{equation}
For band insulators, where the ground state can written as a product of eigenfunctions of the single particle Hamiltonian, the change in polarization can written as
\begin{equation}
\Delta P_x = \Delta \left[(-i) \int \frac{d k_x}{2\pi} \sum_{\alpha \in \text{occ}} \bra{\alpha, k_x} \frac{\partial}{\partial k_x} \ket{\alpha, k_x}\right],
\label{eq:BlochPolarization}
\end{equation}
where $\ket{\alpha, k}$ are the eigenfunctions of the single particle Hamiltonian, $\alpha$ is the band index, and the sum is over the occupied bands. This is the celebrated Berry phase formulation of the polarization of band insulators\cite{king1993, resta1994macroscopic}.

An important feature of this framework is that $\Delta P_x$ is only defined modulo an integer. In Eq. \ref{eq:PolSimp2} this ambiguity is due to the fact that the ground state wavefunction can be multiplied by the gauge invariant Wilson loop $\ket{0}\rightarrow \exp(i \int dx A_x )\ket{0}$ (we set $e=\hbar=1$). In Eq. \ref{eq:BlochPolarization} this ambiguity is due to the fact that the eigenfunctions of the single particle Hamiltonian can be shifted by the $k_x$ dependent phase $\ket{\alpha, k_x} \rightarrow e^{ik_x}\ket{\alpha, k_x} $. In both cases, the redefinition of the ground state wavefunction does not change any physical properties of the ground state, but it will shift the polarization by $+1$. Physically, this phase shift corresponds to moving every particle in the system over by one unit cell. For a system with periodic boundary conditions, such a translation is trivial. For a system with boundaries, translating each particle over by one unit cell will add or remove a single unit of charge from the boundaries. This effect can be canceled by adding an integer number of electrons to the boundaries and hence does not meaningfully affect the bulk polarization of the system.

The integer ambiguity in the polarization enables the identification of an SPT protected by inversion symmetry. The polarization response in Eq. \ref{eq:PolResponse} is odd under inversion symmetry, so in an inversion symmetric system $P_x = -P_x$. Since $P_x$ is only defined modulo an integer, $P_x = -P_x$ is satisfied when $P_x = 0$ or $P_x = 1/2$. The former corresponds to a trivial insulator with no fractional boundary charges, while the latter corresponds to a $1$D SPT with half-integer boundary charges. It is well known that this SPT can be realized in the Su-Schrieffer–Heeger (SSH) model for polyacetylene\cite{su1979solitons}. 

\begin{figure}\centering
\includegraphics[width=2\textwidth/5]{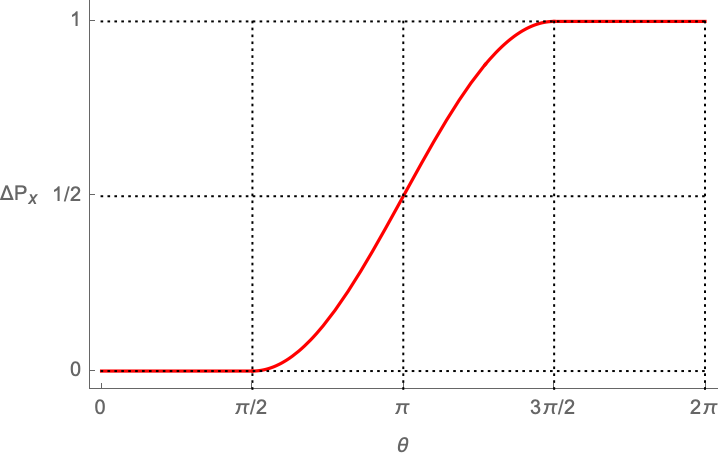}
\caption{The change in polarization  $\Delta P_{x} \equiv P_{x}(\theta)-P_{x}(0)$ of Eq. \ref{eq:SSHHam} as a function of the adiabatic parameter $\theta $ in Eq. \ref{eq:PolParameters}. }\label{fig:PolPlot}
\end{figure}

Having established the linear response theory of polarization, it will be useful to apply this formalism to a lattice model. We would also like to verify that the polarization is only defined modulo an integer, and that the polarization is quantized in units of $1/2$ in an inversion symmetric system. To do this, we shall consider the SSH Hamiltonian with an additional onsite potential term:
\begin{equation}
\begin{split}
H^{\text{SSH}} &= \sum_{k_x} \bm{c}^\dagger(k_x) h^{\text{SSH}}(k_x) \bm{c}(k_x),\\
h^{\text{SSH}}(k_z) &= (u + v\cos(k_z))\tau_1 + v\sin(k_z)\tau_2 + \mu \tau_3,
\end{split}\label{eq:SSHHam}
\end{equation} where $\bm{c} = (c_1,c_2)$, and $\tau_i$ are the Pauli matrices. Here, $u$ and $v$ are the amplitudes for intracell and intercell hopping respectively, and $\mu$ is the strength of the onsite potential. When $\mu = 0$, the model has inversion symmetry. Here, we are interested in the change in polarization during an adiabatic process, and so we will make the parameters $u$, $v$ and $\mu$ functions of an adiabatic parameter $\theta$. Specifically, we will set
\begin{equation}
\begin{split}
&u = \max(\cos(\theta),0),\\
& v = \max(-\cos(\theta),0),\\
&\mu = -\sin(\theta).\label{eq:PolParameters}
\end{split}
\end{equation}
This process is periodic with respect to $\theta$, and the Hamiltonian at $\theta = 0$ is the same as the Hamiltonian at $\theta = 2\pi$. Additionally, the system has inversion symmetry when $\theta = 0(2\pi)$ and $\pi$. Using Eq. \ref{eq:BlochPolarization} the polarization of this system is given by
\begin{equation}
\Delta P_x = 
\begin{cases}
    0 & \text{for } 0 \leq \theta  \leq \pi/2\\
    \frac{1}{2}[1-\sin(\theta)] & \text{for } \pi/2 \leq \theta  \leq 3 \pi/2\\
    1 & \text{for } 3\pi/2 \leq \theta  \leq 2\pi
\end{cases},\label{eq:PValues}
\end{equation}
 where $\Delta P_x \equiv P_x(\theta)-P_x(0)$. The polarization as a function of $\theta$ during this process is shown in Fig. \ref{fig:PolPlot}. From this calculation we can clearly see that when $\theta$ is increased from $0$ to $2\pi$, the polarization of the system increases from $0$ to $1$. As noted before, this process is periodic, confirming that the polarization is only defined modulo an integer. Additionally, the model we are considering has inversion symmetry when $\theta = 0 (2\pi)$ and $\theta = \pi$, and at these points the polarization is $0(1)$ and $1/2$ respectively. At $\theta = 0(2\pi)$ the system only has intracell terms, and hence is a trivial inversion symmetric insulator. At $\theta = \pi$ the we therefore expect that the system is in the SPT phase with polarization $P_x = 1/2$. In this phase, the model has half-integer charges localized at it boundaries. 

It is also useful to consider the fully continuous parameterization of the SSH chain (as opposed to the piecewise continuous parameterization  in Eq. \ref{eq:PolParameters}), 
\begin{equation}
\begin{split}
&u = \gamma[1+\cos(\theta)],\\
& v = v\text{(const.)},\\
&\mu = -\gamma \sin(\theta). \label{eq:PolParametersCont}
\end{split}
\end{equation}
For $v/\gamma < 2$, this pumping process is topologically equivalent to the one given in Eq. \ref{eq:PolParameters}. In particular, at $\theta = \pi$, the parameterization in Eq. \ref{eq:PolParametersCont} leads to the same inversion symmetric SPT discussed earlier, and as $\theta$ is increased from $0$ to $2\pi$, the polarization $P_x$ increases by $+1$. Additionally, since the parameterization depends smoothly on $\theta$, we can relate the $1$D model to a $2$D model by identifying $\theta$ with the lattice momentum $k_y$. This $2$D model is an insulator with Chern number $+1$. For a more detailed discussion of this mapping see Ref. \onlinecite{qi2008}.

\section{Quadrupole Moment in Dipole Conserving Systems}\label{sec:DimRed}
Having reviewed the linear response formulation of polarization, we can now turn our attention to determining the rank-2 quadrupole response of a system. This formulation will largely parallel the formulation of polarization we reviewed in Sec. \ref{sec:PolReview}. Here, we shall start by considering the quadrupole response term. For a $2D$ system coupled to a background rank-2 gauge field, the $Q_{xy}$ quadrupole response term is given by\cite{you2019}:
\begin{equation}
S_{Q} = \int dt d\bm{r} Q_{xy} [\partial_x \partial_y A_{0}-\partial_t A_{xy}].
\label{eq:EffectiveQuad}
\end{equation}
This term can naturally be interpreted as the rank-2 generalization of the polarization response given in Eq. \ref{eq:PolResponse}. In $2$D, there are also $xx$ and $yy$ quadrupole responses, which can be written analogously to Eq. \ref{eq:EffectiveQuad}. Here, we shall focus on the $xy$ quadrupole response. It is straightforward to generalize the results of this section to other quadrupole responses.

\begin{figure}
\includegraphics[width=2\textwidth/5]{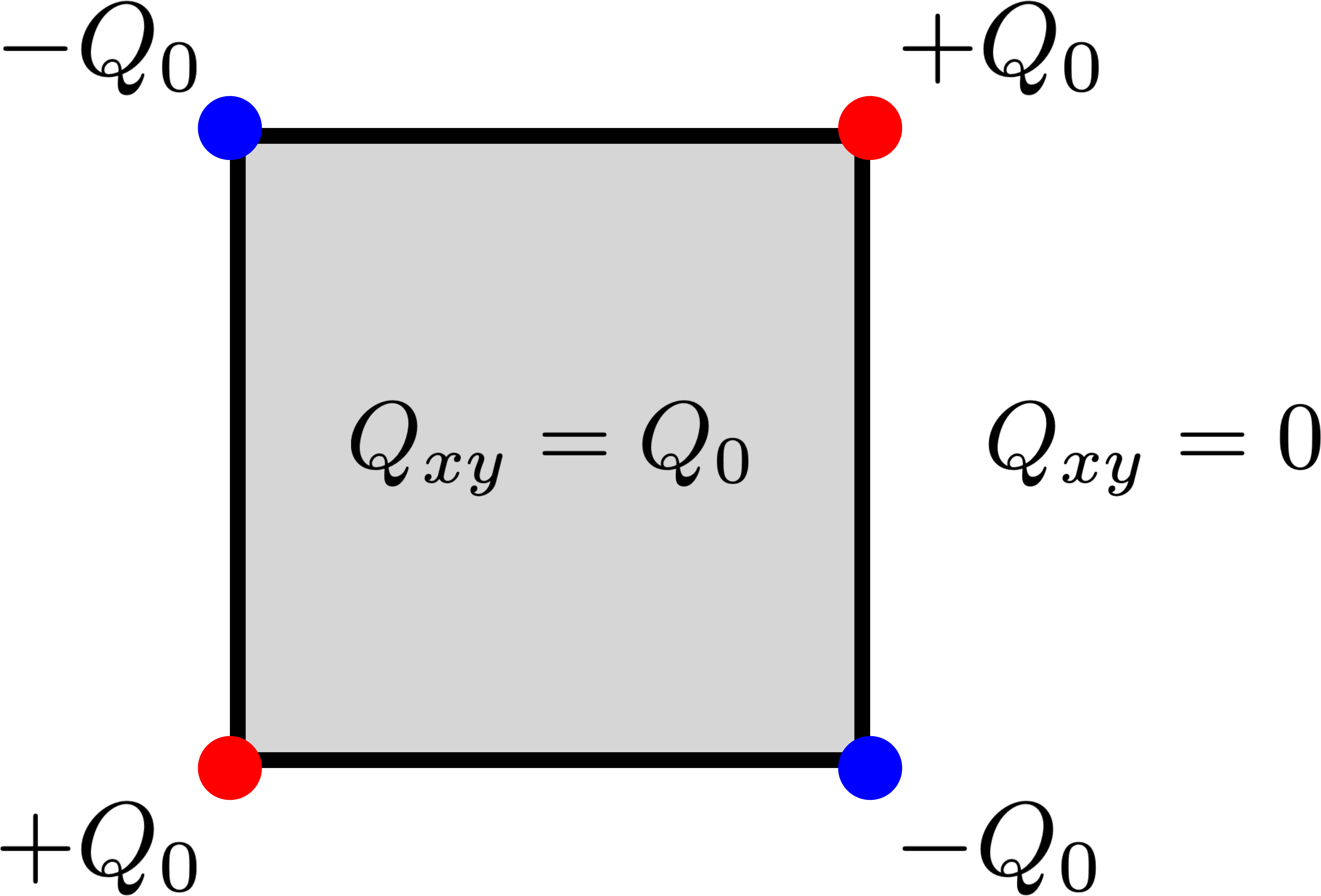}
\caption{The bound charges of an open boundary $2$D sample with quadrupole moment $Q_{xy} = Q_0$ (gray) embedded in a $2$D vacuum with $Q_{xy} = 0$ (white).}\label{fig:QuadDiagram}
\end{figure}

To show that Eq. \ref{eq:EffectiveQuad} does indeed give the desired physics of a model with non-vanishing quadrupole moment, let us consider an infinite 2D plane, where $Q_{xy} = Q_0$ for $0\leq  x \leq L_x$ and  $0\leq  y \leq L_y,$ and  $Q_{xy} = 0$ everywhere else. This corresponds to a rectangular system with quadrupole moment $Q_0$ embedded in a $2$D vacuum. Using $j_0 = \frac{\delta}{\delta A_0} S_Q = \partial_x \partial_y Q_{xy}$, we find that this system has charge $+Q_0$ localized at $(x,y) = (0,0)$ and $(L_x,L_y)$, and charge $-Q_0$  localized at $(x,y) = (0,L_y)$ and $(L_x,0)$ (see Fig. \ref{fig:QuadDiagram}). This is exactly the corner charge distribution we expect for a system with quadrupole moment $Q_0$.

Since $\partial_x \partial_yA_{0} - \partial_t A_{xy} $ is a total derivative, Eq. \ref{eq:EffectiveQuad} is only non-trivial when $Q_{xy}$ varies in space or time, e.g.,  when boundaries are present. Because of this, only \textit{changes} in the quadrupole moment are observable. In this framework, the "quadrupole moment" of a given system is properly defined as the quadrupole moment of a system relative to a reference state. Based on this, we will consider the change in the quadrupole moment during an adiabatic process.  Analogously to how a change in polarization is caused by pumping charges, here the change in the quadrupole moment can be attributed to pumping dipoles. If we parameterize the adiabatic process by $\theta$, then the change in quadrupole moment $Q_{xy}$ for a system that is coupled to a flat background gauge field $A_{xy}$ is given by
\begin{equation}
\begin{split}
\frac{\partial}{\partial \theta} Q_{xy} = \lim_{\epsilon \rightarrow 0}\frac{i}{\epsilon L_x L_y}\sum_{n\neq 0}&\left[\frac{\bra{0} \frac{\partial H}{\partial\theta}\ket{n}\bra{n} \frac{\partial H}{\partial{A_{xy}}} \ket{0}}{\epsilon + E_0 - E_n}\right. \\ & - \left. \frac{\bra{0} \frac{\partial H}{\partial{A_{xy}}}\ket{n}\bra{n} \frac{\partial H}{\partial\theta} \ket{0}}{\epsilon + E_n - E_0}\right]. \label{eq:KuboSpectralQuad}
\end{split}
\end{equation}
As before, this expression can be simplified and written in terms of a Berry curvature as
\begin{equation}
\partial_\theta Q_{xy} = \frac{i}{L_x L_y} [\partial_{A_{xy}}(\bra{0}\partial_{\theta} \ket{0}) -  \partial_{\theta}(\bra{0}\partial_{A_{xy}} \ket{0})],
\label{eq:Berry0}
\end{equation}
where $\ket{0}$ is the ground state of the microscopic model we are considering, and $L_i$ is the length of the system in the $i$-direction. Here, we are implicitly treating the ground state wavefunction as a function of both the adiabatic parameter $\theta$ and the rank-2 gauge field $A_{xy}$. Eq. \ref{eq:Berry0}  is invariant under shifting the wavefunction by an arbitrary phase that depends on $\theta$ and $A_{xy}$. Because of this, we can choose a overall phase for the ground state wavefunction such that $\partial_{A_{xy}}(\bra{0}\partial_{\theta} \ket{0}) = 0$. For this choice of phase Eq. \ref{eq:Berry0} can reduce to 
\begin{equation}
\Delta Q_{xy} = \Delta\left[\frac{-i}{L_x L_y} \bra{0}\partial_{A_{xy}} \ket{0}\right].
\label{eq:Berry1}
\end{equation} A similar expression for the quadrupole moment was presented in Ref. \onlinecite{dubinkin2019a}. 

Based on Eq. \ref{eq:Berry1}, we can infer that the quadrupole moment $Q_{xy}$ is only defined modulo $1$. This is because the ground state wavefunction can be multiplied by the gauge invariant "Wilson surface" $\ket{0} \rightarrow \exp(i \int dx dy A_{xy})\ket{0}$. Shifting the ground state wavefunction by such a term will increase the quadrupole moment $Q_{xy}$ by $+1$. Physically, this will add/remove an integer amount of charge from the corners of the system, while leaving the bulk of the system unchanged. An integer ambiguity of this form was also seen when considering the polarization in Sec. \ref{sec:PolReview}.

Due to this integer ambiguity, we can predict the existence of a HOTI with half-integer quadrupole moment protected by $C_4$ rotation symmetry. Since the quadrupole response in Eq. \ref{eq:EffectiveQuad} is odd under $C_4$ rotations, a $C_4$ invariant insulator must have an $xy$ quadrupole moment satisfying $Q_{xy}=-Q_{xy}$. Since $Q_{xy}$ is only defined modulo an integer $Q_{xy}=-Q_{xy}$ is satisfied by $Q_{xy} = 0$ and $Q_{xy} = 1/2$. The former is a trivial $C_4$ symmetric insulator, while the latter is a HOTI with quantized half-integer quadrupole moment. This HOTI will have half-integer corner charges, similar to those found in the non-interacting quadrupole insulator of Ref. \onlinecite{benalcazar2017a}. This logic also predicts a similar quantized quadrupole insulator with half-integer quadrupole moment protected by $M_x$ mirror symmetry and $M_y$ mirror symmetry, since the $xy$ quadrupole response is odd under both of these symmetries.

Having established a linear response formalism, we can now turn our attention to calculating the change in the quadrupole moment in a microscopic lattice model. Based on the rank-2 quadrupole response action in Eq. \ref{eq:EffectiveQuad}, we are interested in lattice models that couple to the scalar potential $A_0$, and the rank-2 gauge field $A_{xy}$. In order for a lattice model to couple to these gauge fields it must conserve both charge and dipole in the $x$ and $y$-directions. For a system of lattice fermions, global charge conservation corresponds to the symmetry that sends $c(\bm{r}) \rightarrow c(\bm{r})e^{i\alpha}$, where $c(\bm{r})$ is the lattice fermion annihilation operator, and $\alpha$ is a constant. Similarly, dipole conservation in the $x$ and $y$-directions corresponds to the symmetry that sends $c(\bm{r})\rightarrow c(\bm{r}) e^{ i \bm{\beta}\cdot\bm{r}}$, where $\bm{\beta}$ is a constant two component vector.

To show why these symmetries are necessary, let us consider an arbitrary lattice model that is composed of fermion operators $c(\bm{r}),$ and the background gauge fields $A_0$ and $A_{xy}.$
Under gauge transformations $\Lambda({\bf{r}})$ these fields transform as
\begin{equation}
\begin{split}
&A_{0}(\bm{r}) \rightarrow A_{0}(\bm{r}) + \partial_t \Lambda(\bm{r}),\\
&A_{xy}(\bm{r}) \rightarrow A_{xy}(\bm{r}) + \Lambda(\bm{r}) - \Lambda(\bm{r}+\hat{x}) \\ &\phantom{======}- \Lambda(\bm{r}+\hat{y}) + \Lambda(\bm{r}+\hat{x}+\hat{y})\\
&\phantom{A_{xy}(\bm{r}) }\equiv A_{xy}(\bm{r}) + \Delta_x \Delta_y \Lambda(\bm{r}),\\
&c(\bm{r}) \rightarrow  c(\bm{r}) e^{ i \Lambda(\bm{r})},
\end{split}\label{eq:GaugeTransDef}
\end{equation}
where $\Delta_i$ is the lattice derivative in the $i$-direction, and we have suppressed any dependence on $t$. Let us now consider a gauge transformation of the form $\Lambda = \alpha$ (const.). The gauge fields $A_0$ and $A_{xy}$ are invariant under such transformations, while the fermions transform as $c(\bm{r}) \rightarrow c(\bm{r})e^{i\alpha}$. So in order for the system to be gauge invariant, it must be invariant under shifting the phase of the fermions by a constant amount. Similarly, we can also consider a gauge transformation of the form $\Lambda(\bm{r}) = \bm{\beta}\cdot \bm{r}$. Under this gauge transformation both $A_0$ and $A_{xy}$ are invariant, while the fermions transform as $c(\bm{r})\rightarrow c(\bm{r}) e^{ i \bm{\beta}\cdot\bm{r}}$. So in order for the system to be gauge invariant, it must also be invariant under shifting the phase of the fermions by an amount that depends linearly on position.

While charge conservation is fairly common in lattice models, dipole conservation is more unusual, and places strong constraints on the types of terms that can appear in a lattice Hamiltonian. Importantly, because of dipole conservation, single-particle hopping terms such as  $c^{\dagger}(\bm{r}+\hat{x})c(\bm{r})$ and $c^{\dagger}(\bm{r}+\hat{y})c(\bm{r})$ are not allowed. However if a system has multiple degrees of freedom within a unit cell, single-particle intracell terms are allowed, since they do not change the dipole moment of the system. This means one can include terms like $c_{i}^{\dagger}(\bm{r})c_{j}(\bm{r})$ where $i$ and $j$ label the different fermionic degrees of freedom within a given unit cell. 

Although single-particle hopping terms are not allowed, quartic interactions can allow for pairs of electrons to have dynamics. A simple term of this form is the ring exchange term: $c^{\dagger}(\bm{r})c^{\dagger}(\bm{r}+\hat{y}+\hat{x})c(\bm{r}+\hat{x})c(\bm{r}+\hat{y})$. A quick calculation confirms that this term is indeed invariant under linear phase shifts, and does not change the dipole moment in the $x$ or $y$-direction. Physically, this term can be thought of as a dipole hopping term. To see this, we note that $c^{\dagger}(\bm{r}+\hat{y}+\hat{x})c(\bm{r}+\hat{y})$ can be interpreted as creating a dipole with dipole vector $\hat{x}$ centered at $\bm{r}+\hat{y}+\tfrac{\hat{x}}{2}$ (the inclusion of $\tfrac{\hat{x}}{2}$ indicates that the dipole is defined on the link between $\bm{r}+\hat{y}+\hat{x}$ and $\bm{r}+\hat{y}$). Similarly, $c(\bm{r}+\hat{x})c^{\dagger}(\bm{r})$ creates a dipole with dipole vector $-\hat{x}$ (equiv. annihilates a dipole with dipole vector $\hat{x}$) centered at $\bm{r}+\tfrac{\hat{x}}{2}$. The ring exchange term thereby hops an $\hat{x}$-oriented dipole from $\bm{r}+\tfrac{\hat{x}}{2}$ to $\bm{r}+\hat{y}+\tfrac{\hat{x}}{2}$. This process can also be interpreted as hopping a $\hat{y}$-oriented dipole one unit in the $\hat{x}$-direction, from $\bm{r}+\frac{\hat{y}}{2}$ to $\bm{r}+\hat{x}+\frac{\hat{y}}{2}$. The ring exchange term minimally couples to the rank-2 gauge field $A_{xy}$ via a rank-2 Peierls factor of the form $c^{\dagger}(\bm{r})c^{\dagger}(\bm{r}+\hat{y}+\hat{x})c(\bm{r}+\hat{x})c(\bm{r}+\hat{y})e^{iA_{xy}(\bm{r})}$\cite{you2019, dubinkin2019a}. With the rank-2 Peierls factor included, this term is invariant under the gauge transformations given in Eq. \ref{eq:GaugeTransDef}.

\begin{figure}
\includegraphics[width=2\textwidth/5]{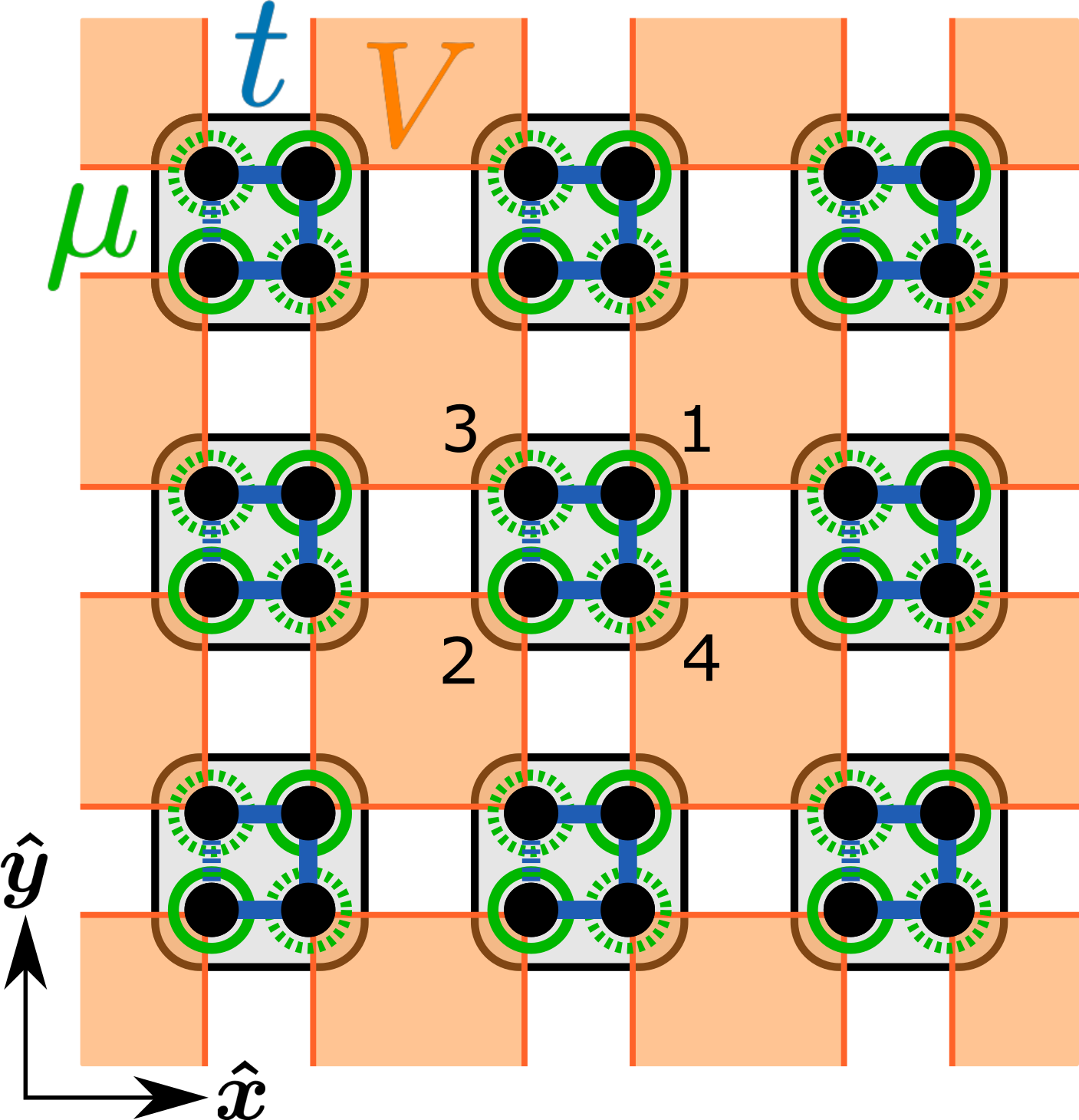}
\caption{Schematic of the dipole conserving lattice model Eq. \ref{eq:LatticeHam2}. Orange squares correspond to the ring exchange interactions with amplitude $V$. Solid (dashed) blue lines correspond to intracell couplings with amplitude $t$ ($-t$). Solid (dashed) green circles correspond to onsite potential with strength $\mu$ ($-\mu$). }\label{fig:2DLatDia}
\end{figure}

Based on these considerations, we introduce the following $2D$ dipole conserving lattice model with 4 fermionic degrees of freedom per unit cell:
\begin{equation}
\begin{split}
H^{Q} &= \sum_{\bm{r}} \bm{c}^\dagger(\bm{r}) h^o \bm{c}(\bm{r}) - A_0(\bm{r}) \bm{c}^\dagger(\bm{r}) \bm{c}(\bm{r})\\&\phantom{=} - V c_1^\dagger(\bm{r})c^\dagger_2(\bm{r}+\hat{x}+\hat{y})c_3(\bm{r}+\hat{x})c_4(\bm{r}+\hat{y})e^{i A_{xy}(\bm{r})} \\&\phantom{=}+ h.c.,\\
h^o &= \mu \Gamma_0 +t (\Gamma_2+\Gamma_4).
\end{split}\label{eq:LatticeHam2}
\end{equation}
Here, $c_i({\bf{r}})$ ($i = 1...4$) are the four lattice fermion operators for a unit cell $\bm{r} = (x,y)$ in the $2$D square lattice (see Fig. \ref{fig:2DLatDia}). The $\Gamma$ matrices are defined as $\Gamma_0 = \tau_3 \otimes \tau_0$, $\Gamma_k = -\tau_2 \otimes \tau_k $, and $\Gamma_4 = \tau_1 \otimes \tau_0,$ for $k = 1,2,3$, where $\tau_{1,2,3}$ are the Pauli matrices. The $4\times 4$ matrix  $h^o$ contains the single-particle intracell terms, and $V$ is the amplitude of the ring exchange term. Since the square lattice is bipartite, the sign of $V$ can be changed by an appropriate unitary transformation. Because of this, we will take $V>0$ without loss of generality. Within the intracell term $h^o$,  $t$ is the amplitude for intracell hopping, while $\mu$ is a staggered onsite potential. We have also included the coupling to the rank-2 gauge field $A_{xy},$ as well as the scalar potential $A_0$. We will restrict our attention to the case where the model is half filled ($2$ fermions per unit cell). 

As desired, this model is gauge invariant, and invariant under constant and linear phase shifts of the fermion operators $c_i({\bf{r}})$. When $\mu = 0$, this model also has $C_4$ rotation symmetry. This symmetry acts on the internal fermionic degrees of freedom as $\bm{c} \rightarrow U_{\mathcal{C}_4}  \bm{c}$ in Eq. \ref{eq:LatticeHam2}, where 
\begin{equation}
\begin{split}
U_{\mathcal{C}_4} & = \begin{bmatrix}
0 & 0 & 1 & 0\\
0 & 0 & 0 & 1\\
0 & -1 & 0 & 0\\
1 & 0 & 0 & 0\\
\end{bmatrix}.\label{eq:C4SymDef}
\end{split}
\end{equation}
 When $\mu = 0$, this model also has $M_x$ and $M_y$ mirror symmetries. The $M_x$ and $M_y$ symmetries act on the internal fermionic degrees of freedom as $\bm{c} \rightarrow U_{M_x}  \bm{c}$ and $\bm{c} \rightarrow U_{M_y}  \bm{c}$ respectively, where $U_{M_x} = \tau_1 \otimes \tau_3$ and $U_{M_y} = \tau_1 \otimes \tau_1$ ($\tau_i$ are the Pauli matrices). When $\mu\neq 0$ The staggered onsite potential explicitly breaks the $C_4$ and mirror symmetries. The model we consider here also has a subsystem symmetry that acts on the fermionic degrees of freedom as $\bm{c}(\bm{r})\rightarrow \bm{c}(\bm{r})e^{i f_x(x)+i f_y(t)}$, where $f_x(x)$ is an arbitrary function of the $x$-coordinate \textit{only}, and $f_y(y)$ is an arbitrary function of the $y$-coordinate \textit{only}. This symmetry correspond to charge being conserved along every row and column of the $2$D lattice.

As we shall show, by adiabatically changing the parameters in this model ($t$, $V$ and $\mu$), it is possible to pump dipole and change the quadrupole moment. In the following subsections, we will present two related parameterizations of this pumping process. First, we shall present a piecewise continuous periodic parameterization of Eq. \ref{eq:LatticeHam2}, where the quadrupole moment can be found exactly. Second, we shall present a fully continuous periodic parameterization, where the quadrupole moment can be found via self-consistent mean field theory.

\subsection{Exactly Solvable Dipole Pumping Process}\label{ssec:ExactSolvablePump}
\begin{figure}
\includegraphics[width=2\textwidth/5]{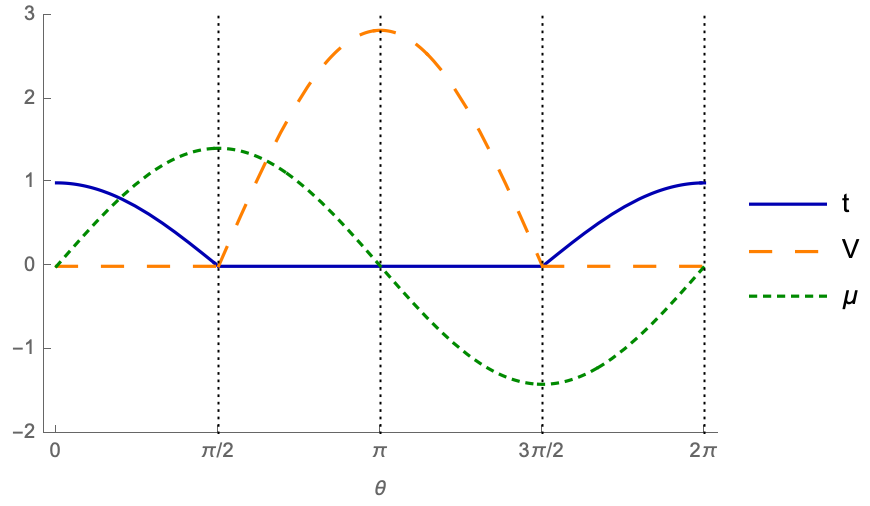}
\caption{The evolution of $t$, $V$, and $\mu$ in Eq. \ref{eq:LatticeHam2} as a function of $\theta$ using the parameterization in Eq. \ref{eq:adParameters}}\label{fig:parameterplot1}
\end{figure}

A simple and illustrative example of dipole pumping in the lattice model Eq. \ref{eq:LatticeHam2} is found by considering the parameterization
\begin{equation}
\begin{split}
& t(\theta ) = \max(\cos(\theta ),0),\\
&V(\theta ) = 2\sqrt{2}\max(-\cos(\theta ),0),\\
&\mu(\theta ) = \sqrt{2}\sin(\theta ).
\end{split}
\label{eq:adParameters}
\end{equation}  The values of $t$, $V$, and $\mu$, for this adiabatic evolution, are shown in Fig \ref{fig:parameterplot1}. This system has $C_4$ symmetry when $\theta = 0$ and $\theta = \pi$. At $\theta  = 0$ the only non-zero terms are the intracell hopping terms $t$, and at $\theta = \pi$, the only non-zero terms are the ring exchange terms $V$. Based on this we can identify the system at $\theta = 0$ is a trivial $C_4$ symmetric insulator. We also expect that at $\theta = \pi$, the system is a HOTI protected by $C_4$ symmetry, with half-integer quadrupole moment (relative to the trivial insulator at $\theta = 0$). 

To show that this expectation is correct, we will use Eq. \ref{eq:Berry1} to calculate the quadrupole moment. The ground state of Eq. \ref{eq:LatticeHam2} with the parameterization given in Eq. \ref{eq:adParameters} can be found exactly in three steps. First, for $0 \leq \theta \leq \pi/2$ the amplitude of the ring exchange term $V$ vanishes, and the Hamiltonian consists only of the intracell terms $t$ and $\mu$. The ground state for this range of $\theta$ can thereby be expressed as a of product single-particle wavefunctions defined on a single site. For $0 \leq \theta \leq \pi/2$ the ground state is
\begin{equation}
\begin{split}
\ket{0} = &\prod_{\bm{r}} \Big[a_1(\theta)c^\dagger_1(\bm{r}) + a_2(\theta)c^\dagger_2(\bm{r})+a_3(\theta)c^\dagger_3(\bm{r})\Big]\\ &\times \Big[a_1(\theta)c^\dagger_1(\bm{r}) - a_2(\theta)c^\dagger_2(\bm{r})+a_4(\theta)c^\dagger_4(\bm{r})\Big] \ket{vac},\\
&a_1(\theta) =a_2(\theta)= \sqrt{\frac{1-\sin(\theta)}{4}}, \\
&a_3(\theta) =a_4(\theta) = -\sqrt{\frac{1+\sin(\theta)}{2}}. 
\end{split}\label{eq:exactGS1}
\end{equation}

Second, for $\pi/2 \leq \theta \leq 3\pi/2$, the amplitude of the intracell hopping $t$ vanishes, and only the ring exchange term $V$ and on-site potential term $\mu$ remain. In this case, the different plaquettes spanned by $c_1(\bm{r})$, $c_2(\bm{r}+\hat{x}+\hat{y})$, $c_3(\bm{r}+\hat{x})$, and $c_4(\bm{r}+\hat{y})$ decouple from each other. The resulting 4 site interacting problem can be solved using exact diagonalization, and the ground state of the system can be expressed as a product of 2 particle wavefunctions defined each plaquette. For $\pi/2 \leq \theta \leq 3\pi/2$, the ground state is given by
\begin{equation}
\begin{split}
\ket{0} = \prod_{\bm{r}} \Big[& e^{iA_{xy}} a_{12}(\theta)c^\dagger_1(\bm{r})c^\dagger_2(\bm{r}
+\hat{x}+\hat{y})\\& + a_{34}(\theta) c^\dagger_3(\bm{r}+\hat{x})c^\dagger_4(\bm{r}+\hat{y})\Big]\ket{vac},\\
&a_{12}(\theta) = \sqrt{\frac{1-\sin(\theta)}{2}},\\
&a_{34}(\theta) = \sqrt{\frac{1+\sin(\theta)}{2}}.
\end{split}\label{eq:exactGS2}
\end{equation}

Third, for $3\pi/2 \leq \theta \leq 2\pi$, the ring exchange terms vanish, and the ground state can again be written as a product of single-particle wavefunctions. Similar to Eq. \ref{eq:exactGS1} the ground state for $3\pi/2 \leq \theta \leq 2\pi$ is given by 
\begin{equation}
\begin{split}
\ket{0} = &\prod_{\bm{r}} e^{iA_{xy}}\Big[a_1(\theta)c^\dagger_1(\bm{r}) + a_2(\theta)c^\dagger_2(\bm{r})+a_3(\theta)c^\dagger_3(\bm{r})\Big]\\ &\times \Big[a_1(\theta)c^\dagger_1(\bm{r}) - a_2(\theta)c^\dagger_2(\bm{r})+a_4(\theta)c^\dagger_4(\bm{r})\Big] \ket{vac},\\
&a_1(\theta) =a_2(\theta)= \sqrt{\frac{1-\sin(\theta)}{4}}, \\
&a_3(\theta) =a_4(\theta) = -\sqrt{\frac{1+\sin(\theta)}{2}}. 
\end{split}\label{eq:exactGS3}
\end{equation}
Compared to Eq. \ref{eq:exactGS1}, Eq. \ref{eq:exactGS3} differs by the addition of the phase $e^{iA_{xy}}$. This phase is needed in order for the wavefunctions in Eq. \ref{eq:exactGS2} and Eq. \ref{eq:exactGS3} to match at $\theta = 3\pi /2$. 

\begin{figure}\centering
\includegraphics[width=2\textwidth/5]{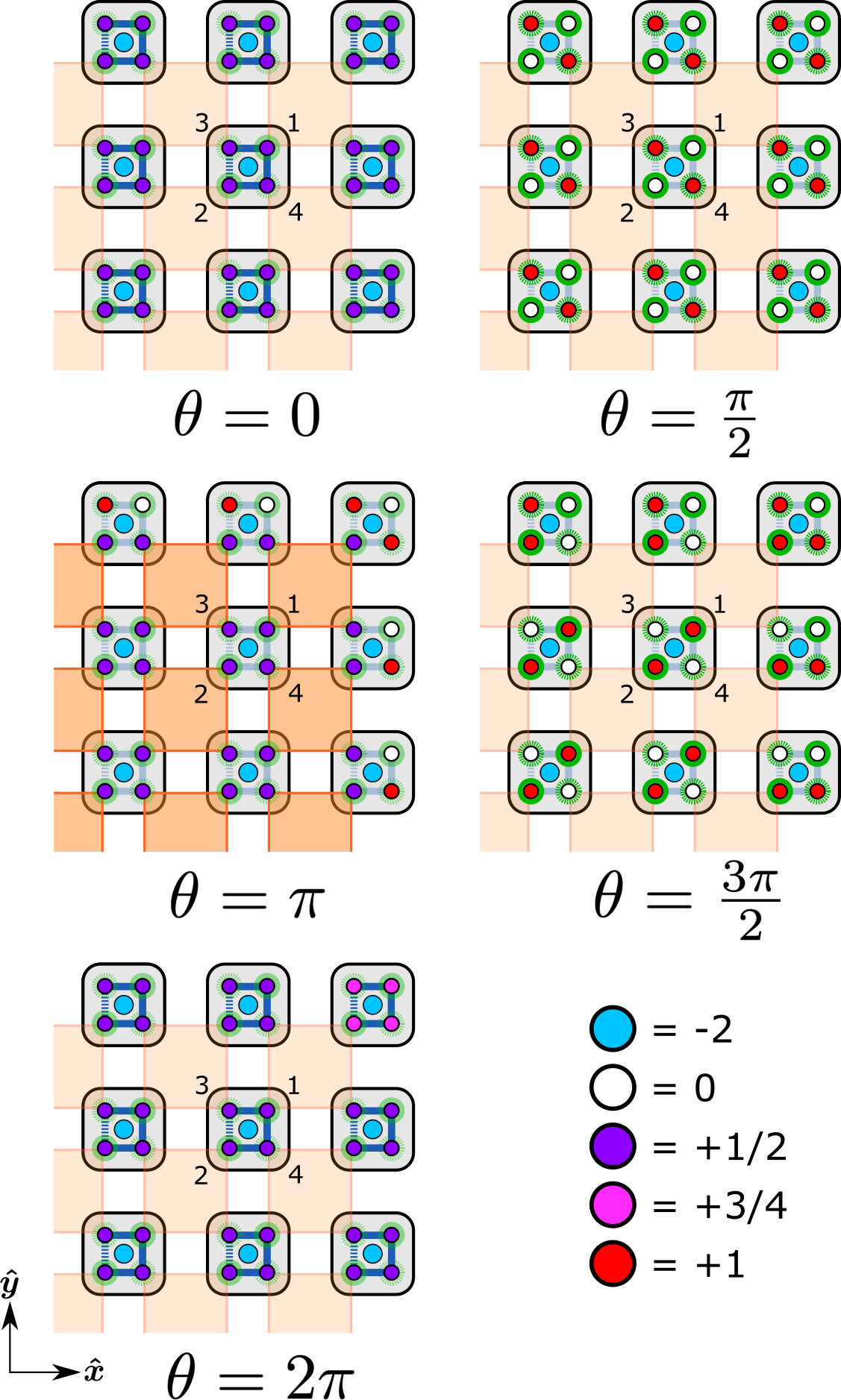}
\caption{Schematic of the change in the charge distribution near the corner of a lattice during the pumping process from Eq. \ref{eq:adParameters}. Here we have added a charge $-2$ ion (in units of the electron charge) to each unit cell to make the system charge neutral. }\label{fig:DipolePumping}
\end{figure}

\begin{figure}\centering
\includegraphics[width=2\textwidth/5]{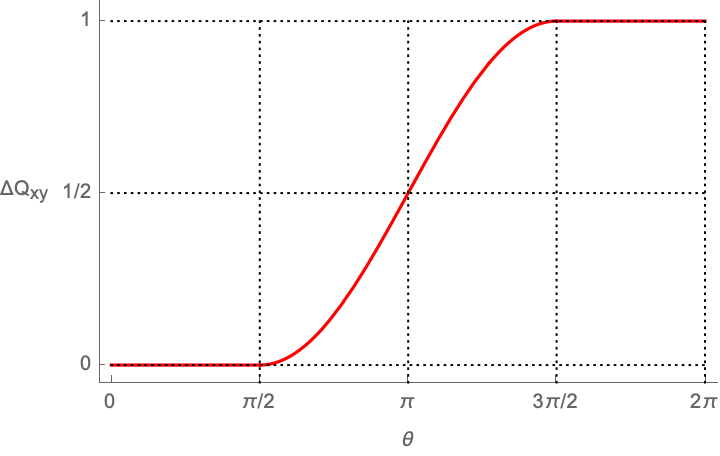}
\caption{The quadrupole moment $\Delta Q_{xy} \equiv Q_{xy}(\theta)-Q_{xy}(0)$ as a function of the adiabatic parameter $\theta $ from Eq. \ref{eq:adValues2}. }\label{fig:quadrupolePlot}
\end{figure}

During this process, the quadrupole moment can be calculated using Eq. \ref{eq:Berry1}, and the wavefunctions in Eq. \ref{eq:exactGS1}-\ref{eq:exactGS3}. As a function of $\theta$, the quadrupole moment is given by
\begin{equation}
\Delta Q_{xy} = 
\begin{cases}
    0 & \text{for } 0 \leq \theta  \leq \pi/2\\
    \frac{1}{2}(1-\sin(\theta)) & \text{for } \pi/2 \leq \theta  \leq 3 \pi/2\\
    1 & \text{for } 3\pi/2 \leq \theta  \leq 2\pi
\end{cases},\label{eq:adValues2}
\end{equation}
where $\Delta Q_{xy}\equiv Q_{xy} (\theta ) - Q_{xy} (0)$. These values of $\Delta Q_{xy} $ are shown in Fig. \ref{fig:quadrupolePlot}. From Eq. \ref{eq:adValues2} we can clearly see that after a full period, the quadrupole increases by $1$. This corresponds to an integer amount of charge being pumped to the corners, and agrees with our earlier claim that the quadrupole moment of the system is only defined modulo $1$. Furthermore, we can also confirm that $Q_{xy} (\pi) - Q_{xy} (0) = 1/2$, and that when $\theta = \pi$ the dipole conserving model is a HOTI protected by $C_4$ symmetry.  

Since the model is exactly solvable over the full range of $\theta$ between $0$ and $2\pi$ for the parameterization given in Eq. \ref{eq:adParameters}, we can also exactly calculate the change in the corner charge of this system as $\theta$ is varied. For a lattice of size $N_x \times N_y$ with open boundaries, the ground state can be found by following the same steps we used in Eq. \ref{eq:exactGS1}-\ref{eq:exactGS3}. For this geometry, there is a net charge of $+\Delta Q_{xy}$ located at the $(1,1)$ and $(N_x,N_y)$ sites, and a net charge of $-\Delta Q_{xy}$ located at the $(N_x,1)$ and $(1,N_y)$ sites, where $\Delta Q_{xy}$ is defined as in Eq. \ref{eq:adValues2} (the charge remains constant at all other sites). So, we find that the quadrupole response does indeed predict the correct corner charges. Additionally, we can confirm that the $C_4$ symmetric HOTI has half-integer corner charges, as expected from the quadrupole response in Eq. \ref{eq:EffectiveQuad}. The change in the charge distribution as $\theta$ is increased from $0$ to $2\pi$ is illustrated in Fig. \ref{fig:DipolePumping}.

\subsection{Continuous Dipole Pumping and Mean Field Theory}\label{ssec:ContDipolePumping}
In the previous section, we established how an adiabatic process can lead to a change in the quadrupole moment.  However, despite being exactly solvable, the pumping process in Eq. \ref{eq:adParameters} is not continuous (it is only piecewise continuous). Preparing for our eventual replacement of the adiabatic parameter $\theta$ with a new momentum $k_z$, we also want to analyze a parameterization of Eq. \ref{eq:LatticeHam2} that is fully continuous with respect to $\theta$, and displays the same essential phenomenology we found using the piece-wise continuous parameterization in Sec. \ref{ssec:ExactSolvablePump}. Specifically, we want a periodic parameterization where, as a function of $\theta,$ the model evolves from a $C_4$ symmetric trivial insulator through a $C_4$ symmetric HOTI, and then back to a trivial insulator. Based on Sec. \ref{ssec:ExactSolvablePump}, we should find that the quadrupole moment of the model should increase by $1/2$ during this evolution from trivial to HOTI, and should change by $1$ after a full period. 

\begin{figure}
\includegraphics[width=2\textwidth/5]{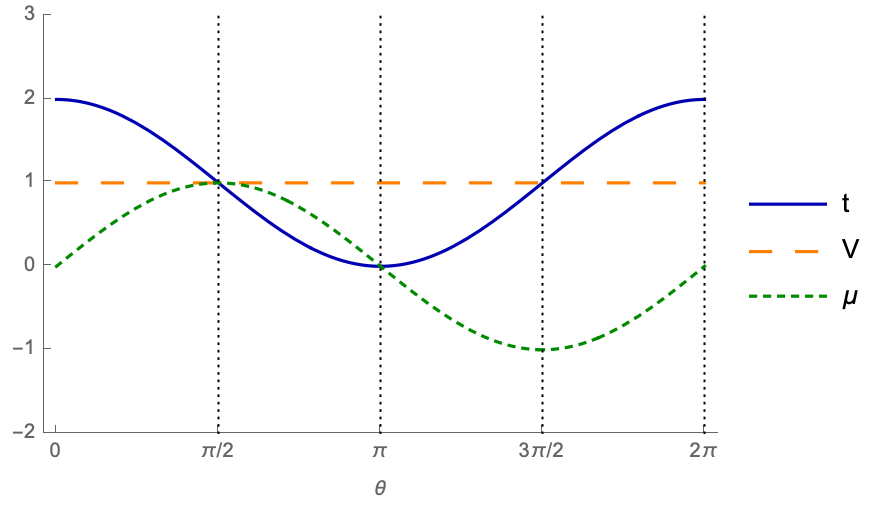}
\caption{The evolution of $t$, $V$, and $\mu$ in Eq. \ref{eq:LatticeHam2} as a function of $\theta$ using the parameterization in Eq. \ref{eq:adParameters2} with $V = \gamma = 1$.}\label{fig:parameterplot2}
\end{figure}

With this in mind, we will consider the parameterization 
\begin{equation}
\begin{split}
&t(\theta ) = \gamma[1+\cos(\theta )],\\
&V(\theta ) = V(\text{const.}),\\
&\mu(\theta ) = \gamma\sin(\theta ),
\end{split}
\label{eq:adParameters2}
\end{equation}
where $\gamma$ is a constant. The values of $t$, $V$ and $\mu$ as a function of $\theta$ are plotted in Fig \ref{fig:parameterplot2}. As before, the model has $C_4$ symmetry when $\mu = 0$, which occurs when $\theta = 0,\pi \mod(2\pi)$. When $\theta = \pi$, the parameters in Eq. \ref{eq:adParameters} and \ref{eq:adParameters2} are the same, and the ground state of the model can be found exactly (see Eq. \ref{eq:exactGS2}) Based on our results from Sec. \ref{ssec:ExactSolvablePump}, at $\theta = \pi$ the model is a HOTI with half-integer corner charges for all values of $V$. Similarly, when $\theta = 0$ and $V=0$, the parameters in Eq. \ref{eq:adParameters} and \ref{eq:adParameters2} are the same and at this point the system is a trivial $C_4$ symmetric insulator. Since this system is gapped the system should remain in this trivial phase up to some finite value of $V$. Beyond this value of $V$, we expect that the will undergo a phase transition and become a $C_4$ symmetric HOTI.

\begin{figure}\centering
\includegraphics[width=.5\textwidth]{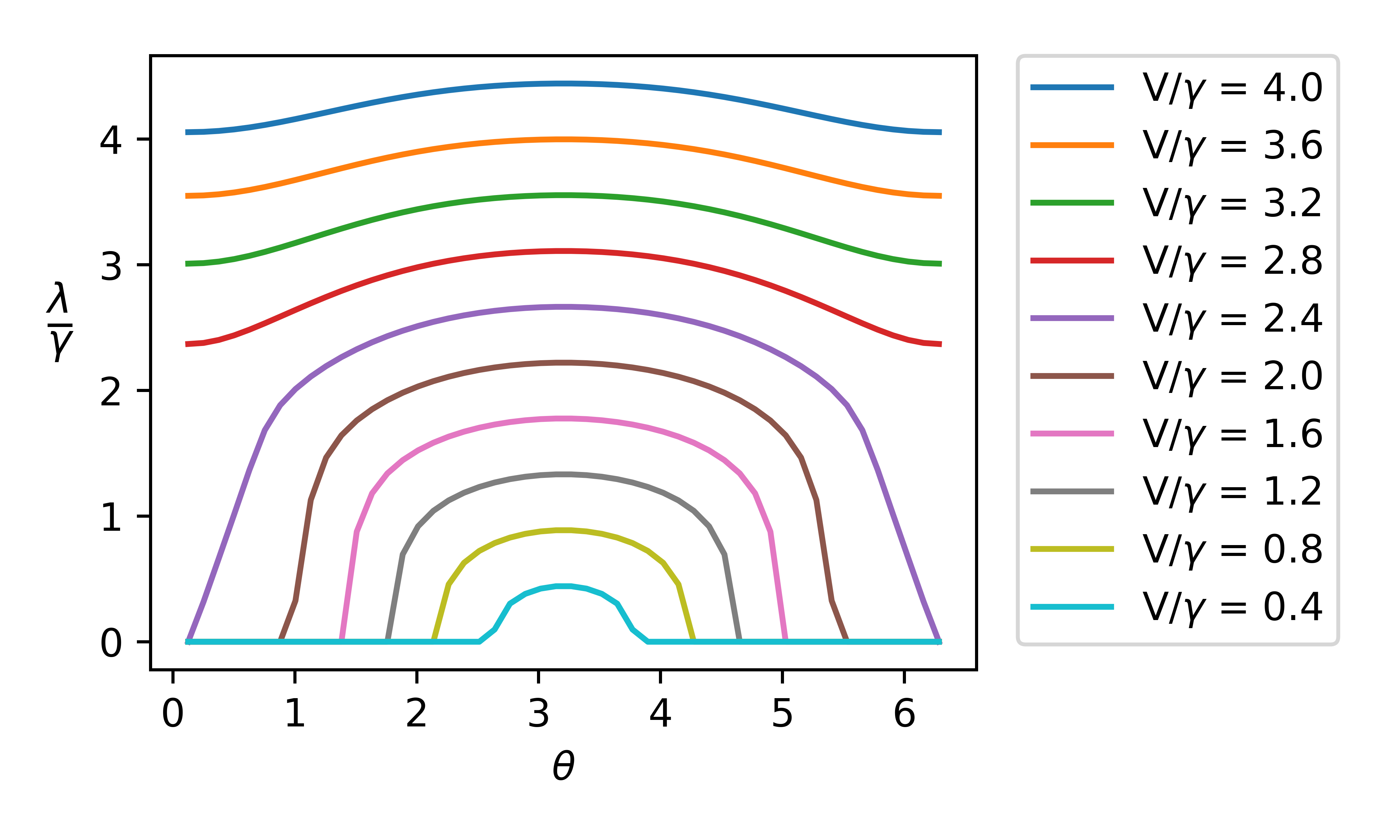}
\caption{The self-consistent values of $\lambda/\gamma$ as a function of $\theta$ for $V/\gamma$ between $0.4$ and $4.0$.}\label{fig:consistentplot2}
\end{figure}

For $V\neq 0$, this model cannot be exactly solved (except at $\theta = \pi$). Because of this, we will use self-consistent mean field theory to analyze the interacting model. The first step in this approximation is to decompose the ring exchange interaction into terms that are quadratic in the lattice fermion operators using a Hubbard-Stratonovich transformation. Here, we will choose to use the following decomposition of the ring exchange term:
\begin{equation}
\begin{split}
&-V c_1^\dagger(\bm{r})c_2^\dagger(\bm{r}+\hat{x}+\hat{y})c_3(\bm{r}+\hat{x})c_4(\bm{r}+\hat{y})e^{iA_{xy}(\bm{r})} \\ &\rightarrow \lambda_{1x}(\bm{r})c_2^\dagger(\bm{r}+\hat{x}+\hat{y})c_4(\bm{r}+\hat{y}) + \lambda_{2x}(\bm{r}) c_1^\dagger(\bm{r})c_3(\bm{r}+\hat{x}) \\&\phantom{\rightarrow}+\lambda_{1y}(\bm{r})c_2^\dagger(\bm{r}+\hat{x}+\hat{y})c_3(\bm{r}+\hat{x})- \lambda_{2y}(\bm{r})c_1^\dagger(\bm{r})c_4(\bm{r}+\hat{y})\\ &\phantom{\rightarrow}- \frac{2}{V}\lambda_{1x}(\bm{r})\lambda_{2x}(\bm{r})e^{-iA_{xy}(\bm{r})} - \frac{2}{V}\lambda_{1y}(\bm{r})\lambda_{2y}(\bm{r})e^{-iA_{xy}(\bm{r})}. 
\end{split}\label{eq:LatticeHS}
\end{equation}
The equations of motion for the Hubbard-Stratonovich fields $\lambda$ are
\begin{equation}
\begin{split}
&\lambda_{1x}(\bm{r}) = \frac{V}{2}e^{iA_{xy}(\bm{r})} c_1^\dagger(\bm{r})c_3(\bm{r}+\hat{x}), \\
&\lambda_{2x}(\bm{r}) = \frac{V}{2}e^{iA_{xy}(\bm{r})} c_2^\dagger(\bm{r}+\hat{x}+\hat{y})c_4(\bm{r}+\hat{y}), \\
&\lambda_{1y}(\bm{r}) = -\frac{V}{2}e^{iA_{xy}(\bm{r})} c_1^\dagger(\bm{r})c_4(\bm{r}+\hat{y}), \\
&\lambda_{1x}(\bm{r}) = \frac{V}{2}e^{iA_{xy}(\bm{r})} c_2^\dagger(\bm{r}+\hat{x}+\hat{y})c_3(\bm{r}+\hat{x}).
\end{split}\label{eq:HSEqOM}
\end{equation}
As can be seen from the equations of motion, under a linear phase shift $c(\bm{r})\rightarrow c(\bm{r})e^{i\bm{\beta}\cdot \bm{r}}$, the Hubbard-Stratonovich fields transform as 
\begin{equation}
\begin{split}
& \lambda_{1x}(\bm{r})\rightarrow \lambda_{1x}(\bm{r})e^{i\bm{\beta}\cdot \hat{x}}, \phantom{==} \lambda_{2x}(\bm{r})\rightarrow \lambda_{2x}(\bm{r})e^{-i\bm{\beta}\cdot \hat{x}},\\ 
&\lambda_{1y}(\bm{r})\rightarrow \lambda_{1y}(\bm{r})e^{i\bm{\beta}\cdot \hat{y}}, \phantom{==} \lambda_{2y}(\bm{r})\rightarrow \lambda_{2y}(\bm{r})e^{-i\bm{\beta}\cdot \hat{y}}.
\end{split}\label{eq:HSLinearTransformation}
\end{equation}
In the self-consistent mean field theory approximation, we assume that the Hubbard-Stratonovich fields acquire an expectation value that satisfies the equations of motion Eq. \ref{eq:HSEqOM}. Equivalently, this approximation can be interpreted as increasing the number of flavors of lattice fermions that couple to the Hubbard-Stratonovich fields from $1$ to $N$, and taking the $N \rightarrow \infty$ limit\cite{zinn2002quantum}.
As we show in Appendix \ref{app:selfCon}, the self-consistent values of $\lambda_{ai}$ ($a = 1,2$, $i=x,y$) can be written as
\begin{equation}
\begin{split}
&\lambda_{1x}(\bm{r})=  \lambda e^{i\phi_x(\bm{r})+iA_{xy}(\bm{r})},\\
&\lambda_{2x}(\bm{r})=  \lambda e^{-i\phi_x(\bm{r})},\\
&\lambda_{1y}(\bm{r})=  \lambda e^{i\phi_y(\bm{r})+iA_{xy}(\bm{r})},\\
&\lambda_{2y}(\bm{r})=  \lambda e^{-i\phi_y(\bm{r})},
\end{split}\label{eq:ConsistentHS2DC}
\end{equation}
where the phase fields $\phi_x$ and $\phi_x$ satisfy the relationships
\begin{equation}
\begin{split}
\Delta_y\phi_x(\bm{r}) = \Delta_x\phi_y(\bm{r}) = A_{xy}(\bm{r}).
\end{split}\label{eq:PhaseCont2DC}
\end{equation}
The self-consistent values of $\lambda$ can be found numerically as a function of $\theta$ and $V/\gamma,$ and are shown in Fig. \ref{fig:consistentplot2}. Due to the equations of motion in Eq. \ref{eq:HSEqOM}, the phase fields $\phi_x$ and $\phi_y$ transform under a gauge transformation $\Lambda$ as 
\begin{equation}
\begin{split}
&\phi_x(\bm{r})\rightarrow \phi_x(\bm{r}) + \Delta_x \Lambda(\bm{r}),\\
&\phi_y(\bm{r})\rightarrow \phi_y(\bm{r}) + \Delta_y \Lambda(\bm{r}),
\end{split}\label{eq:HSGaugeTransform}
\end{equation}
and Eq. \ref{eq:PhaseCont2DC} is consistent with the rank-2 gauge symmetry. 

If we fix the external gauge fields to vanish ($A_0 = A_{xy} = 0$),  the quadratic mean field Hamiltonian can be written in Fourier space as 
\begin{equation}
\begin{split}
&H^{Q}_{\text{mf}} = \sum_{\vec{k}} \Big[ \bm{c}^\dagger(\vec{k}) h^Q_{\text{mf}}(\vec{k})\bm{c}(\vec{k}) \Big],\\
&h^Q_{\text{mf}}(\vec{k}) =\mu \Gamma_0 +t (\Gamma_2+\Gamma_4) \\
&\phantom{===}+ \lambda \big[\cos(k_x+\phi_{x})\Gamma_4 + \sin(k_x + \phi_{x}) \Gamma_3\\
&\phantom{===}+ \cos(k_y+\phi_{y})\Gamma_2 + \sin(k_y + \phi_{y}) \Gamma_1],\\
\end{split}\label{eq:DimRedLatticeHSGauge}
\end{equation}
where $\vec{k} = (k_x, k_y)$ is the $2$D lattice momentum. Here, we have left the $\theta$-dependence of $t$, $\mu$ and $\lambda$ implicit. Up to a shift in momentum due to $\phi_x$ and $\phi_y$, this mean field Hamiltonian is equivalent to that of the non-interacting quadrupole insulator that was analyzed in Ref. \onlinecite{benalcazar2017b}. We note that this mean-field  Hamiltonian can be coupled to $A_{xy}$ fields using the values for the $\lambda_{ai}$ in Eq. \ref{eq:ConsistentHS2DC}. 

The spectrum of the mean field Hamiltonian is given by
\begin{equation}
\epsilon(\bm{k}) = \pm\sqrt{\epsilon_x^2(k_x) + \epsilon_y^2(k_y) + \mu^2},
\end{equation}
where $\epsilon_i(k_i) = \sqrt{\lambda^2 + t^2 + 2\lambda t\cos(k_i + \phi_i)},$ for $i = x,y$. The band structure of Eq. \ref{eq:DimRedLatticeHSGauge} consists of two upper bands and two lower bands. There is a gap between these bands when $\lambda/t \neq 1$ and/or $\mu \neq 0$.  When $\lambda/t = 1$ and $\mu = 0,$ a gap closing occurs at $(k_x,k_y) = (\pi - \phi_x,\pi - \phi_y)$. In Ref. \onlinecite{benalcazar2017b} it was shown that when $\mu = 0$, this Hamiltonian describes a trivial $C_4$ symmetric insulator for $\lambda/t < 1,$ and a $C_4$ symmetric HOTI with half-integer corner charges for $\lambda/t > 1$.  

With this in mind, we shall now consider the quadrupole response of the mean field Hamiltonian, using our linear response formalism. To do this, we will couple the mean field Hamiltonian to the gauge fields according to Eq. \ref{eq:ConsistentHS2DC}. Here, we are primarily interested in finding the change in the quadrupole moment as $\theta$ is increased from $0$ to $\pi$, as well as the total change in the quadrupole moment as $\theta$ is increased from $0$ to $2\pi$. At these values of $\theta$, the lattice model has $C_4$ symmetry, and the change in quadrupole moment is quantized as an integer or half-integer. Indeed within the mean field framework, we find that
\begin{equation}\begin{split}
Q_{xy}(\theta = \pi) - Q_{xy}(\theta = 0) = \begin{cases} \frac{1}{2} &\mbox{if }  V < V_c \\
0 & \mbox{if }  V > V_c \end{cases},\\
Q_{xy}(\theta = 2\pi) - Q_{xy}(\theta = 0) = \begin{cases} 1 &\mbox{if }  V < V_c \\
0 & \mbox{if } V > V_c \end{cases},
\label{eq:MeanFieldQuadResult}\end{split}
\end{equation}
where $V_c \sim 2.5 \gamma$. The details of this calculation are presented in Appendix \ref{app:QuadMF}. 

For $V<V_c$ these results are consistent with an adiabatic process where a trivial $C_4$ symmetric insulator adiabatically evolves into a $C_4$ symmetric HOTI as $\theta$ is increased from $0$ to $\pi$. Indeed, we see in Fig. \ref{fig:consistentplot2} that $\lambda(\theta = 0) = 0$ for $V<V_C$, and the mean field Hamiltonian at $\theta = 0$ is a trivial insulator. As noted before, at $\theta = \pi$, the model can be solved exactly, and we can confirm that it is indeed a $C_4$ symmetric HOTI. Additionally, we find that for this range of $V$ the quadrupole moment of the model increases by $1$ after a full period. This agrees with the results of Sec. \ref{ssec:ExactSolvablePump}. 

\begin{figure}
\includegraphics[width=.5\textwidth]{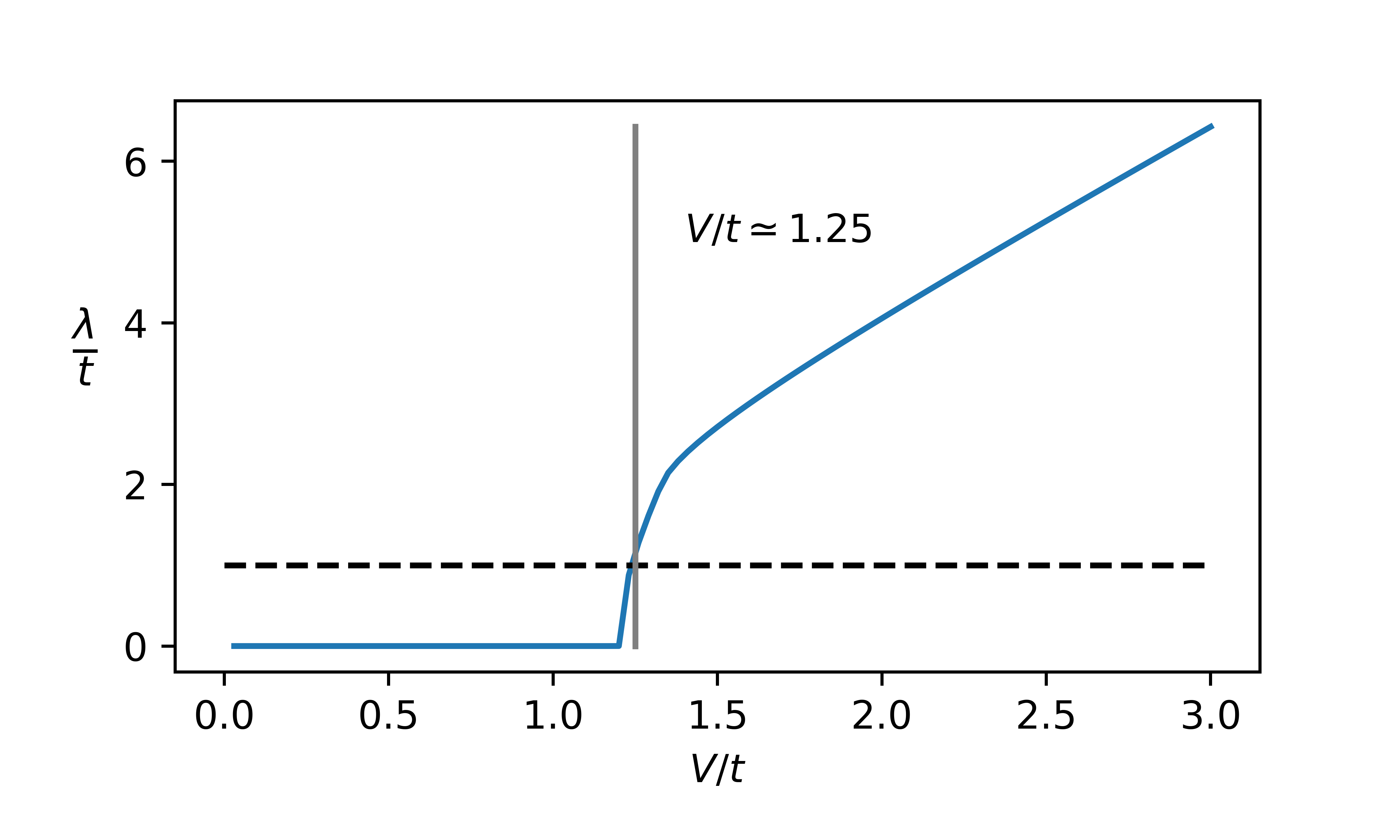}
\includegraphics[width=.4\textwidth]{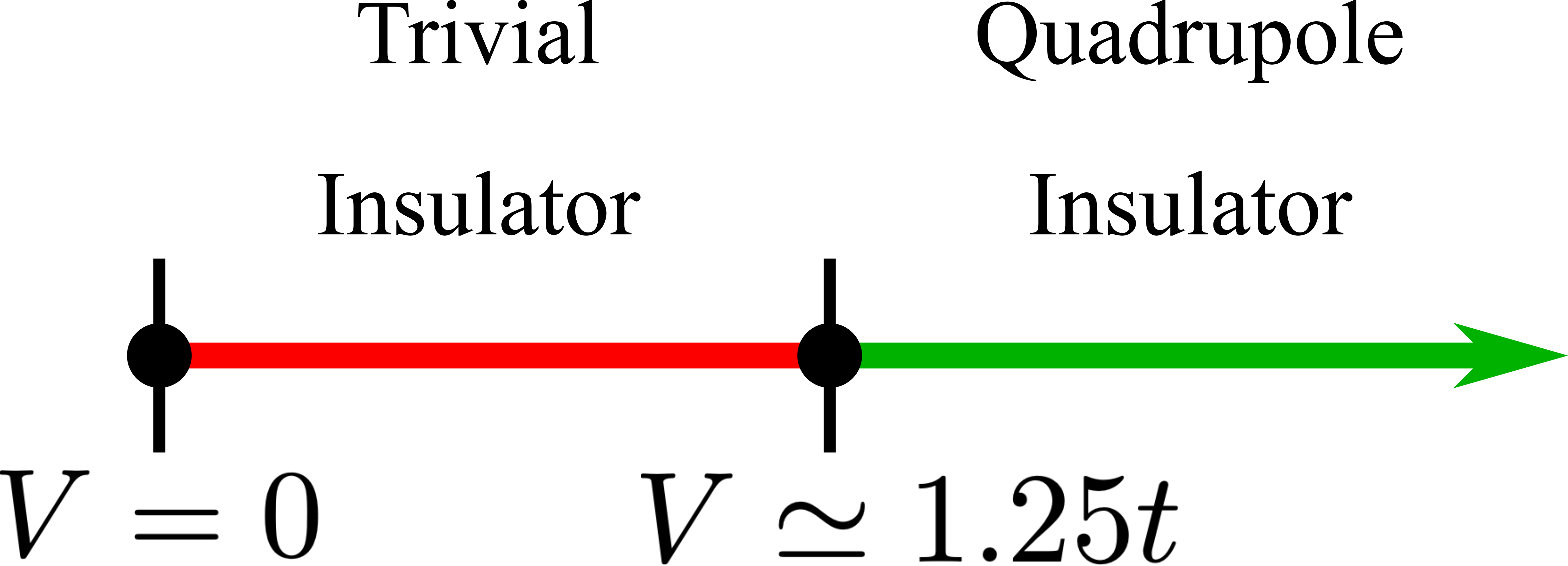}
\caption{Top: The solutions to the self-consistent solutions for $\lambda/t$ as a function of $V/t$ at $\mu = 0$. When $V/t \simeq 1.25$ (grey line), the quadrupole moment of the model changes by $1/2$.  Bottom: The mean field phase diagram of Eq. \ref{eq:LatticeHam2} at $\mu = 0$. Note that the relation between $t$ and $\gamma$ is given in Eq. \ref{eq:adParameters2}.}
\label{fig:SelfConsistentPlotPD2D}
\end{figure}

For $V > V_c$ the quadrupole moment of the model at $\theta = 0$ is equal to the quadrupole moment of the model at $\theta = \pi$. As noted before, at $\theta = \pi$, the model is a HOTI for all values of $V$. This means that at $\theta = 0$ the model transitions from being a trivial insulator to being a HOTI at $ V = V_c$. In Fig. \ref{fig:consistentplot2} we can see that at this transition the value of $\lambda(\theta = 0)$ jumps from $0$ to $\sim 2\gamma$. Indeed, for $\lambda(\theta = 0) > 2\gamma$, the mean field Hamiltonian describes a non-interacting HOTI with half-integer corner charges\cite{benalcazar2017a}. In terms of the parameters of the lattice model in Eq. \ref{eq:LatticeHam2}, this means that the dipole conserving model is a $C_4$ symmetric trivial insulator for $\mu = 0$, $V/t \lesssim 1.25,$ and a $C_4$ symmetric HOTI for $\mu = 0$, $V/t \gtrsim 1.25$. This phase diagram is shown in Fig. \ref{fig:SelfConsistentPlotPD2D}.

As a final point, we would like to address the role of the phase fields $\phi_{x}$ and $\phi_{y}$.
Based on Eq. \ref{eq:HSGaugeTransform}, the phase fields can be considered as dynamic gauge fields. Since these fields are dynamic, they must be integrated over. This integration will project out any states that transform non-trivially under shifts in $\phi_i$. To see what states transform non-trivially, we note that in Eq. \ref{eq:DimRedLatticeHSGauge} a shift in $\phi_i$ is equivalent to shifting the lattice momentum $k_i$. Integrating over $\phi_i$ therefore projects out any many-body states that transform non-trivial under a shift in momentum. Momentum is dual to position, i.e., it is the position operator that acts to shift the momentum. From our earlier discussion of polarization in Sec. \ref{sec:PolReview}, we can conclude that the many-body states that transform non-trivially under a shift in momentum are the states with non-vanishing polarization. So, the integration over the phase fields projects out any many-body states with a non-vanishing polarization, and will ensure that the many-body polarization is a good quantum number in the ground state of the mean field Hamiltonian, as it is in the full interacting Hamiltonian. This is crucial for applying the rank-2 formalism which relies on dipole conservation. This can be shown explicitly, by considering the mean field Hamiltonian when $\mu=t= 0$ (this corresponds to $\theta = \pi$ in Eq. \ref{eq:adParameters2}). For these parameters, the mean field ground state is given by
\begin{equation}\begin{split}
\ket{0} = &\prod_{\bm{r}} \Big[\frac{e^{-i\phi_y} c^\dagger_1(\bm{r})}{2} + \frac{e^{i\phi_x} c^\dagger_2(\bm{r}+\hat{x}+\hat{y})}{2}  - \frac{c^\dagger_4(\bm{r}+\hat{y})}{\sqrt{2}}\Big]\\&\times \Big[ \frac{e^{-i\phi_x} c^\dagger_1(\bm{r})}{2} - \frac{e^{i\phi_y} c^\dagger_2(\bm{r}+\hat{x}+\hat{y})}{2}  - \frac{c^\dagger_3(\bm{r}+\hat{x})}{\sqrt{2}}]\ket{vac}.
\end{split}\label{eq:MFGround1}
\end{equation}
After integrating over the phase field $\phi_x$ and $\phi_y$, Eq. \ref{eq:MFGround1} reduces to
\begin{equation}
\ket{0} = \prod_{\bm{r}} \frac{1}{\sqrt{2}}\Big[ c^\dagger_1(\bm{r})c^\dagger_2(\bm{r}-\hat{x}+\hat{y}) + c^\dagger_3(\bm{r}+\hat{x}) c^\dagger_4(\bm{r}+\hat{y})\Big]\ket{vac}.\label{eq:IntGroundPi}
\end{equation} 
This is exactly the ground state of the full interacting model when $\mu=t= 0$ (see Eq. \ref{eq:exactGS2}). Here we can directly see that the integration over the phase fields is necessary in order for the ground state of the non-dipole conserving mean field Hamiltonian to match that of the dipole conserving interacting Hamiltonian. We address this topic in more detail in Appendix \ref{app:PhaseFields}. 

We note that this is a very interesting outcome, i.e., if we take the ground state of the free fermion quadrupole model which has vanishing polarization in the ground state because of symmetry quantization, we can couple it to phase fields and after integrating them out we recover the ground state for the ring-exchange quadrupole model which has microscopic dipole conservation at the Hamiltonian level. Ref. \onlinecite{dubinkin2020} showed these two ground states were adiabatically connected in the presence of $C_4$ symmetry if the dipole conservation was relaxed, and so our results forge a new connection between these models. These results open the possibility to use a rank-2 quadrupole calculation to determine the quadrupole moment of free-fermion systems if the polarization-free projection can be carried out as we did above. We leave such a program to future work.

\section{Dipolar Chern-Simons Response Action}\label{sec:ResponseAction}
It is well known that the polarization response of a $1$D system is related to the Chern-Simons response of a $2$D system via dimensional reduction\cite{qi2008}.  Based on this, one would expect that the $2$D quadrupole response in Eq. \ref{eq:EffectiveQuad} is related to a topological response in $3$D. Such a $3$D topological response was proposed in Ref. \onlinecite{you2019}. This response describes a system with anomalous chiral hinge modes, and can be written in terms of three background gauge fields, two rank-1 gauge fields $A_0$, $A_z$, and 1 rank-2 gauge field $A_{xy}$,
\begin{equation}
\begin{split}
S_{\text{dCS}} = \frac{1}{4\pi} \int d^4x[&A_{xy}\partial_z A_0 +A_{0}\partial_z A_{xy} -  A_{xy}\partial_t A_z \\ &-   A_{z}\partial_t A_{xy} + A_z \partial_x \partial_y A_0 - A_0 \partial_x \partial_y A_z ].
\end{split}
\label{eq:ResponseAction}\end{equation}
This action breaks both $C_4$ rotation symmetry around the $z$-axis, and time reversal symmetry $\mathcal{T}$, but is invariant under their product, which we will refer to as $C_4\mathcal{T}$ symmetry. Eq. \ref{eq:ResponseAction} is gauge invariant up to surface terms, which shall be discussed later. Provided that the gauge fields are non-singular, Eq. \ref{eq:ResponseAction} can be written as a sum of total derivatives. Because of this, the response action can be reduced to a sum of boundary terms when defined on a manifold with boundary. The Lagrangians for these boundary terms are
\begin{equation}
\begin{split}
&\mathcal{L}_{dCS,\pm t}= \mp\frac{1}{4\pi} A_z A_{xy}, \phantom{==}\mathcal{L}_{dCS,\pm z}= \pm\frac{1}{4\pi} A_0 A_{xy},\\
&\mathcal{L}_{dCS,\pm x}= \pm\frac{1}{4\pi} A_z \partial_y A_0,\phantom{==} \mathcal{L}_{dCS,\pm y}= \pm\frac{1}{4\pi} A_z \partial_x A_0,\\
&\mathcal{L}_{dCS,\pm x,\pm y} = -\frac{1}{4\pi} A_z A_0,\phantom{==}\mathcal{L}_{dCS,\pm x,\mp y} = \frac{1}{4\pi} A_z A_0.
\end{split}
\label{eq:BoundResponseAction}\end{equation}
Here, $\mathcal{L}_{dCS,\pm\mu}$ ($\mu = x,y,z,t$) is the Lagrangian for the boundaries oriented normal to the $\pm\mu$-direction. Similarly, $\mathcal{L}_{dCS,\pm x,\pm y} $ is the Lagrangian for the hinges between boundaries normal to the $\pm x$ and $\pm y$-direction, while $\mathcal{L}_{dCS,\pm x,\mp y} $ is the Lagrangian for the hinges between boundaries normal to the $\pm x$ and $\mp y$-direction. 

Since the bulk action is a total derivative, all bulk currents vanish, and there are only non-vanishing currents at the boundaries. These boundary currents are anomalous, and their anomalous conservation laws are given by
\begin{equation}
\begin{split}
&\partial_t j^{\pm x}_0 + \partial_z j^{\pm x}_z = \pm \frac{1}{4\pi} [\partial_y \partial_z A_0 - \partial_y\partial_t A_z]= \pm \frac{1}{4\pi} \partial_y E_z,\\
&\partial_t j^{\pm y}_0 + \partial_z j^{\pm y}_z = \pm \frac{1}{4\pi} [\partial_x \partial_z A_0 - \partial_x\partial_t A_z]= \pm \frac{1}{4\pi} \partial_x E_z,\\
&\partial_t j^{\pm z}_t + \partial_x \partial_y  j^{\pm z}_{xy} =\mp \frac{1}{4\pi} [\partial_x \partial_y A_0 - \partial_t A_{xy}] =\mp \frac{1}{4\pi} E_{xy},\\
&\partial_z j^{\pm t}_z + \partial_x \partial_y  j^{\pm t}_{xy} =\pm \frac{1}{4\pi} [\partial_x \partial_y A_z - \partial_z A_{xy}]= \pm \frac{1}{4\pi} B,\\
&\partial_t j^{\pm x\pm y}_0 + \partial_z j^{\pm x \pm y}_{z} = \pm \frac{1}{4\pi}[\partial_z A_0 - \partial_t A_z]= \pm \frac{1}{4\pi} E_z,\\
&\partial_t j^{\pm x\mp y}_0 + \partial_z j^{\pm x \mp y}_{z} = \mp \frac{1}{4\pi}[\partial_z A_0 - \partial_t A_z]= \mp \frac{1}{4\pi} E_z,
\end{split}\label{eq:responseAnomaly}
\end{equation}
where we have introduced the fields $E_{xy} \equiv \partial_x \partial_y A_0 - \partial_t A_{xy}$ and $B \equiv \partial_x \partial_y A_z - \partial_z A_{xy}$, which can be thought of as rank-2 electric and magnetic fields respectively. Here, the superscript $\pm \mu$ ($\mu = x,y,z,t$) indicates that the current is defined on boundaries oriented normal to the $\pm\mu$-direction. Similarly, the superscript $\pm x \pm y$ indicates that the current is defined on hinges between boundaries normal to the $\pm x$ and $\pm y$-direction, while $\pm x \mp  y$ indicates that the current is defined on hinges between boundaries normal to the $\pm x$ and $\mp y$-direction. From Eq. \ref{eq:responseAnomaly}, we see that a rank-2 electric field $E_{xy}$ produces an anomalous dipole current $j_{xy}$ on the boundaries normal to the $\pm z$-direction, and a gradient of the electric field $\partial_i E_z$ ($i=y, x$) produces an anomalous current in the $z$-direction on boundaries normal to the $\pm x$ and $\pm y$-directions respectively. There is also a chiral anomaly-like response on the hinges.

As noted before, Eq. \ref{eq:ResponseAction} is only gauge invariant up to boundary terms. After a gauge transformation $\Lambda$, the gauge variation at the boundaries is given by
\begin{equation}
\begin{split}
&\delta\mathcal{L}_{dCS,\pm t}= \mp\frac{1}{4\pi} \Lambda [\partial_x \partial_y A_{z}-\partial_z A_{xy}],\\
&\delta \mathcal{L}_{dCS,\pm z}= \pm\frac{1}{4\pi} \Lambda[\partial_x \partial_y A_0 - \partial_t A_{xy}],\\
&\delta \mathcal{L}_{dCS,\pm x}= \pm\frac{1}{4\pi} \Lambda[\partial_y \partial_t A_z - \partial_y \partial_z A_0 ],\\&\delta \mathcal{L}_{dCS,\pm y}= \pm\frac{1}{4\pi} \Lambda[\partial_x \partial_t A_z - \partial_x \partial_z A_0 ],\\
&\delta\mathcal{L}_{dCS,\pm x,\pm y} = -\frac{1}{4\pi} \Lambda [\partial_t A_z - \partial_z A_0],\\&
\delta \mathcal{L}_{dCS,\pm x,\mp y} = \frac{1}{4\pi}  \Lambda [\partial_t A_z - \partial_z A_0].
\end{split}
\label{eq:BoundGaugeVariation}\end{equation}
In order for the full theory to be gauge invariant, there must be additional degrees of freedom located at the boundaries and hinges in order to restore gauge invariance. This is similar to what occurs when the $2$D Chern-Simons action is defined on a manifold with boundary. Specifically, due to the hinge terms $\delta \mathcal{L}_{dCD,\pm x \pm y}$, there must be chiral modes that propagate along hinges between boundaries normal to the $\pm x$ and $\pm y$-direction in order for the theory to be gauge invariant. Similarly, due to $\delta\mathcal{L}_{dCD,\pm x \mp y}$   there must be anti-chiral modes that propagate along hinges between boundaries normal to the $\pm x$ and $\mp y$-direction.
We can therefore conclude that a consistent theory described by Eq. \ref{eq:ResponseAction} must have chiral hinge modes. Additionally, due to the boundary terms, $\mathcal{L}_{\pm x}$ and $\mathcal{L}_{\pm y}$, there must also be additional modes at boundaries normal to the $\pm x$ and $\pm y$-directions in order for the theory to be gauge invariant.
 
The dipolar Chern-Simons response also predicts a quadrupole analog of the Laughlin pump. The typical quantum Hall Laughlin pumping process can be observed by considering a $2$D Chern-Simons theory defined on cylinder. When a unit of flux is inserted through this cylinder, the Hall current pumps charge from one end of the cylinder to the other, and changes the dipole moment by $1$. A similar process occurs when the dipolar Chern-Simons term is defined on an annulus with periodic boundary conditions in the $z$-direction. In this case, when a unit of flux is inserted in the $z$-direction, Eq. \ref{eq:ResponseAction} predicts that the $xy$ quadrupole moment of the system will increase by $1$. This process can analogously be thought of as dipole pumping, and will change the amount of charge located at the hinges of the system.

Finally, we would like to confirm that the dipolar Chern-Simons action is in fact related to the $2$D quadrupole response by dimensional reduction. To do this, we will consider an arbitrarily thin annulus with periodic boundary conditions in the $z$-direction. In this limit, we can treat the flux $\theta$ passing through the $z$-direction as an adiabatic parameter of a $2$D theory, and the dipole analog of the Laughlin pump becomes a dipole analog of a Thouless pump. As $\theta$ is increased, the quadrupole moment of the $2$D system is shifted due to an adiabatic pumping of dipole moment across the system. This is exactly the phenomenology we saw before when considering the quadrupole moment in Sec. \ref{sec:DimRed}. 

To show this formally, we can dimensionally reduce Eq. \ref{eq:ResponseAction} by setting $A_{z} = \Theta/L_z$, and taking the limit $L_z \rightarrow 0$. If we take $\Theta$ to be constant, the dimensionally reduced action is given by
\begin{equation}
\begin{split}
S_{dCS,2D} = \int d^3x \frac{\Theta}{4\pi} [ \partial_x \partial_y A_0- \partial_t A_{xy}].
\end{split}
\label{eq:2dResponseAction}\end{equation}
However, as discussed in Ref. \onlinecite{you2019}, the term in Eq. \ref{eq:2dResponseAction} is only half the quadrupole response of the dimensionally reduced system. The other half comes from the boundary degrees of freedom we discussed earlier. Taking both of these contributions into account, the full response of the dimensionally reduced system is given by
\begin{equation}
\begin{split}
S_{dCS,2D} = \int d^3x \frac{\Theta}{2\pi} [ \partial_x \partial_y A_0- \partial_t A_{xy}],
\end{split}
\label{eq:2dResponseAction2}\end{equation}
which is the quadrupole response from Eq. \ref{eq:EffectiveQuad} with $Q_{xy} = \Theta /2\pi$.

\section{$C_4\mathcal{T}$ Symmetric Lattice model} \label{sec:LatticeC}
\begin{figure}\centering
\includegraphics[width=.4\textwidth]{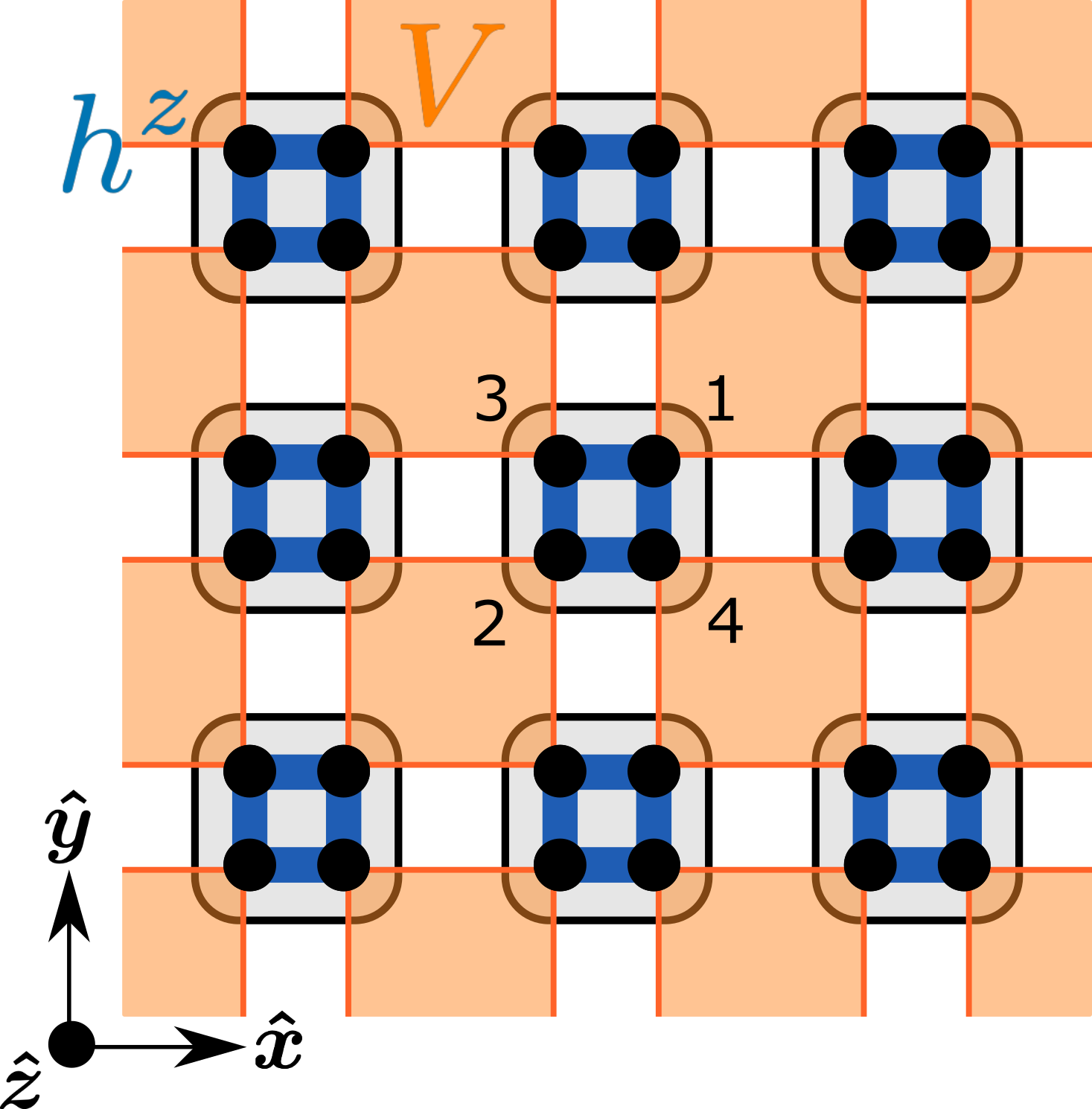}
\caption{A cross section of the $3$D lattice model Eq. \ref{eq:LatticeHamC}. The Hamiltonian consists of intracell and $z$-direction hopping terms $h^z$ (blue) and a ring exchange term $V$ in the $xy$ plane (orange).}\label{fig:LatDia}
\end{figure}

Since the $2$D quadrupole response is related to the $3$D dipolar Chern-Simons response via dimensional reduction, the $2$D dipole conserving insulator presented in Sec. \ref{sec:DimRed} should be similarly related to a chiral hinge insulator in $3$D that realizes the dipolar Chern-Simons response. To go from the $2$D model to the $3$D model, we will proceed in the usual fashion of identifying the adiabatic parameter $\theta$ (which controls the dipole pumping of Eq. \ref{eq:LatticeHam2}) with the momentum along the $z$-direction, $k_z$. A local Hamiltonian should depend smoothly on the momentum $k_z$, and so we will consider the parameterization given in Eq. \ref{eq:adParameters2}, which is fully continuous with respect to $\theta$. If we substitute $\theta \rightarrow k_z$, the resulting $3$D Hamiltonian is 
\begin{equation}
\begin{split}
H^{dCS} &= \sum_{k_z, \bm{r}} \bm{c}^\dagger(\bm{r}) h^z(k_z) \bm{c}(\bm{r})\\&\phantom{=} - V c_1^\dagger(\bm{r})c^\dagger_2(\bm{r}+\hat{x}+\hat{y})c_3(\bm{r}+\hat{x})c_4(\bm{r}+\hat{y}) \\&\phantom{=}+ h.c.,\\
h^z(k_z) &= \gamma \sin(k_z) \Gamma_0 + \gamma[1+\cos(k_z)](\Gamma_2+\Gamma_4).
\end{split}\label{eq:LatticeHam3}
\end{equation}
An illustration of this Hamiltonian is shown in Fig. \ref{fig:LatDia}. Eq. \ref{eq:LatticeHam3} is invariant under phase shifts that depend linearly on the $x$ and $y$ coordinates, $\bm{c}(\bm{R})\rightarrow \bm{c}(\bm{R}) e^{ i (\beta_1 x + \beta_2 y )}$, and conserves dipole in both the $x$ and $y$-directions. However, due to the explicit dependence of $h^z$ on $k_z$, the Hamiltonian does not conserve dipole in the $z$-direction. Eq. \ref{eq:LatticeHam3} also has $C_4\mathcal{T}$ symmetry and $C_2$ spatial rotations in the $xy$ plane. 

As expected from our discussion of the dipolar Chern-Simons response, Eq. \ref{eq:LatticeHam3} can be coupled to the  background gauge fields $A_0$, $A_z$, and $A_{xy}$. In real space, the minimally coupled Hamiltonian is given by 
\begin{equation}
\begin{split}
H^{dCS} =\sum_{\bm{R}} &[\gamma \bm{c}^\dagger(\bm{R}) T \bm{c}(\bm{R}) + \gamma \bm{c}^\dagger(\bm{R}+\hat{z}) T^z \bm{c}(\bm{R})e^{i A_z(\bm{R})} \\
  -& V c_1^\dagger(\bm{R})c_2^\dagger(\bm{R}+\hat{x}+\hat{y})c_3(\bm{R}+\hat{x})c_4(\bm{R}+\hat{y})e^{i A_{xy}(\bm{R})} \\
-& A_0(\bm{R})  \bm{c}^\dagger(\bm{R}) \bm{c}(\bm{R})   +h.c. ],\\
T =&  \frac{1}{2}[\Gamma_2 + \Gamma_4], \phantom{==} T^z = \frac{1}{2}[i \Gamma_0 +  \Gamma_2 + \Gamma_4],
\end{split}\label{eq:LatticeHamC}
\end{equation}
where $\bm{R} \equiv (x,y,z)$ labels the points of a $3$D cubic lattice. The matrix $T$ is made up of the intracell terms, and the matrix $T^z$ is made up of intercell hopping terms in the $z$-direction. In this form, it is clear that $T^z$ violates dipole conservation in the $z$-direction.

\subsection{Mean Field Analysis}\label{ssec:MFC}
To analyze the physics of the interacting $C_4\mathcal{T}$ symmetric lattice model in Eq. \ref{eq:LatticeHam3}, we will employ a self-consistent mean field theory approach. This will largely mirror the mean field analysis of the $2$D dipole conserving model in Sec. \ref{ssec:ContDipolePumping}. As we shall show, within the self-consistent mean field framework the 4 particle ring exchange interaction $V$ causes a (Mott) gap to form for the lattice fermions. Additionally, the resulting mean field Hamiltonian exactly maps on to a known non-interacting HOTI with chiral hinge modes. 

\begin{figure}
\includegraphics[width=.5\textwidth]{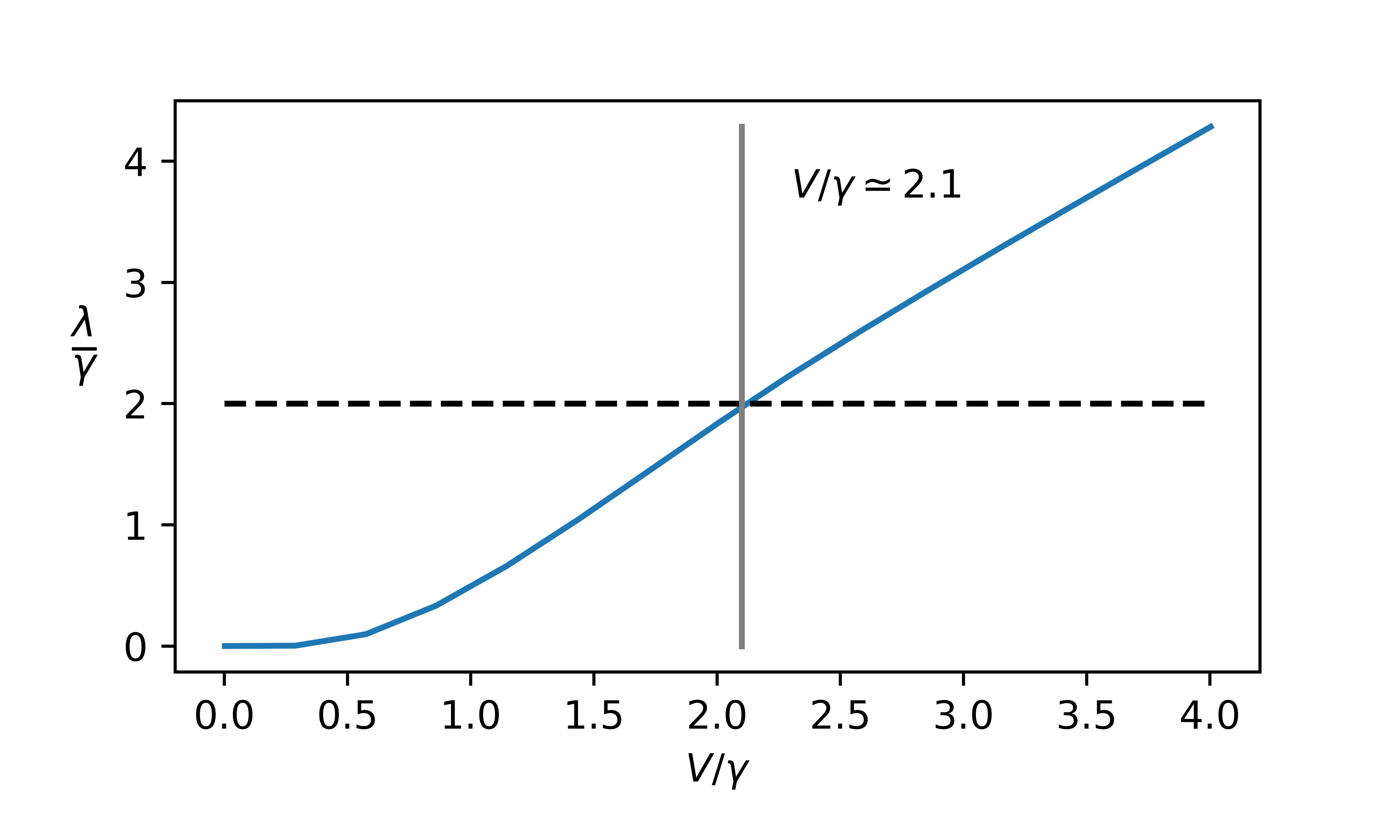}
\includegraphics[width=.4\textwidth]{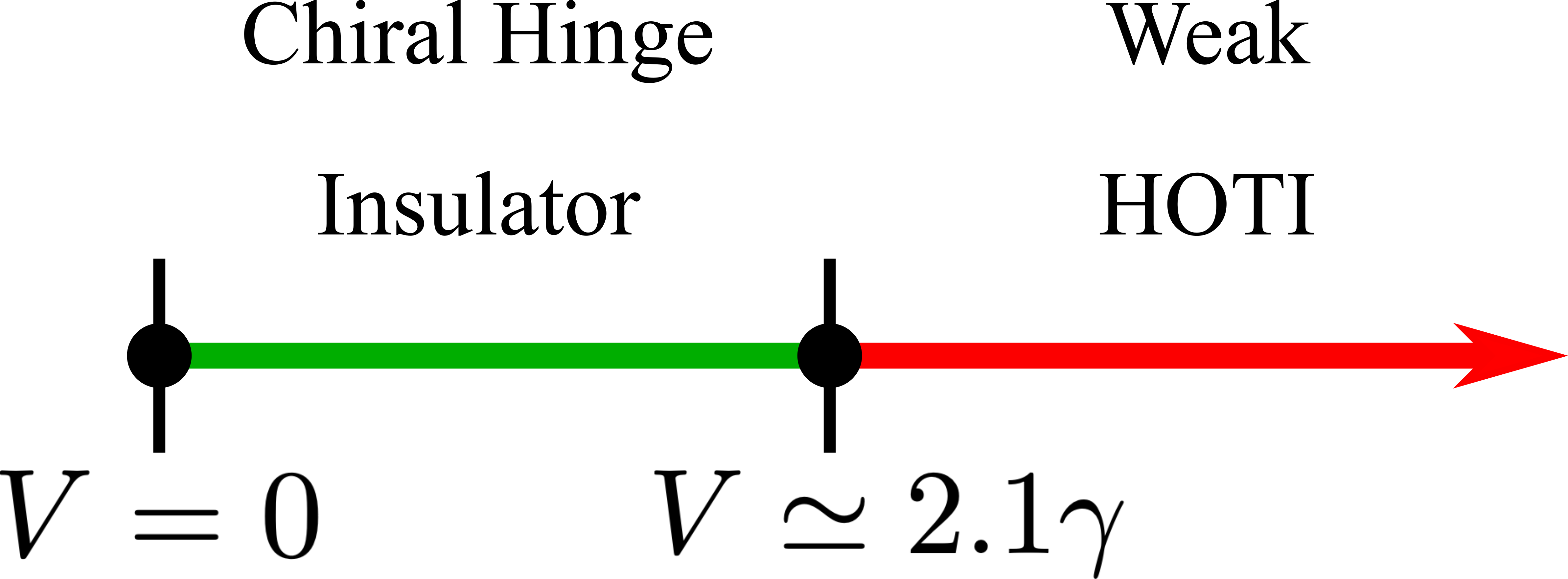}
\caption{Top: The solutions to the self-consistent solutions for $\lambda/\gamma$ as a function of $V/\gamma$. When $V/\gamma \simeq 2.1$ (grey line), $\lambda/\gamma = 2$ (dashed line) and the mean field model undergoes a phase transition from a topological chiral hinge insulator to a weak HOTI. Bottom: The mean field phase diagram of Eq. \ref{eq:LatticeHamC}.}
\label{fig:SelfConsistentPlot}
\end{figure}

Similar to Sec. \ref{ssec:ContDipolePumping}, we will decompose the quartic ring exchange terms via a Hubbard-Stratonovich transformation,
\begin{equation}
\begin{split}
&-V c_1^\dagger(\bm{R})c_2^\dagger(\bm{R}+\hat{x}+\hat{y})c_3(\bm{R}+\hat{x})c_4(\bm{R}+\hat{y})e^{iA_{xy}(\bm{R})} \\ &\rightarrow \lambda_{1x}(\bm{R})c_2^\dagger(\bm{R}+\hat{x}+\hat{y})c_4(\bm{R}+\hat{y})\\ &\phantom{\rightarrow}+ \lambda_{2x}(\bm{R}) c_1^\dagger(\bm{R})c_3(\bm{R}+\hat{x}) \\&\phantom{\rightarrow}+\lambda_{1y}(\bm{R})c_2^\dagger(\bm{R}+\hat{x}+\hat{y})c_3(\bm{R}+\hat{x})\\ &\phantom{\rightarrow}- \lambda_{2y}(\bm{R})c_1^\dagger(\bm{R})c_4(\bm{R}+\hat{y})\\ &\phantom{\rightarrow}- \frac{2}{V}\lambda_{1x}(\bm{R})\lambda_{2x}(\bm{R})e^{-iA_{xy}(\bm{R})}\\ &\phantom{\rightarrow} - \frac{2}{V}\lambda_{1y}(\bm{R})\lambda_{2y}(\bm{R})e^{-iA_{xy}(\bm{R})}. 
\end{split}\label{eq:LatticeHS3D}
\end{equation}
The equations of motion are exactly those in Eq. \ref{eq:HSEqOM} (upon exchange the $2$D lattice coordinate $\bm{r}$ with the $3$D lattice coordinate $\bm{R}$). Under a linear phase shift $\bm{c}(\bm{R})\rightarrow \bm{c}(\bm{R}) e^{ i (\beta_1 x + \beta_2 y )}$, the Hubbard-Stratonovich fields transform as
\begin{equation}
\begin{split}
& \lambda_{1x}(\bm{R})\rightarrow \lambda_{1x}(\bm{R})e^{i\beta_1}, \phantom{==} \lambda_{2x}(\bm{R})\rightarrow \lambda_{2x}(\bm{R})e^{-i\beta_1},\\ 
&\lambda_{1y}(\bm{R})\rightarrow \lambda_{1y}(\bm{R})e^{i\beta_2}, \phantom{==} \lambda_{2y}(\bm{R})\rightarrow \lambda_{2y}(\bm{R})e^{-i\beta_2}.
\end{split}\label{eq:HSLinearTransformation2}
\end{equation}
As shown in Appendix \ref{app:selfCon}, in the self-consistent mean field theory approximation the Hubbard-Stratonovich fields acquire expectation values of the form
\begin{equation}
\begin{split}
&\lambda_{1x}(\bm{R})=  \lambda e^{i\phi_x(\bm{R})+iA_{xy}(\bm{R})},\\
&\lambda_{2x}(\bm{R})=  \lambda e^{-i\phi_x(\bm{R})},\\
&\lambda_{1y}(\bm{R})=  \lambda e^{i\phi_y(\bm{R})+iA_{xy}(\bm{R})},\\
&\lambda_{2y}(\bm{R})=  \lambda e^{-i\phi_y(\bm{R})},
\end{split}\label{eq:HSSelfConsistent3DC}
\end{equation}
where the phase fields $\phi_x$ and $\phi_x$ satisfy the relationship
\begin{equation}
\begin{split}
\Delta_y\phi_x(\bm{R}) = \Delta_x\phi_y(\bm{R}) = A_{xy}(\bm{R}).
\end{split}\label{eq:PhaseCont3DC}
\end{equation}
The self-consistent values of $\lambda$ depend on $V/\gamma$, and are shown in Fig. \ref{fig:SelfConsistentPlot}. As before, under a gauge transformation $\Lambda$, the phase fields gauge transform as $\phi_i \rightarrow \phi_i + \Delta_i \Lambda$.

For vanishing gauge fields ($A_0 = A_z = A_{xy} = 0$) the mean field Hamiltonian can be written in Fourier space as
\begin{equation}
\begin{split}
&H^{dCS}_{\text{mf}} = \sum_{\bm{k}} \Big[\bm{c}^\dagger(\bm{k}) h^{dCS}_{\text{mf}}(\bm{k}) \bm{c}(\bm{k})\Big], \\
&h^{dCS}_{\text{mf}}(\bm{k}) = \gamma \sin(k_z) \Gamma_0 +\gamma[1+\cos(k_z)] (\Gamma_2+\Gamma_4) \\
&\phantom{====}+ \lambda \big[\cos(k_x+\phi_{x})\Gamma_4 + \sin(k_x + \phi_{x}) \Gamma_3\\
&\phantom{====}+ \cos(k_y+\phi_{y})\Gamma_2 + \sin(k_y + \phi_{y}) \Gamma_1\big],
\end{split}\label{eq:LatticeHSGauge}
\end{equation}
where $\bm{k} = (k_x, k_y,k_z)$ is the momentum of the $3$D lattice model. As a consistency check, we note that Eq. \ref{eq:LatticeHSGauge} is related to the $2$D mean field model in Eq. \ref{eq:DimRedLatticeHSGauge} by dimensional reduction.  The single-particle energy spectrum of the mean field Hamiltonian is given by 
\begin{equation}
\begin{split}
\epsilon(\bm{k}) = &\pm \big[ 7\gamma^2 + 4\lambda^2 + 8\gamma^2\cos(k_z) + \gamma^2\cos(2k_z) \phantom{\big) }\\ &\phantom{\big( } + 4
\lambda \gamma \cos(k_x+\phi_x)(\cos(k_z)+1)\phantom{\big) }\\ &\phantom{\big( } + 4
\lambda \gamma \cos(k_y+\phi_y)(\cos(k_z)+1) \big]^{1/2}.
\end{split}
\end{equation} 
When $\lambda/\gamma = 2$, the system is gapless at $(k_x,k_y,k_z) = (\pi-\phi_x,\pi-\phi_y,0)$, and when $\lambda = 0$, the system is gapless at $k_z = \pi$ for all values of $k_x$ and $k_y$. As noted in Sec. \ref{ssec:ContDipolePumping}, integrating over the dynamic phase fields $\phi_x$ and $\phi_y$ projects out any states with non-vanishing dipole moment in the $x$ and $y$-directions.

In Sec. \ref{ssec:ContDipolePumping}, we found that in the mean field limit the $2$D dipole conserving model is equivalent to the non-interacting quadrupole insulator. Here, we see that the mean field Hamiltonian in Eq. \ref{eq:LatticeHSGauge} is equivalent to a non-interacting chiral hinge insulator, which was also considered in Ref. \onlinecite{benalcazar2017b}, and is related to the non-interacting quadrupole insulator by dimensional reduction. Based on this, we shall quote several key results concerning the mean field model. When $0<\lambda/\gamma < 2$ the mean field Hamiltonian describes a topological chiral hinge insulator. In this phase, the model has modes that propagate in the $z$-direction on hinges between boundaries normal to the $x$  and $y$-directions. In the corresponding non-interacting HOTI, these chiral hinge modes correspond to the non-trivial Wannier-band Chern number of the system. For $2<\lambda/\gamma$, there are no chiral hinge modes. This phase is connected to the $V\rightarrow \infty$ limit of Eq. \ref{eq:LatticeHam3}. In this limit, the ring exchange term gaps out each $xy$-layer of the model individually. This suppresses all tunneling in the $z$-direction and the $3D$ system becomes a  stack of $2$D insulators. Based on our analysis in Sec. \ref{ssec:ExactSolvablePump}, each of these layers will have quadrupole moment $1/2$, and the system can be adiabatically connected to an atomic insulator by breaking translational symmetry along the $z$-direction, so we might refer to it as a weak higher order topological insulator. At the phase transition connecting these two phases ($\lambda/\gamma = 2$), there is a bulk band crossing and the system is gapless. Using the numerical solutions to the self-consistent mean field equations, we find that this phase transition occurs at $V \equiv V_c \simeq 2.1\gamma$. The mean field phase diagram is shown in Fig. \ref{fig:SelfConsistentPlot}.

Having determined the mean field structure of the lattice model, we can now turn our attention to determining the effective response action for this system. As noted in Sec. \ref{sec:ResponseAction}, the dipolar Chern-Simons action can be written as a sum of total derivatives, and the only non-trivial responses occur at the boundaries. Because of this, we will need to define our model on a lattice with boundary in order to show that it exhibits a dipolar Chern Simons response. As we will show, analyzing the boundaries of our model is made tractable upon passing to the continuum in the $z$ and $t$-directions. This analysis will be the main topic of the next subsection.

\subsection{Continuum Analysis} \label{ssec:ContAnalysis}
In this subsection, we will study the $3$D model given in Eq. \ref{eq:LatticeHamC} in the continuum, near $V = 0$.  At $V = 0$, the system is quasi-1d, i.e., a decoupled $3$D array of $1$D wires oriented in the $z$-direction. If we pass to the continuum along the $z$-direction, each of these wires consists of 4 massless Dirac fermions (which correspond to the fluctuations of the lattice fermions with momentum near $k_z = \pi$), and 4 massive fermions (which correspond to the fluctuations of the lattice fermions with momentum near $k_z = 0$). We can split lattice fermions into these two contributions using the identification 
\begin{equation}
c_i(\bm{R}) \rightarrow \Psi_i(\bm{R}) + (-1)^z \psi_{i}(\bm{R}),
\label{eq:Lat2ContFermions}
\end{equation}
where $\Psi_i$ are the heavy fermions and $\psi_i$ are the light fermions. At $V = 0$, the continuum Lagrangian for these fields is 
\begin{equation}
\begin{split}
&\mathcal{L}^{dCS} = \sum_{\bm{r}} \left[\bm{\psi}^\dagger(\bm{r})G^{-1}_0(\omega+A_0(\bm{r}),p_z+A_z(\bm{r}))\bm{\psi}(\bm{r})\right.\\ &\phantom{====} +\left.\bm{\Psi}^\dagger(\bm{r})G^{-1}_M(\omega+A_0(\bm{r}),p_z+A_z(\bm{r}))\bm{\Psi}(\bm{r})\right], \\
&G^{-1}_0(\omega,p_z) =  \omega \text{I} + p_z \Gamma^0,\\
&G^{-1}_M(\omega,p_z) =  \omega \text{I} - p_z \Gamma^0 - \tfrac{M}{\sqrt{2}}(\Gamma^2 + \Gamma^4),
\end{split}\label{eq:ContLag1}
\end{equation}
where $\text{I}$ is the $4\times 4$ identity matrix, $\bm{\psi} = (\psi_1,\psi_2,\psi_3,\psi_4),$ and $\bm{\Psi} = (\Psi_1,\Psi_2,\Psi_3,\Psi_4)$. Here, we have only passed to the continuum in the $z$-direction. The $x$ and $y$ coordinates are still lattice coordinates. Based on Eq. \ref{eq:ContLag1}, when $V = 0$, $\psi_1$ and $\psi_2$ are anti-chiral fermions and $\psi_3$ and $\psi_4$ are chiral fermions (with respect to their propagation along the $z$-direction). Additionally, the heavy fermions $\Psi$ are massive with mass $M$. We will also take $M$ to be the UV cutoff for this theory. 

We can now consider the interactions in this theory. As before, we will use the Hubbard-Stratonovich transformation in Eq. \ref{eq:LatticeHS3D} to decompose the lattice ring exchange interaction. Using Eq. \ref{eq:Lat2ContFermions} we find that the Hubbard-Stratonovich fields $\lambda_{1/2,x}$ $\lambda_{1/2,y}$ couple to both the light fermions $\psi$ and the heavy fermions $\Psi$. Since the heavy fermions $\Psi$ are gapped, they can be integrated out, leaving a Lagrangian in terms of the light fermions and the Hubbard-Stratonovich fields. After integrating out the heavy fermions, the Lagrangian can be written as
\begin{equation}
\mathcal{L}^{dCS} = \mathcal{L}_{\psi} + \mathcal{L}_{\lambda\lambda}, 
\end{equation}
where $\mathcal{L}_{\psi}$ contains all terms involving the light fermions $\psi,$ and $\mathcal{L}_{\lambda\lambda}$ contains all couplings between the Hubbard-Stratonovich fields. If we ignore any terms that osculate like $(-1)^z$, $\mathcal{L}_{\psi}$ is given by
\begin{equation}
\begin{split}
\mathcal{L}_{\psi} = &\sum_{\bm{r}} \left[\bm{\psi}^\dagger(\bm{r})G^{-1}_0(\omega+A_0(\bm{r}),p_z+A_z(\bm{r}))\bm{\psi}(\bm{r})\right.\\& - [ \lambda_{2x}(\bm{r}) \psi_1^\dagger(\bm{r})\psi_3(\bm{r}+\hat{x}) \\
&\phantom{=} + \lambda_{1x}(\bm{r})\psi_2^\dagger(\bm{r}+\hat{x}+\hat{y})\psi_4(\bm{r}+\hat{y}) \\
&\phantom{=}+\lambda_{2y}(\bm{r})\psi_1^\dagger(\bm{r})\psi_4(\bm{r}+\hat{y})\\
&\phantom{=}+\left.\lambda_{1y}(\bm{r})\psi_2^\dagger(\bm{r}+\hat{x}+\hat{y})\psi_3(\bm{r}+\hat{x}) + h.c.]\right].
\end{split}\label{eq:ContHS}
\end{equation}
To determine $\mathcal{L}_{\lambda\lambda}$ we must integrate out the heavy fermions $\Psi$. At one loop order, $\mathcal{L}_{\lambda\lambda}$ is given by
\begin{equation}
\begin{split}
\mathcal{L}_{\lambda\lambda} = &\sum_{\bm{r}} -u_1 [\lambda_{1x}(\bm{r})\lambda^*_{1x}(\bm{r})+ \lambda_{2x}(\bm{r})\lambda^*_{2x}(\bm{r}) \\
&+ \lambda_{1y}(\bm{r})\lambda^*_{1y}(\bm{r})+ \lambda_{2y}(\bm{r})\lambda^*_{2y}(\bm{r})] \\
&+ [u_2 \lambda_{1x}(\bm{r})\lambda_{2x}(\bm{r}+\hat{y}) + u_2 \lambda_{1y}(\bm{r})\lambda_{2y}(\bm{r}+\hat{x}) \\
&\phantom{=} + \frac{2}{V}\lambda_{1x}(\bm{r})\lambda_{2x}(\bm{r})e^{-iA_{xy}(\bm{r})}\\ &\phantom{=} + \frac{2}{V}\lambda_{1y}(\bm{r})\lambda_{2y}(\bm{r})e^{-iA_{xy}(\bm{r})}
+ h.c.],
\end{split}\label{eq:AdditionalInteractions}
\end{equation}
where $u_1 = \frac{\log(4)-1}{16\pi},$ and $u_2 = \frac{1}{16\pi}$. 

As before, we shall employ the self-consistent mean field theory approximation. We find the Hubbard-Stratonovich fields acquire expectation values of the form
\begin{equation}
\begin{split}
\lambda_{1x}(\bm{r}) &= \lambda e^{i\phi_x(\bm{r}) + iA_{xy}(\bm{r})},\\
\lambda_{2x}(\bm{r}) &= \lambda e^{-i\phi_x(\bm{r}) },\\
\lambda_{1y}(\bm{r}) &= \lambda e^{i\phi_y(\bm{r}) + iA_{xy}(\bm{r})},\\
\lambda_{2y}(\bm{r}) &= \lambda e^{-i\phi_y(\bm{r})}.\\
\end{split}\label{eq:ContSelfConsist}
\end{equation}
The value of $\lambda$ is determined by the effective potential
\begin{equation}
\begin{split}
H_{\lambda\lambda} = &\frac{8}{V'} \lambda^2 + \frac{1}{2\pi}\lambda^2\Big(\log(\frac{2\lambda^2}{M^2})-1\Big).
\end{split}
\end{equation} 
where $V' = (V^{-1} + \frac{u_1+u_2}{2} )^{-1}$. The effective potential is minimized by $\lambda = \frac{1}{\sqrt{2}} M e^{-\frac{8\pi}{V'}}$. In agreement with the numerical results, we find that $\lambda$ vanishes when $V \rightarrow 0,$ and it increases monotonically with increasing $V$. Additionally, due to the $u_2$ term in Eq. \ref{eq:AdditionalInteractions}, at low energies the phase fields $\phi_i$ satisfy
\begin{equation}
\begin{split}
 \Delta_y\phi_x(\bm{r})  =\Delta_x\phi_y(\bm{r}) = A_{xy}(\bm{r}).\\
\end{split}\label{eq:ContPhaseConstraint}
\end{equation}
These results agree with the lattice mean field results from Sec. \ref{ssec:MFC}. Here, as before, under a gauge transformation $\Lambda$, the phase fields $\phi_i$ transform as $\phi_i(\bm{r}) \rightarrow \phi_i(\bm{r}) + \Delta_i \Lambda(\bm{r})$. 

\begin{figure}\centering
\includegraphics[width=.4\textwidth]{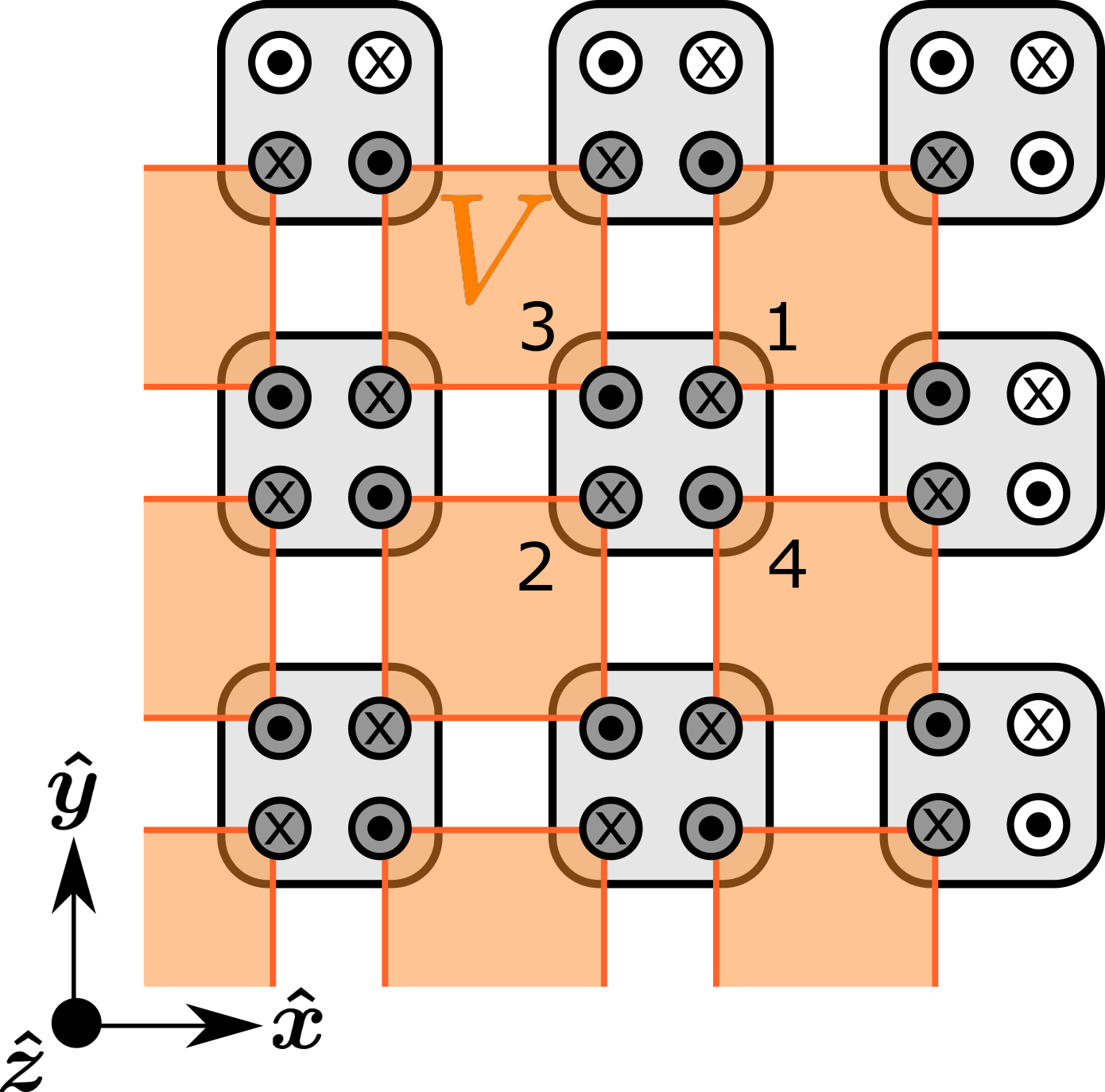}
\caption{The continuum model Eq. \ref{eq:ContLag1}. In each unit cell there are 2 chiral ($\bullet$) and 2 anti-chiral ($\times$) modes. The ring exchange interaction (orange), gaps out these 4 modes (grey circles) at each plaquette. The modes along the edges remain gapless (white circles), but can be gapped when symmetry-preserving perturbations are added. An odd number of chiral hinge modes is stabilized on each corner of the $xy$ plane.}\label{fig:ContinuumDia}
\end{figure}

Based on Eq. \ref{eq:ContHS} we see that in the mean field limit, the quartet fermion cluster $\psi_1(\bm{r})$, $\psi_2(\bm{r}+\hat{x}+\hat{y})$, $\psi_3(\bm{r} + \hat{x})$, and $\psi_4(\bm{r}+\hat{y})$ couples to one another for each value of in plane lattice coordinate ${\bm{r}}.$ Hence, the continuum Lagrangian decouples into 4-fermion clusters, which are defined on each plaquette of the $xy$-plane. This feature will make it possible for us to analytically consider boundaries normal to the $x$ and $y$-directions, which we  do in the next section. For $\lambda \neq 0$, these fermions all become massive (see Eq. \ref{eq:ContHS}). From this, we can explicitly confirm that this model has chiral hinge modes. As we show in Fig \ref{fig:ContinuumDia}, at the top right hinge there is a net chiral mode. There are also similar hinge modes at the other hinges of the system. In addition, there is both a chiral mode and an anti-chiral mode located at each at each lattice site along the boundaries normal to the $x$ and $y$-direction (see Fig \ref{fig:ContinuumDia}). Since these modes all come in pairs they can be gapped out with local symmetry-persevering perturbations. 

\subsection{Effective Response Theory}\label{ssec:ContEffectiveResponse}
We will now turn our attention to finding the bulk and boundary responses for the $3$D $C_4\mathcal{T}$ symmetric chiral hinge insulator. The response action for this system is composed of terms that depend on the three background gauge fields, $A_0$, $A_z$ and $A_{xy},$ as well as the two phase fields $\phi_x$ and $\phi_y$. The phase fields must be included since they depend on the background gauge field $A_{xy}$ (see Eq. \ref{eq:ContPhaseConstraint}). The coefficients for the various terms in the response action are determined by the current-current correlation functions\cite{fradkin2013}. Here the current-current correlation functions are determined using the continuum mean field Lagrangian from Sec. \ref{ssec:ContAnalysis}.

First, we shall consider the effective response action for the bulk of the system. In the limit of low frequency and momentum, the effective Lagrangian for the bulk is
\begin{equation}
\begin{split}
\mathcal{L}^{dCS}_{eff, \text{bulk}} =&  (2A_{xy} -  \partial_y \phi_x - \partial_x \phi_y  ) \partial_t A_z\\ -& (2A_{xy} -  \partial_y \phi_x - \partial_x \phi_y  ) \partial_z A_0\\
=& 0,
\end{split}\label{eq:EffectiveActionBulk}
\end{equation}
where we have passed to the continuum in the $x$ and $y$-directions, and used  Eq. \ref{eq:ContPhaseConstraint} in the last line. This result is consistent with the dipolar Chern-Simons response action given in Eq. \ref{eq:ResponseAction}, since that response action is a total derivative.

To probe the non-vanishing the boundary effects, we can consider the mean field Lagrangian defined on a lattice with boundary. As noted in Sec. \ref{ssec:ContAnalysis}, there are gapless fermions at the boundaries, but they can be gapped out with symmetry preserving perturbations. Upon doing so, we can find that the effective response Lagrangian for the background gauge fields and the phase fields at low frequency and momentum is given by
\begin{equation}
\begin{split}
L^{dCS}_{eff,\pm y} = &\pm \frac{1}{4\pi} [A_0\partial_x A_z +\phi_x\partial_zA_0  - \phi_x \partial_tA_z ], \\
L^{dCS}_{eff,\pm x} = &\pm \frac{1}{4\pi} [A_0\partial_y A_z + \phi_y \partial_zA_0  - \phi_y\partial_tA_z ].
\end{split}\label{eq:EffectiveActionBound}
\end{equation}
The first term on the right hand sides of each line of Eq. \ref{eq:EffectiveActionBound} is exactly the boundary term of the dipolar Chern-Simons response action. The remaining two terms are the couplings between the gauge fields $A_0$ and $A_z$ and the phase fields $\phi_x$ and $\phi_y$. With the addition of the phase fields, the boundary Lagrangian is gauge invariant up to hinge terms. 

Finally, we can consider the action on the hinges. As noted before, there is a gapless chiral mode at the $\pm x, \pm y$ hinges and a gapless anti-chiral mode at the $\pm x, \mp y$ hinges.  If we ignore these gapless fermions, the effective hinge Lagrangian is given by
\begin{equation}
\begin{split}
L^{dCS}_{eff,\pm x, \pm y} =& -\frac{1}{4\pi}A_0 A_z, \\
L^{dCS}_{eff,\pm x, \mp y} =& + \frac{1}{4\pi}A_0 A_z .\\
\end{split}\label{eq:EffectiveActionHinge}
\end{equation}
This is exactly the hinge term from the dipolar Chern-Simons response. From Eq. \ref{eq:EffectiveActionBulk}-\ref{eq:EffectiveActionHinge} we can confirm that the effective action of Eq. \ref{eq:LatticeHam3} matches the dipolar Chern-Simons action in the bulk as well as boundaries normal to the $x$ and $y$-directions, and the hinges that separate them. 

It will also be useful to consider how the effective Lagrangian transforms under a gauge transformation. Using Eq. \ref{eq:EffectiveActionBound} and \ref{eq:EffectiveActionHinge}, we find that under a gauge transformation $\Lambda$, the effective Lagrangian is shifted by the hinge term
\begin{equation}
\begin{split}
&\delta\mathcal{L}^{dCS}_{eff,\pm x,\pm y} = -\frac{1}{4\pi} \Lambda [\partial_t A_z - \partial_z A_0],\\&
\delta \mathcal{L}^{dCS}_{eff,\pm x,\mp y} = \frac{1}{4\pi}  \Lambda [\partial_t A_z - \partial_z A_0],
\end{split}
\label{eq:EffectiveHingeGaugeVariation}\end{equation}
which is the same hinge term given in Eq. \ref{eq:BoundGaugeVariation}. Gauge invariance is restored by the aforementioned chiral hinge modes of the system (which were ignored in our derivation of the effective Lagrangian). It is well known that a single chiral fermion is not gauge invariant, and the gauge variation of a single chiral fermion exactly cancels out the gauge variation $\delta L_{eff,\pm x, \pm y} $\cite{naculich1988axionic,chandrasekharan1994}. Similarly, the gauge variation of a single anti-chiral fermion exactly cancels out the gauge variation of $L_{eff,\pm x, \mp y} $. From this, we can conclude that the lattice degrees of freedom (the phase fields and the chiral hinge modes) cancel out the gauge anomalies of the dipolar Chern-Simons action (Eq. \ref{eq:BoundGaugeVariation}) and make the full theory gauge invariant, as desired.

As a final point, we note that in this analysis we have relied on translational invariance in the $t$ and $z$-directions.  Because of this we cannot make any statements about the boundaries normal the $t$ or $z$-directions. Determining how to analyze these boundaries is an interesting topic for further research. In particular, due to the coupling to the rank-2 gauge field $A_{xy}$, boundaries normal to the $z$-direction may host interesting and exotic physics.

\section{$M \mathcal{T}$ Symmetric Lattice model} \label{sec:LatticeM}
In Sec. \ref{sec:LatticeC} we considered a dipole conserving $C_4 \mathcal{T}$ symmetric $3$D model with chiral hinge modes. In this section, we will consider a related dipole conserving $3$D model, which breaks $C_4 \mathcal{T}$ symmetry, but is instead invariant under $M_x \mathcal{T}$ and $M_y \mathcal{T}$. For simplicity we shall use the shorthand $M\mathcal{T}$ to refer to both of the symmetries. As we shall show, the $M\mathcal{T}$ model also displays chiral hinge modes, similar to those of the $C_4 \mathcal{T}$ symmetric model, but has other interesting features that distinguish it from the previous case. 

Here, we shall consider the Hamiltonian
\begin{equation}
\begin{split}
H^{M\mathcal{T}} &= \sum_{k_z, \bm{r}}\left[ \bm{c}^\dagger(\bm{r}) h^{M\mathcal{T}}(k_z)  \bm{c}(\bm{r})\right. \\&\phantom{=} - V c_1^\dagger(\bm{r})c^\dagger_2(\bm{r}+\hat{x}+\hat{y})c_3(\bm{r}+\hat{x})c_4(\bm{r}+\hat{y}) \\&\phantom{=}+\left. h.c.\right],\\
h^{M\mathcal{T}}(k_z) &= \gamma \sin(k_z) \Gamma_0 + [1+\cos(k_z)](\gamma' \Gamma_2+\gamma \Gamma_4).
\end{split}\label{eq:LatticeHamM}
\end{equation}
As desired, this model is invariant under phase shifts that depend linearly on the $x$ and $y$ coordinates, $\bm{c}(\bm{R})\rightarrow \bm{c}(\bm{R}) e^{ i (\beta_1 x + \beta_2 y )}$.
For $\gamma' = \gamma,$ Eq. \ref{eq:LatticeHSFourierM} has $C_4 \mathcal{T}$ and is the same as Eq. \ref{eq:LatticeHam3}. When $\gamma' \neq \gamma$, the $C_4 \mathcal{T}$ symmetry is explicitly broken, and Eq. \ref{eq:LatticeHamM} only has $M \mathcal{T}$ symmetry. The $M_x \mathcal{T}$ and $M_y \mathcal{T}$ symmetries act on the fermionic degrees of freedom in each unit cell as $\bm{c} \rightarrow U_{M_x}  \bm{c}$ and $\bm{c} \rightarrow U_{M_y}  \bm{c}$ respectively, where $U_{M_x} = \tau_1 \otimes \tau_3,$ and $U_{M_y} = \tau_1 \otimes \tau_1$. Here we are interested in the case where $C_4 \mathcal{T}$ symmetry is explicitly broken, and without loss of generality we will take $\gamma' < \gamma$.

\subsection{Mean Field Analysis}\label{ssec:MFM}
As before, we shall use self-consistent mean field theory to analyze the interacting $M\mathcal{T}$ symmetric lattice model. To do this, we shall decompose the ring exchange interaction using the decomposition from Eq. \ref{eq:LatticeHS3D}. 
In the self-consistent mean field limit, the Hubbard-Stratonovich fields acquire the expectation values of the form
\begin{equation}
\begin{split}
&\lambda_{1x}(\bm{R}) = \lambda_x e^{i\phi_x(\bm{R}) + i A_{xy}(\bm{R})},\\
&\lambda_{2x}(\bm{R})  = \lambda_x e^{-i \phi_x(\bm{R})},\\
& \lambda_{1y}(\bm{R})=\lambda_{2y}(\bm{R})=0,\\
 &\Delta_y\phi_x(\bm{R}) = A_{xy}(\bm{R}).
\end{split}\label{eq:HSSolutionsM3D}
\end{equation}
The self-consistent values of $\lambda_x$ depend on both $V/\gamma$ and $\gamma'/\gamma,$ and can be found numerically. The self-consistent values of $\lambda_x/\gamma$ as a function of $V/\gamma$ for $\gamma'/\gamma  = 1/2$ are shown in Fig. \ref{fig:SelfConsistentPlotM} (details of this calculation are given in Appendix \ref{app:selfCon}). Unlike the self-consistent solutions for the $C_4\mathcal{T}$ model in Eq. \ref{eq:HSSelfConsistent3DC}, these solutions break $C_4 \mathcal{T}$ symmetry, but remain invariant under $M\mathcal{T}$ symmetry. Since $\lambda_{1y}(\bm{R})=\lambda_{2y}(\bm{R})=0$ in the mean field limit, there are no terms that hop fermions along the $y$-direction, although there are terms that hop fermions along the $x$-direction. Because of this, in the mean-field limit, this model will be quasi-$2$D, i.e., a decoupled array of $2$D planes stacked along the $y$-direction.Here, the planes are stacked along the $y$-direction because we chose to set $\gamma'<\gamma$ in Eq. \ref{eq:LatticeHamM}. If instead we had chosen to set $\gamma'>\gamma$, the planes would be stacked along the $x$-direction.

\begin{figure}
\includegraphics[width=.5\textwidth]{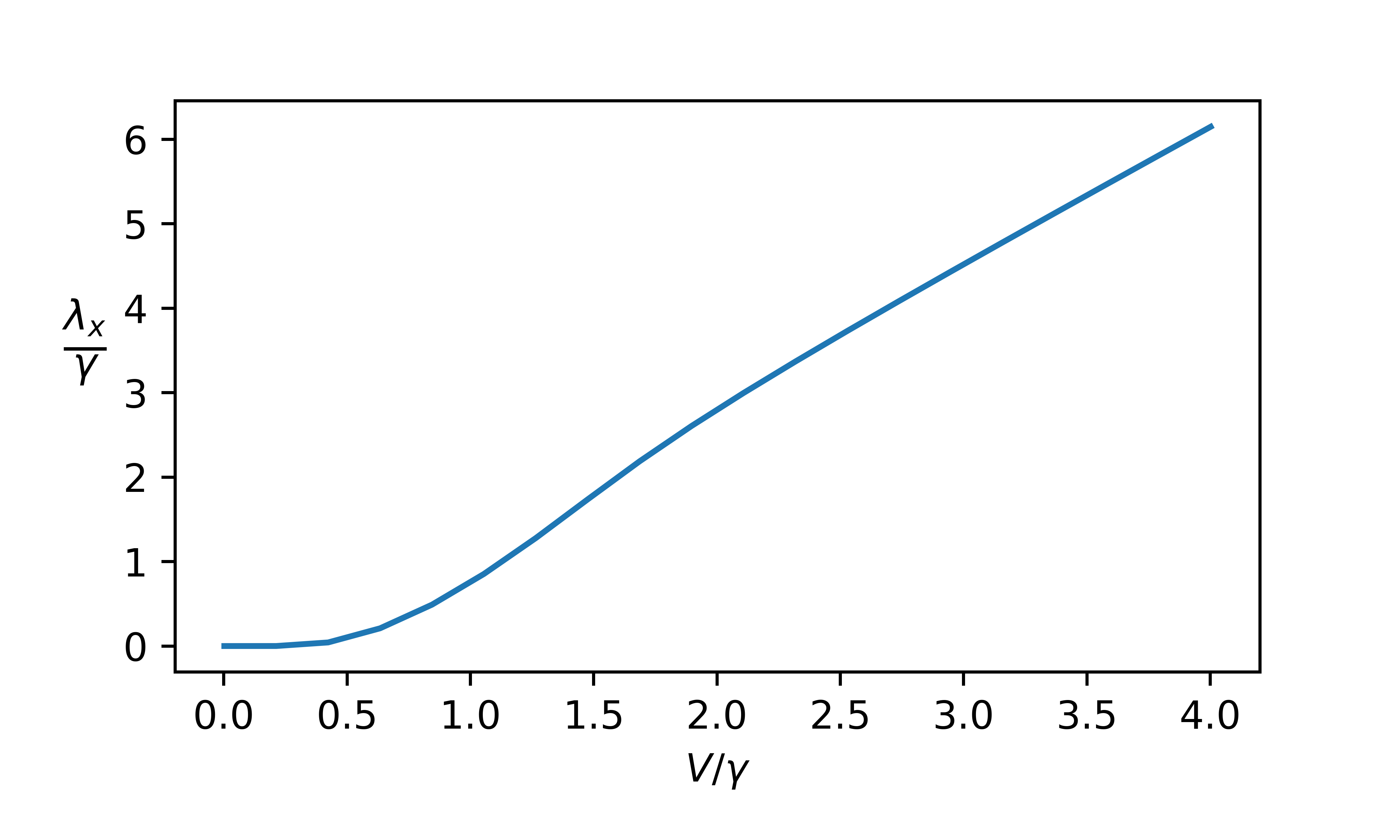}
\caption{The self-consistent solutions of $\lambda_x/\gamma$ as a function of $V/\gamma$ for $\gamma'/\gamma = 1/2$. At the mean field level, the bulk of the system remains gapped for all values of  $\lambda_x$. }
\label{fig:SelfConsistentPlotM}
\end{figure}

For periodic geometries, the quasi-2D nature of the mean field Hamiltonian can be made manifest by writing it in the following form:
\begin{equation}
\begin{split}
H^{M\mathcal{T}}_{\text{mf}} =& \sum_{\bm{k}',y}  \bm{c}^\dagger(\bm{k}',y) h^{M\mathcal{T}}_{\text{mf}}(\bm{k}',y) \bm{c}(\bm{k}',y),  \\
h^{M\mathcal{T}}_{\text{mf}}(\bm{k}',y) =&\gamma \sin(k_z )\Gamma_0 +[1+\cos(k_z )](\gamma'\Gamma_2+\gamma \Gamma_4) \big)\\
&+ \lambda_x \big(\cos(k_x+\phi_{x})\Gamma_4 + \sin(k_x + \phi_{x}) \Gamma_3\big),\\
\end{split}\label{eq:LatticeHSFourierM}
\end{equation}
where $\bm{k}' = (k_x,k_z)$, and we have omitted the gauge fields for simplicity. Here, we can directly see that the mean field Hamiltonian describes a layered system where each layer is an $xz$-plane, and the $y$-coordinate labels the different layers. Because the different layers are fully decoupled, the mean field Hamiltonian explicitly conserves dipole in the $y$-direction. As before, the phase field $\phi_x$ projects out states with non-vanishing polarization in the $x$-direction. 

When $\gamma' = 0$, each layer in Eq. \ref{eq:LatticeHSFourierM} is composed of two decoupled $2$D insulators, one of which  depends only on the $c_1$ and $c_3$ fermions, and one of which depends only on the $c_2$ and $c_4$ fermions. When $0< \lambda_x/\gamma < 2,$ the first of these insulators has Chern number $-1,$ and the second has Chern number $+1$. Since the two insulators have opposite Chern number, there are no net chiral modes associated with such a layer, and any surface modes can be gapped out with symmetry preserving perturbations. The $\gamma'$ term couples these two insulators and turns them into a single insulator with vanishing Chern-number.   When $\lambda_x/\gamma > 2$ both the insulators have Chern number 0. 

The energy spectrum of the mean field Hamiltonian is given by 
\begin{equation}
\begin{split}
\epsilon(\bm{k}') = &\Big(\frac{3}{2}\gamma'^2 + 2\gamma^2 + \lambda_x^2+4[\gamma'^2 + \gamma^2 ]\cos(k_z)  \\
&+2\gamma\lambda_x [1+\cos(k_z)] \cos(k_x+\phi_x)+ \gamma'\cos(2k_z)\Big)^{1/2}.
\end{split}\label{eq:MeanFieldEnergyM}
\end{equation} 
For $\gamma' \neq 0$, the spectrum is gapped for all values of $\lambda_x/\gamma >0$. When $\gamma' = 0$, and $\lambda_x/\gamma = 2$ there is a gap closing at $(k_x,k_z) = (\pi-\phi_x,0)$. As noted before, at this point the two decoupled insulators in each layer transition from having Chern number $\pm1$ to having Chern number 0. 

To determine the existence of any hinge modes in this model, we will consider a lattice with open boundaries along the $y$-direction. Such a geometry is straightforward to analyze given the quasi-$2$D nature of the mean field Hamiltonian. Specifically, let us consider a lattice of length $N_y$ in the $y$-direction. In the mean field limit, this corresponds to $N_y$ decoupled $2$D systems, which are indexed by their $y$-coordinate. For the layers away from the boundaries  ($y\neq 1,N_y$), the mean field Hamiltonian is the same as $h(\bm{k}',y)$ given in Eq. \ref{eq:LatticeHSFourierM}. Based on Eq. \ref{eq:HSSolutionsM3D}, in the $y = 1$ layer (i.e., the boundary normal to the $-\hat{y}$-direction), there are no Hubbard-Stratonovich fields coupling the $c_2$ and $c_4$ lattice fermions, and the mean field Hamiltonian is given by 
\begin{equation}
\begin{split}
h^{M\mathcal{T}}_{\text{mf}}(\bm{k}',1) =&\gamma \big(\sin(k_z )\Gamma_0 +[1+\cos(k_z )](\gamma'\Gamma_2+\gamma \Gamma_4) \big)\\
&+ \lambda \big(\cos(k_x+\phi_{x})\Gamma^+_4 + \sin(k_x + \phi_{x}) \Gamma^+_3\big),\\
\end{split}\label{eq:LatticeMBoun1}
\end{equation}
where $\Gamma^{\pm}_i = (\Gamma_i \pm \eta \Gamma_i \eta)/2$, and $\eta = \text{diag}(-1,1,1,1)$. For $\lambda_x < \lambda^c_x \equiv 2(\gamma + 2\gamma'^2/\gamma)$, this layer has Chern number $-1$. At $\lambda_x = \lambda^c_x$ this boundary layer is gapless, and for $\lambda_x > \lambda^c_x$ the boundary layer is an insulator with Chern number $0$. At the $y = L_y$ layer (i.e., the boundary normal to the $+\hat{y}$-direction), there are no Hubbard-Stratonovich fields coupling the $c_1$ and $c_3$ lattice fermions, and the mean field Hamiltonian is given by 
\begin{equation}
\begin{split}
h^{M\mathcal{T}}_{\text{mf}}(\bm{k}',N_y) =&\gamma \big(\sin(k_z )\Gamma_0 +[1+\cos(k_z )](\gamma'\Gamma_2+\gamma \Gamma_4) \big)\\
&+ \lambda \big(\cos(k_x+\phi_{x})\Gamma^-_4 + \sin(k_x + \phi_{x}) \Gamma^-_3\big).\\
\end{split}\label{eq:LatticeMBoun2}
\end{equation}
Similar to before, for $0<\lambda_x < \lambda_x^c$, this layer has Chern number $+1$. At $\lambda_x = \lambda_x^c$ this boundary layer is also gapless, and for $\lambda_x > \lambda_x^c$ this boundary layer is an insulator with Chern number $0$. It is worth noting that the boundary Hamiltonians in Eq. \ref{eq:LatticeMBoun1} and \ref{eq:LatticeMBoun2} can be modified by the addition of additional of symmetry preserving terms that are localized at the boundaries of the system. Because of this, the value of value of $\lambda^c_x$ is not uniquely determined by the bulk of the system. 

\begin{figure}\centering
\includegraphics[width=.4\textwidth]{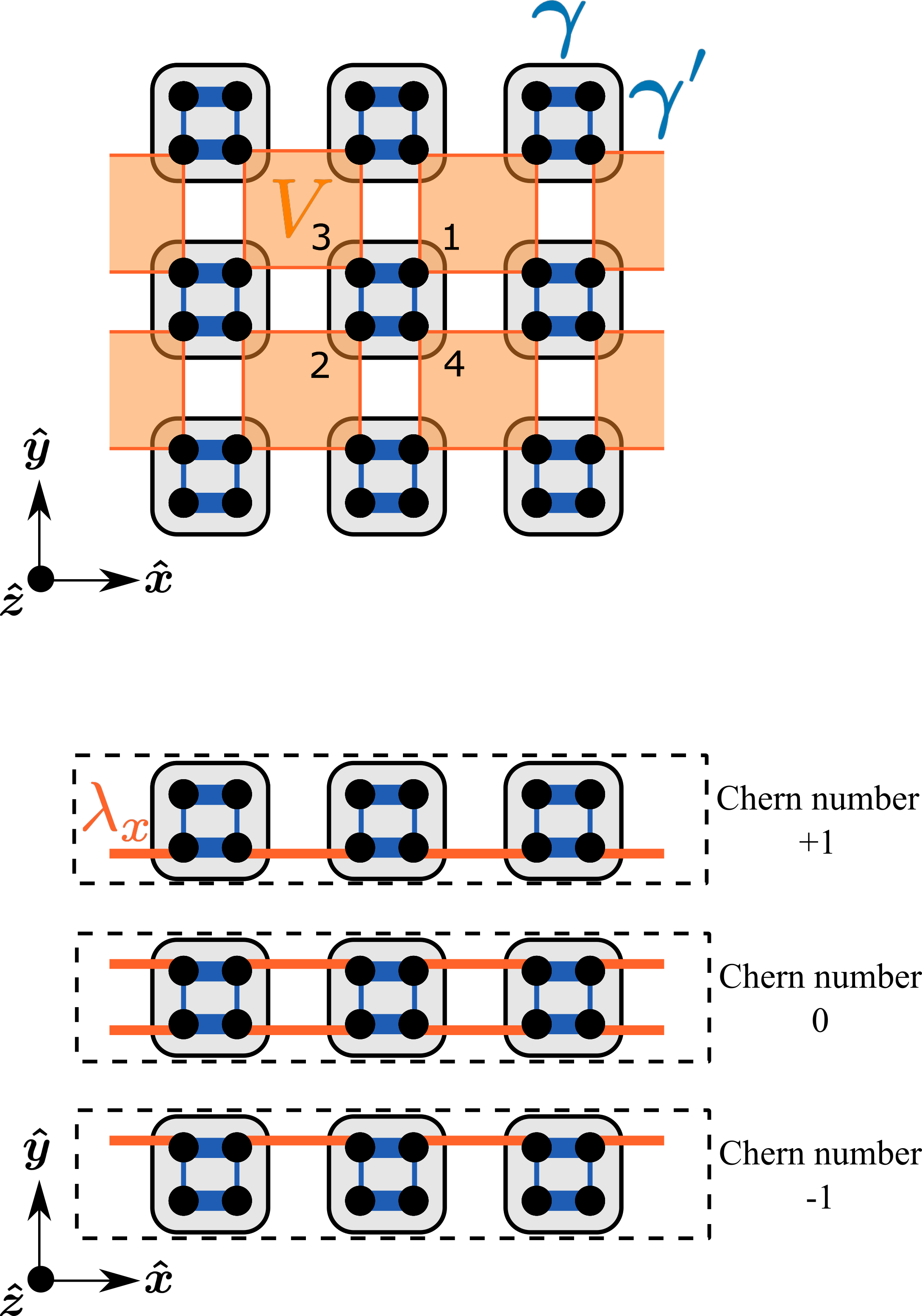}
\caption{Schematic of the interacting $M\mathcal{T}$ symmetric model in Eq. \ref{eq:LatticeHamM} with open boundary conditions (top), and the corresponding mean field Hamiltonian (bottom). The mean field Hamiltonian describes a system composed of $2$D layers stacked along the $y$-direction. In the bulk, these layers have Chern number $0$, while the layers at the boundaries normal to the $\pm y$-direction have Chern number $\pm 1$.}\label{fig:LatDiaM1}
\end{figure}

As we have seen, the mean field model can be interpreted as a layered system, where for $0<\lambda_x < \lambda_x^c$ each layer in the bulk consists of two $2$D insulators with opposite Chern numbers, which are coupled via $\gamma$. The layers at boundaries normal to the $\pm \hat{y}$-direction, have Chern number $\pm 1$. This is shown schematically in Fig. \ref{fig:LatDiaM1}. Based on this, we can confirm the existence of chiral hinge modes in the $M \mathcal{T}$ symmetric model for $0<\lambda_x<\lambda_x^c$. In terms of the lattice parameters, this means that the chiral hinge modes persist up to a critical value of $V$ that depends on the value of $\gamma'/\gamma$, as well as any additional boundary terms. At $\lambda_x = \lambda_x^c$ the mean field model transitions from a chiral hinge insulator to a weak HOTI. Provided that $\gamma'\neq 0$, the energy-gap of the bulk of the system remains open during this transition at the mean field level. Only the single-particle energy gap of the boundary closes. In contrast, we found that when the $C_4\mathcal{T}$ model transitions from being a chiral hinge insulator to a weak HOTI, the bulk energy-gap closes. Phase transitions where only the boundary energy-gap closes have been previously studied in the context of boundary obstructed topological phases\cite{benalcazar2017a,khalaf2019boundary}. For these systems, it has been shown that while the bulk energy gap does not close during such a phase transition, the Wannier gap in the bulk does close, and signals a transition between the topologically distinct states. 

While the mean-field Hamiltonian we have derived is only a boundary obstructed topological phase, let us comment on the initial interacting model. It is worth remarking that in non-dipole conserving models, e.g., free-fermion band theories, chiral hinge modes cannot be protected by $M\mathcal{T}$ symmetry alone in $3$D. This can be concluded from the fact that if we consider a $3$D model with chiral hinge modes, it is possible to add $2$D insulators with Chern number $\pm 1$, to the boundaries normal to the $\mp \hat{y}$ direction without breaking $M\mathcal{T}$ symmetry. This will cause there to be both a chiral and anti-chiral fermion at each hinge, which can be gapped out with symmetry preserving perturbations. However, adding a $2$D insulator with Chern number $\pm 1$ to the boundaries of a system necessarily violates dipole conservation. This can be concluded from the fact that adiabatically shifting the momentum of an insulator with non-vanishing Chern number causes the insulator to polarize, via the Laughlin pumping process. This clearly violates dipole conservation (lattice models of Chern insulators also have single-particle tunneling terms and tunable orbital magnetization, both of which violate dipole conservation). Because of this, the chiral hinge modes in the $M\mathcal{T}$ model we have considered here are protected, and cannot be gapped out without breaking dipole conservation in either the $x$ or $y$-direction.

Having established the existence of the chiral hinge modes, would now like to use the mean field Hamiltonian to find the response action for the chiral hinge insulator phase of the $M\mathcal{T}$ model. This can be done efficiently due to the quasi-2D nature of the mean field model. If we reintroduce the gauge fields to the mean field Hamiltonian, we find that in the bulk (Eq. \ref{eq:HSSolutionsM3D}) the response action vanishes. At the boundaries normal to the $x$ and $y$-directions, we find that in the topological phase, the boundary and hinge responses are given by
\begin{equation}
\begin{split}
L_{eff,\pm x} = 0,\\
L_{eff,\pm y} =\pm & \frac{1}{2\pi}\big[ A_0 \partial_x A_z + \phi_x \partial_z A_0 - \phi_x\partial_t A_y \big],\\
L_{eff,\pm x , \pm y} =- & \frac{1}{4\pi} A_0A_z. \\
L_{eff,\pm x, \mp y} = + & \frac{1}{4\pi} A_0A_z.
\end{split}\label{eq:EffectiveActionBoundM}
\end{equation}
In Appendix \ref{app:MContAnalysis}, we show that this response can also be derived from a continuum analysis, similar to what was done in Sec. \ref{ssec:ContAnalysis}. Here, we can see that the effective response vanishes for boundaries normal to the $x$-direction, while for boundaries normal to the $y$-direction, the response action is equivalent to a $2$D Chern-Simons action, if we identify the phase field $\phi_x$ with the $x$-direction component of a rank-1 gauge field, $A_x$ (recall that the phase field gauge transforms as $\phi_x \rightarrow \phi_x + \partial_x \Lambda$). This $2$D Chern-Simons term can be understood from the fact that in the mean field limit, the layers normal to the $\pm \hat{y}$-direction have Chern number $\pm 1$. As expected, the anomalies at the hinges are  canceled by the inclusion of chiral fermionic modes. 

Comparing the mean field analysis of the $M\mathcal{T}$ model to that of the $C_4\mathcal{T}$ model in Sec. \ref{sec:LatticeC}, we find that both of these models have a topological phase with chiral hinge modes that persists up to a finite value of ring exchange amplitude $V$, after which they transition into a weak HOTI phase with no chiral hinge modes (though it does have non-chiral hinge modes if we preserve translation symmetry along the hinge). Despite this, we find that the effective surface Lagrangian of the $M\mathcal{T}$ model (Eq. \ref{eq:EffectiveActionBoundM}) differs significantly from that of the $C_4\mathcal{T}$ model (Eq. \ref{eq:EffectiveActionBound}). For the $C_4\mathcal{T}$ model, the boundary Lagrangian is non-vanishing for all four boundaries, and all boundary terms have a prefactor of $\frac{1}{4\pi}$. However, for the $M\mathcal{T}$ model, the boundary Lagrangian is non-vanishing for the boundaries normal only to the $\pm \hat{y}$-directions, and these terms have a prefactor of $\frac{1}{2\pi}$. In terms of the bulk response action, this difference corresponds to the addition of a boundary term
\begin{equation}
\delta S_{eff} = \int \frac{d^4x}{4\pi} [\partial_y(A_0 \partial_x A_z) -\partial_x(A_0 \partial_y A_z) ].\label{eq:AdBound}
\end{equation} 
This term is invariant under $M\mathcal{T}$, but not $C_4\mathcal{T}$, as expected. Importantly, this term does not affect the hinge terms, which are the same for the $M\mathcal{T}$ model and the $C_4\mathcal{T}$ model. So, it appears that the $C_4\mathcal{T}$ and $M\mathcal{T}$ lattice models share the same hinge physics, but have different boundary physics.

\subsection{Dimensional Reduction}
We will now consider dimensionally reducing the $M\mathcal{T}$ model in Eq. \ref{eq:LatticeHamM} from $3$D to $2$D. This should result in a $2$D model parameterized for dipole pumping, analogous to the model given in Sec. \ref{sec:DimRed}. If we use the standard procedure of identifying the $z$-direction momentum $k_z$ with an adiabatic parameter $\theta$, the dimensionally reduced Hamiltonian is given by
\begin{equation}
\begin{split}
H^{QM} &= \sum_{\bm{r}}\left[ \bm{c}^\dagger(\bm{r}) h^{oM} \bm{c}(\bm{r}) \right.\\&\phantom{=} - V c_1^\dagger(\bm{r})c^\dagger_2(\bm{r}+\hat{x}+\hat{y})c_3(\bm{r}+\hat{x})c_4(\bm{r}+\hat{y}) \\&\phantom{=}+\left. h.c.\right],\\
h^{oM} &= \gamma \sin(\theta) \Gamma_0 + [1+\cos(\theta)](\gamma' \Gamma_2+\gamma \Gamma_4).
\end{split}\label{eq:LatticeHamM2D}
\end{equation}
For $\gamma' = \gamma$, this is exactly the model from Eq. \ref{eq:LatticeHam2} with the parameterization given in Eq. \ref{eq:adParameters2}. For $\gamma' \neq \gamma$ the model has $M_x$ and $M_y$ symmetry when $\theta = 0$, but still has $C_4$ symmetry when $\theta = \pi$. As before we will take $\gamma' < \gamma$.

\begin{figure}\centering
\includegraphics[width=.5\textwidth]{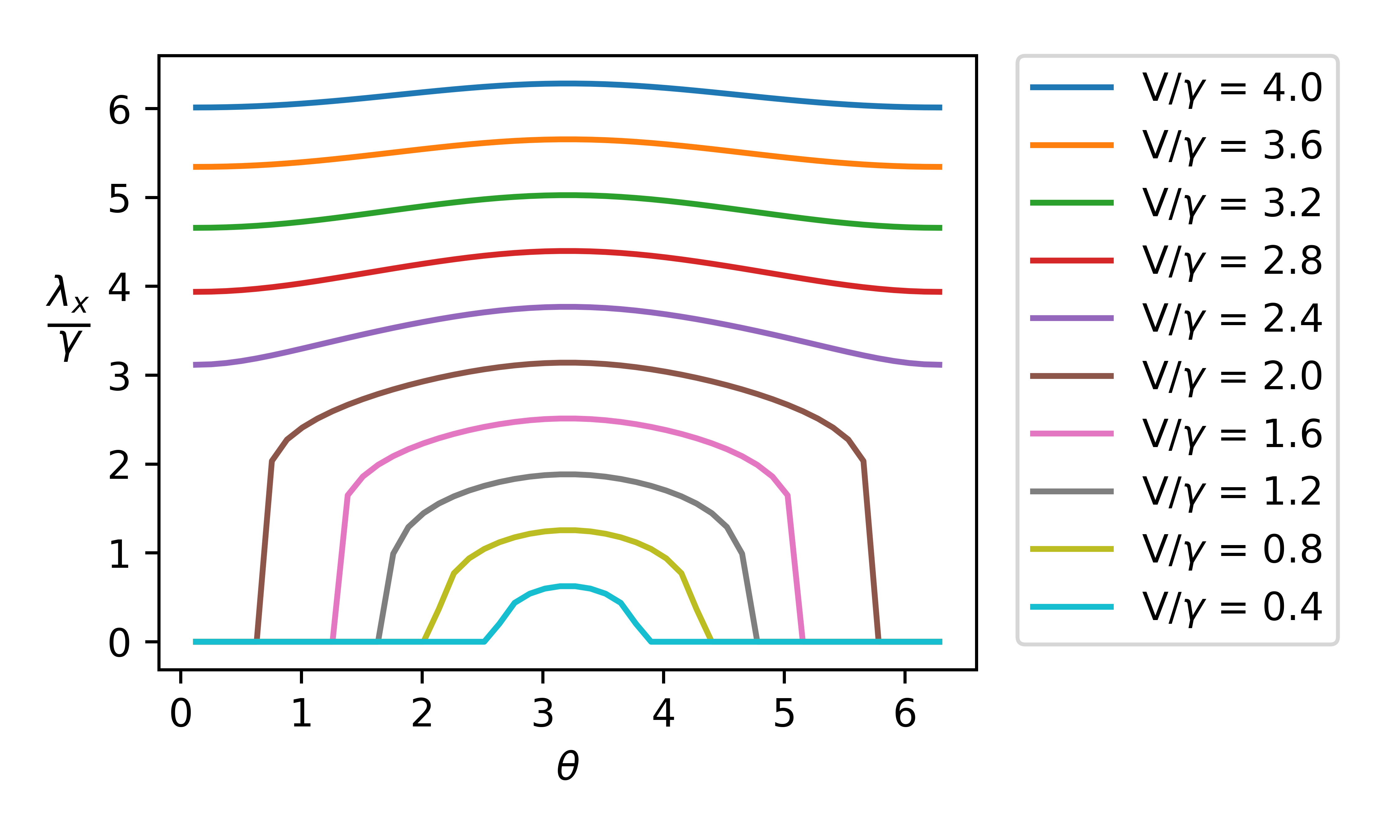}
\caption{The self-consistent values of $\lambda_x/\gamma$ as a function of $\theta$ for $\gamma'/\gamma = 1/2$ and values of $V/\gamma$ between $0.4$ and $4.0.$}\label{fig:consistentplot2DM}
\end{figure}

Following our analysis in Sec. \ref{ssec:ContDipolePumping}, we will use mean field theory to determine the behavior of Eq. \ref{eq:LatticeHamM2D} at various values of $\theta$. Using the same decomposition of the ring exchange term given in Eq. \ref{eq:LatticeHS}, we find that in the mean field limit, the Hubbard-Stratonovich fields acquire expectation values of the form
\begin{equation}
\begin{split}
&\lambda_{1x}(\bm{r}) = \lambda_x e^{i\phi_x(\bm{r}) + i A_{xy}(\bm{r})},\\
&\lambda_{2x}(\bm{r})  = \lambda_xe^{-i \phi_x(\bm{r})} ,\\
&\lambda_{1y}(\bm{r}) = \lambda_{2y}(\bm{r}) = 0,\\
 &\Delta_y \phi_x(\bm{r}) = A_{xy}(\bm{r}).
\end{split}\label{eq:selfconsistentM2D}
\end{equation}
Here, the self-consistent values of $\lambda_x$ depend on $\theta$, $V/\gamma$, and $\gamma'/\gamma$. The self-consistent values of $\lambda_x$ are shown in Fig. \ref{fig:consistentplot2DM}. 

As can be seen from Eq. \ref{eq:selfconsistentM2D}, in the mean field limit there are no terms that hop a single fermion along the $y$-direction. Because of this, the mean field fermionic Hamiltonian is quasi-$1$D. Using the self-consistent values of $\lambda_x,$ the mean field Hamiltonian on periodic geometries can be written in a manifestly quasi-$1$D form as
\begin{equation}
\begin{split}
H^{QM}_{\text{mf}} =& \sum_{k_x,y} \bm{c}^\dagger(k_x,y) h^{QM}_{\text{mf}}(k_x,y) \bm{c}(k_x,y),  \\
h^{QM}_{\text{mf}}(k_x,y) =&\gamma \big(\sin(\theta )\Gamma_0 +[1+\cos(\theta )](\gamma'\Gamma_2+\gamma \Gamma_4) \big)\\
&+ \lambda_x \big(\cos(k_x+\phi_{x})\Gamma_4 + \sin(k_x + \phi_{x}) \Gamma_3\big).\\
\end{split}\label{eq:LatticeHSFourier2DM}
\end{equation}
The energy spectrum of the mean field Hamiltonian is the same as in Eq. \ref{eq:MeanFieldEnergyM}, upon identifying $k_z \rightarrow \theta$ (and treating $\lambda_x$ a function of $\theta$). From this, we see that for $\gamma' \neq 0$, the mean field Hamiltonian remains gapped for all values of $\lambda_x/\gamma$. 

To simplify the description, we note that the mean field Hamiltonian in Eq. \ref{eq:LatticeHSFourier2DM} can be treated as an array of wires which are aligned along the $x$-direction, and indexed by their $y$ coordinate. When $\gamma' = 0$, each wire in Eq. \ref{eq:LatticeHSFourier2DM} is composed of two decoupled $1$D insulators, one which only depends on the $c_1$ and $c_3$ fermion operators, and one which only depends on the $c_2$ and $c_4$ fermion operators. The first of these insulators is equivalent to the SSH Hamiltonian given in Eq. \ref{eq:SSHHam}, with parameterization \begin{equation}
(u,v,\mu) = (\gamma[1+\cos(\theta)], \lambda_x ,\gamma\sin(\theta)),
\end{equation}
while the second is equivalent to the SSH Hamiltonian with parameterization 
\begin{equation}
(u,v,\mu) = (\gamma[1+\cos(\theta)], \lambda_x ,-\gamma\sin(\theta)).
\end{equation}
Using Eq. \ref{eq:BlochPolarization} we find that these two insulators have opposite polarization responses. In particular, as $\theta$ is varied, the two insulators pump an opposite amount of charge, such that the net boundary charge remains unchanged. This is to be expected from dimensional reduction since, in the mean field limit of the $3$D model (Eq. \ref{eq:LatticeHSFourierM}), each layer consists of two decoupled insulators with opposite Chern-numbers when $\gamma' = 0$. The $\gamma'$ term in Eq. \ref{eq:LatticeHSFourier2DM} couples the SSH models at each layer, resulting in a single insulator with vanishing polarization for all values of $\theta$. 

To determine the existence of any corner modes, we will consider a system with open boundaries in the $y$-direction of length $N_y$. For the layers away from the boundaries  ($y\neq 1,N_y$), the mean field Hamiltonian is the same as $h(k_x, y)$ given in Eq. \ref{eq:LatticeHSFourier2DM}. At the $y = 1$ layer (i.e., the boundary normal to the $-\hat{y}$-direction), there are no Hubbard-Stratonovich fields coupling the $c_2$ and $c_4$ lattice fermions, and the mean field Hamiltonian is equivalent to the boundary Hamiltonian from Eq. \ref{eq:LatticeMBoun1} upon identifying $k_z \rightarrow \theta$. Similarly, the mean field Hamiltonian at the $y = N_y$ layer (i.e., the boundary normal to the $+\hat{y}$-direction) is equivalent to the boundary Hamiltonian from Eq. \ref{eq:LatticeMBoun2} upon identifying $k_z \rightarrow \theta$. 

In the quasi-$1$D limit this model takes, the corner charges are contributed only by the polarization of the boundary layers. The boundary polarization can be calculated using Eq. \ref{eq:BlochPolarization}. We find that, as a function of $\theta$, the change in polarization of the $y=1$ layer is equal and opposite to the change in polarization of the $y = L_y$ layer. Provided that $V$ exceeds a finite critical value that depends on $\gamma'/\gamma$, the polarization of the $y = L_y$ ($y=1$ layer) increases (decreases) as $\theta$ is increased. Over a full period, we find that the polarization of the boundary normal to the $\pm \hat{y}$-direction changes by $\pm 1$. Additionally, when the mean field model has $M_x$ and $M_y$ symmetry at $\theta = \pi$, the boundary polarization is $\pm 1/2$. This value of the boundary polarization indicates that there will be half-integer charges localized at the corners of the system. As discussed in Sec. \ref{sec:DimRed}, for systems coupled to the rank-2 gauge fields, half-integer corner charges correspond to a quadrupole moment $Q_{xy} = 1/2$ (relative to a trivial insulator). To confirm that the quadrupole moment at $\theta = \pi$ is in-fact $1/2$, we can couple the mean field Hamiltonian to the rank-2 gauge field and use linear response to determine the quadrupole moment (see Appendix \ref{app:QuadMF} for further details). Upon doing this, we indeed find that $Q_{xy}(\theta = \pi) - Q_{xy}(\theta = 0) = 1/2$ for $V$ less than the aforementioned critical value. 

As a final point, we would like to compare the mean field analysis of the two $2$D dipole conserving Hamiltonians we have considered so far. In Sec. \ref{ssec:ContDipolePumping} we considered a mean field decomposition of Eq. \ref{eq:LatticeHam2} that is manifestly invariant under $C_4$ symmetry, and showed that in the mean field limit, the model remains fully $2$D (see Eq. \ref{eq:DimRedLatticeHSGauge}). In this section, we considered a mean field decomposition of Eq. \ref{eq:LatticeHSFourier2DM} that is invariant under only $M_x$ and $M_y$ symmetry, and showed that in the mean field limit, the model is quasi-$1$D (equivalent to an array of decoupled $1$D wires). However, if we compare the two interacting models, we see that at $\theta = \pi$ Eq. \ref{eq:LatticeHam2} and \ref{eq:LatticeHSFourier2DM} are the same, and both Hamiltonians consist only of the ring exchange term $V$. Because of this, at $\theta=\pi$ the two mean field models should be equivalent. While this is not immediately obvious, it is true that the two mean field models are equivalent at $\theta = \pi$ provided we integrate over the phase fields $\phi_x$ and $\phi_y$. As discussed in Sec. \ref{ssec:ContDipolePumping}, when using the $C_4$ symmetric mean field decomposition, the ground state of the mean field Hamiltonian at $\theta = \pi$ is given by Eq. \ref{eq:MFGround1}, and upon integrating over the phase fields, it is equivalent to the ground state of the full interacting model Eq. \ref{eq:IntGroundPi}. Similarly, when using the $M_x$ and $M_y$ symmetric mean field decomposition, the ground state at $\theta = \pi$ can be written in real space as 
\begin{equation}\begin{split}
\ket{0} = &\prod_{\bm{r}} \Big[\frac{e^{i\phi_x} c^\dagger_2(\bm{r}+\hat{x}+\hat{y})}{\sqrt{2}}  - \frac{c^\dagger_4(\bm{r}+\hat{y})}{\sqrt{2}}\Big]\\&\times \Big[\frac{e^{-i\phi_x} c^\dagger_1(\bm{r})}{\sqrt{2}}  - \frac{c^\dagger_3(\bm{r}+\hat{x})}{\sqrt{2}}\Big]\ket{vac}.
\end{split}\label{eq:MFGround2}
\end{equation}
Upon integrating over the phase fields $\phi_x$ we see that this ground state is also equal to the ground state of the full interacting model in Eq. \ref{eq:IntGroundPi}. We can therefore conclude that although the mean field Hamiltonians appear very different at $\theta = \pi$, they do in fact lead to the same physics, provided that the dynamic phase fields are properly accounted for.

\section{Conclusion and Outlook}\label{sec:conclusion}

In this work we considered how various topological multipolar theories can be realized in dipole conserving lattice models. In $2$D we considered the quadrupole response and used a linear response formalism to determine the quadrupole moment of a dipole conserving many-body system. We showed that during a periodic adiabatic process, the quadrupole moment can change by an integer, and that there is a $C_4$ symmetric HOTI with quadrupole moment $1/2$ relative to a trivial insulator. 
We verified these results by considering an exactly solvable interacting lattice model. We also showed that for a properly chosen parameterization, this $2$D model can be related to a $3$D $C_4 \mathcal{T}$ symmetric model with chiral hinge modes. The boundary responses of this $3$D model match those of the dipolar Chern-Simons action, which is related to the rank-2 quadrupole response via dimensional reduction. We also considered a variation of the $3$D model, that is instead invariant under $M_x \mathcal{T}$ and $M_y \mathcal{T}$ symmetries, and showed that they had similar hinge modes and topological responses. It is worth restating that some of the analyses of these models has relied on self-consistent mean field theory (which is equivalent to taking increasing the number of flavors of fermions from $1$ to $N$ and taking the limit $N \rightarrow \infty$). While we believe that the results of this approximation are valid, it would be useful to find a way to verify these results that do not rely on the mean field approximation.

An interesting facet of our analysis is that the interacting models we considered here map exactly onto known non-interacting HOTIs in the mean field limit. The $2$D dipole conserving $C_4$ symmetric HOTI and the $3$D dipoles conserving $C_4\mathcal{T}$ symmetric HOTI we presented here, correspond to the non-interacting $2$D quadrupole insulator, and $3$D chiral hinge insulator of Ref. \onlinecite{benalcazar2017b} respectively. The dipole conserving models and non-interacting models also display the same topological phenomena (quantized corner charges in $2$D and chiral hinge modes in $3$D). This connection is interesting, since in Ref. \onlinecite{benalcazar2017b}, the topological nature of the non-interacting models is manifest in the non-trivial topology of the Wannier bands. In the interacting models presented here, there are no Wannier bands (outside the mean field limit) and the topological nature of the models is manifest in non-trivial rank-2 topological responses. Based on this, it appears that there may be deeper connections between rank-2 gauge theories and HOTIs than initially expected.

It should be possible to simulate the models we have presented here in cold atom systems. Due to the exceptional degree of control in cold atom systems, it is possible to tune a system such that any single-particle hopping terms are negligible, and the dominant terms are interactions, such as the ring-exchange interaction we considered here\cite{dai2016observation}. This should allow for an explicit construction of the $2$D model we have considered here, and a simulation of the pumping process we discussed. In principle it should also be possible to construct a cold atom analog $C_4\mathcal{T}$ chiral hinge insulator as well. Beyond cold atoms, other metamaterial systems exhibiting tunable non-linear couplings may be able to realize the models we have considered here. In known solid state materials there are no rank-2 gauge fields that couple to the charge degrees of freedom. However, provided we are treating the rank-2 gauge fields as background probes, it is possible to identify the rank-2 gauge fields with the derivatives of physical rank-1 gauge fields in certain situations\cite{dubinkin2019a}. It would be an interesting question for further research to consider how the rank-1 analogs of the phenomena we discussed here could occur in physical materials, and how they could be measured in experiments.

\section*{Acknowledgments}
We thank O. Dubinkin for helpful discussions. JMM is supported by the National Science Foundation Graduate Research Fellowship Program under Grant No. DGE – 1746047. TLH thanks the US Office of Naval Research (ONR) Multidisciplinary University Research Initiative (MURI) grant
N00014-20- 1-2325 on Robust Photonic Materials with
High-Order Topological Protection for support. 

\bibliography{TopologicalDipole}
\bibliographystyle{apsrev4-1}

\appendix

\section{Details of the Self-Consistent Mean Field Theory Calculations} \label{app:selfCon}

In this appendix we shall go over the details of the self-consistent mean field theory approximations we used in the main body of the text. The method we shall discuss is general, although the details of the calculation will depend on the details of the models we have considered. In order to preserve generality, we shall present this calculation in $d$ spatial dimensions. When $d = 2$, this is the approximation that is used to analyze the interacting Hamiltonians in Eq. \ref{eq:LatticeHam2} and Eq. \ref{eq:LatticeHamM2D} in the main body of the text. When $d = 3$, this is the approximation that is used to analyze the interacting Hamiltonians in Eq. \ref{eq:LatticeHam3} and Eq. \ref{eq:LatticeHamM} in the main body of the text. 

To start, we will rewrite the Hubbard-Stratonovich fields as
\begin{equation}
\begin{split}
&\lambda_{1x} = \lambda_0 \cos(\nu)\cos(\theta_x)e^{i\phi_{1x}},\\
&\lambda_{2x} = \lambda_0 \cos(\nu)\sin(\theta_x)e^{i\phi_{2x}},\\
&\lambda_{1y} = \lambda_0 \sin(\nu)\cos(\theta_y)e^{i\phi_{1y}},\\
&\lambda_{2y} = \lambda_0 \sin(\nu)\sin(\theta_y)e^{i\phi_{2y}},\\
\end{split}\label{Aeq:HSDef1}
\end{equation}
where $\lambda_0$ is real and positive, and we have suppressed the dependence on the lattice coordinates for simplicity. In terms of these fields, the Lagrangian can be written as
\begin{equation}
\begin{split}
\mathcal{L} = &\bm{c}^\dagger G_f^{-1} \bm{c} \\ & + \frac{2}{V} \lambda^2_0 \cos(\phi_{1x} + \phi_{2x}- A_{xy})\cos^2(\nu)\sin(2\theta_x)\\&  + \frac{2}{V} \lambda^2_0\cos(\phi_{1y} + \phi_{2y}- A_{xy})\sin^2(\nu)\sin(2\theta_y)],
\end{split}
\end{equation}
where $G_f^{-1} $ is the matrix which contains all terms that are quadratic in the fermion operators. Due to the Hubbard-Stratonovich decomposition of the ring exchange interaction, $G_f^{-1} $ depends on the Hubbard Stratonovich fields in Eq. \ref{Aeq:HSDef1}. 

To proceed, we will make use of the fact that the self-consistent values of the Hubbard-Stratonovich fields are also those that minimize the effective potential for the Hubbard-Stratonovich fields, once the fermions are integrated out. To find the effective potential for the Hubbard-Stratonovich fields, we will assume that they acquire a uniform expectation value. In this approximation, the Lagrangian is quadratic in the fermion operators,
and $G^{-1}_f$ can be written diagonally in momentum and frequency space, $G^{-1}_f \rightarrow G^{-1}_f(\omega, k)$. Using this, the resulting effective potential for the Hubbard Stratonovich fields can be written as
\begin{equation}
\begin{split}
\mathcal{H}_{eff} = -&\log[\text{Det}( G_f^{-1} )]\\ & - \frac{2}{V} \lambda^2_0 \cos(\phi_{1x} + \phi_{2x}- A_{xy})\cos^2(\nu)\sin(2\theta_x)\\&  - \frac{2}{V}\lambda \cos(\phi_{1y} + \phi_{2y}- A_{xy})\sin^2(\nu)\sin(2\theta_y),
\end{split}
\end{equation}
where 
\begin{equation}
\text{Det}( G_f^{-1} ) \equiv \int \frac{d\omega d^d k}{(2\pi)^{d+1}} \text{Det}(G_f^{-1}(\omega,k)).
\end{equation}
In terms of $\mathcal{H}_{eff}$, the self-consistent values of the Hubbard Stratonovich fields satisfy
\begin{equation}
\begin{split}
&\partial_{\phi_{1x}}  \mathcal{H}_{eff} = 0,\\
&\partial_{\phi_{2x}}  \mathcal{H}_{eff} = 0,\\
&\partial_{\phi_{1y}}  \mathcal{H}_{eff} = 0,\\
&\partial_{\phi_{2y}}  \mathcal{H}_{eff} = 0,\\
\end{split}\label{Aeq:SCEq1}
\end{equation}
\begin{equation}
\partial_{\theta_x} \mathcal{H}_{eff} = 0,\label{Aeq:SCEq2}
\end{equation}
\begin{equation}
\partial_{\theta_y} \mathcal{H}_{eff} = 0,\label{Aeq:SCEq3}
\end{equation}
\begin{equation}
\partial_\nu \mathcal{H}_{eff} = 0,\label{Aeq:SCEq4}
\end{equation}
\begin{equation}
\partial_{\lambda_0} \mathcal{H}_{eff} = 0.\label{Aeq:SCEq5}
\end{equation}
For all the models we consider in this paper, the constraints in Eq. \ref{Aeq:SCEq1} are solved by setting 
\begin{equation}
\begin{split}
&\phi_{1x} = \phi_x + A_{xy},\\
&\phi_{2x} = -\phi_x,\\
&\phi_{1y} = \phi_y + A_{xy},\\
&\phi_{2y} = -\phi_y,\\
\end{split}\label{Aeq:HSPhaseCont}
\end{equation}
where $\phi_x$ and $\phi_y$ are two new phase fields we have introduced here. Similarly, for all the models we consider in this paper, the constraints in Eq. \ref{Aeq:SCEq2} and \ref{Aeq:SCEq3} are solved by setting
\begin{equation}
\theta_x = \theta_y = \pi/4.
\end{equation}
The solution to Eq. \ref{Aeq:SCEq4} depends on the details of the model we are considering. For the Hamiltonians in Eq. \ref{eq:LatticeHam2} and \ref{eq:LatticeHam3},  Eq. \ref{Aeq:SCEq4} is solved by 
\begin{equation}
\nu = \pi/4.
\end{equation}
For the Hamiltonians in Eq. \ref{eq:LatticeHamM} and \ref{eq:LatticeHamM2D},  Eq. \ref{Aeq:SCEq4} is solved by 
\begin{equation}
\nu = 0,
\end{equation}
provided that $\gamma > \gamma'$. For all the lattice models, Eq. \ref{Aeq:SCEq5} must be solved numerically as a function of the lattice parameters. In order to avoid any unnecessary numeric constants in our definitions, it will also be useful to define the new variables 
\begin{equation}\begin{split}
&\lambda \equiv \lambda_0/2,\\
&\lambda_x \equiv \lambda_0/\sqrt{2}.
\end{split}\label{Aeq:FinalAp1}
\end{equation}

In addition to the effective potential, we can also consider the effective kinetic term for the Hubbard Stratonovich fields. These terms can be found by evaluating the polarization diagram of the fermions. At one loop order, we find that when $d=2$, the fermions generate a kinetic term of the form
\begin{equation}\begin{split}
\mathcal{H}_{kin-eff} = &-g_x \lambda_{2x}(\bm{r}+\hat{y}) \lambda_{1x}(\bm{r}) \\ &- g_y \lambda_{2y}(\bm{r}+\hat{x}) \lambda_{1y}(\bm{r}) + h.c.,
\end{split}\label{Aeq:KinEff}
\end{equation}
where we have used the original definitions of the Hubbard-Stratonovich fields, and added back in the dependence on the lattice coordinates. When $d=3$, we also find a similar term, which is equivalent to Eq. \ref{Aeq:KinEff} upon exchanging the $2$D lattice coordinate $\bm{r}$ with the $3$D lattice coordinate $\bm{R}$. Combining Eq. \ref{Aeq:KinEff}, with our results from analyzing Eq. \ref{Aeq:SCEq1}-\ref{Aeq:SCEq5}, we find the the kinetic energy is minimized when 
\begin{equation}
\begin{split}
&\Delta_y \phi_x = A_{xy}\\
&\Delta_x \phi_y = A_{xy},\\
\end{split}\label{Aeq:PhaseContDerivative}
\end{equation}
where $\Delta_x$ and $\Delta_y$ are the lattice derivatives.

Using Eq. \ref{Aeq:HSPhaseCont}-\ref{Aeq:FinalAp1} and \ref{Aeq:PhaseContDerivative}, along with the numeric solutions for the self-consistent values of $\lambda_0,$ we are able to determine the self-consistent values of the Hubbard-Stratonovich fields $\lambda_{ai}$.

\section{Quadrupole Moment of the Mean Field Hamiltonian}\label{app:QuadMF}
In this appendix we will discuss how to determine the quadrupole moment for the $2$D mean field models we discussed in the main text. For a Hamiltonian coupled to a spatially varying gauge field $A_{xy}(\bm{r})$, the change in the quadrupole moment during an adiabatic process is given by
\begin{equation}
\begin{split}
\frac{\partial}{\partial \theta} Q_{xy} = \lim_{\epsilon \rightarrow 0}\frac{i}{\epsilon L_x L_y}\sum_{n\neq 0}&\left[\frac{\bra{0} \frac{\partial H}{\partial\theta}\ket{n}\bra{n} \sum_{\bm{r}}\frac{\partial H}{\partial{A_{xy}(\bm{r})}} \ket{0}}{\epsilon + E_0 - E_n}\right. \\ & -  \left.\frac{\bra{0}  \sum_{\bm{r}}\frac{\partial H}{\partial{A_{xy}(\bm{r})}}\ket{n}\bra{n} \frac{\partial H}{\partial\theta} \ket{0}}{\epsilon + E_n - E_0}\right]. \label{Aeq:KuboSpectralQuad}
\end{split}
\end{equation}
In the limit that the gauge field is flat ($A_{xy}(\bm{r}) =$const.), Eq. \ref{Aeq:KuboSpectralQuad} reduces to Eq. \ref{eq:KuboSpectralQuad}. The expectation values in Eq. \ref{Aeq:KuboSpectralQuad} are calculated with vanishing gauge field $A_0(\bm{r}) = A_{xy}(\bm{r}) = 0$. 

Here, we shall consider the mean field limit of the $2$D interacting model 
\begin{equation}
\begin{split}
H^{Q} &= \sum_{\bm{r}} \left\{\bm{c}^\dagger(\bm{r}) h^o \bm{c}(\bm{r}) - A_0(\bm{r}) \bm{c}^\dagger(\bm{r}) \bm{c}(\bm{r})\right.\\&\phantom{=} - V c_1^\dagger(\bm{r})c^\dagger_2(\bm{r}+\hat{x}+\hat{y})c_3(\bm{r}+\hat{x})c_4(\bm{r}+\hat{y})e^{i A_{xy}(\bm{r})} \\&\phantom{=}+\left. h.c.\right\},\\
h^o &= \mu \Gamma_0 +t (\Gamma_2+\Gamma_4).
\end{split}\label{Aeq:LatticeHam2}
\end{equation}
For this Hamiltonian 
\begin{equation}
\begin{split}
\frac{\partial H^Q}{\partial{A_{xy}(\bm{r})}} =& - i V c_1^\dagger(\bm{r})c^\dagger_2(\bm{r}+\hat{x}+\hat{y})c_3(\bm{r}+\hat{x})c_4(\bm{r}+\hat{y}) \\&+ h.c.,
\end{split}
\end{equation}
where we have evaluated the derivative at $A_{xy}(\bm{r}) = 0$. To use mean field theory, we shall decompose the ring exchange term using the following Hubbard-Stratonovich transformation
\begin{equation}
\begin{split}
&-V c_1^\dagger(\bm{r})c_2^\dagger(\bm{r}+\hat{x}+\hat{y})c_3(\bm{r}+\hat{x})c_4(\bm{r}+\hat{y})e^{iA_{xy}(\bm{r})} \\ &\rightarrow \lambda_{1x}(\bm{r})c_2^\dagger(\bm{r}+\hat{x}+\hat{y})c_4(\bm{r}+\hat{y})e^{iA_{xy}(\bm{r})}\\&\phantom{\rightarrow} + \lambda_{2x}(\bm{r}) c_1^\dagger(\bm{r})c_3(\bm{r}+\hat{x}) \\&\phantom{\rightarrow}+\lambda_{1y}(\bm{r})c_2^\dagger(\bm{r}+\hat{x}+\hat{y})c_3(\bm{r}+\hat{x})e^{iA_{xy}(\bm{r})}\\&\phantom{\rightarrow}- \lambda_{2y}(\bm{r})c_1^\dagger(\bm{r})c_4(\bm{r}+\hat{y})\\ &\phantom{\rightarrow}- \frac{2}{V}\lambda_{1x}(\bm{r})\lambda_{2x}(\bm{r}) - \frac{2}{V}\lambda_{1y}(\bm{r})\lambda_{2y}(\bm{r}). 
\end{split}\label{Aeq:LatticeHS}
\end{equation}
This is equivalent to the Hubbard-Stratonovich transformation used in the main text upon shifting $\lambda_{1x} \rightarrow \lambda_{1x} e^{iA_{xy}}$ and $\lambda_{1y} \rightarrow \lambda_{1y} e^{iA_{xy}}$. After this transformation, 
\begin{equation}
\begin{split}
\frac{\partial H^Q}{\partial{A_{xy}(\bm{r})}} =&  [i \lambda_{1x}(\bm{r})c_2^\dagger(\bm{r}+\hat{x}+\hat{y})c_4(\bm{r}+\hat{y}) \\&+ i \lambda_{1y}(\bm{r})c_2^\dagger(\bm{r}+\hat{x}+\hat{y})c_3(\bm{r}+\hat{x}) h.c.] .
\end{split}\label{Aeq:CurrDef1}
\end{equation}
In the mean field limit, the Hubbard-Stratonovich fields acquire expectation values of the form
\begin{equation}
\begin{split}
&\lambda_{1x}(\bm{r})=  \lambda e^{i\phi_x(\bm{r})},\\
&\lambda_{2x}(\bm{r})=  \lambda e^{-i\phi_x(\bm{r})},\\
&\lambda_{1y}(\bm{r})=  \lambda e^{i\phi_y(\bm{r})},\\
&\lambda_{2y}(\bm{r})=  \lambda e^{-i\phi_y(\bm{r})},
\end{split}\label{Aeq:HSSelfConsistent2DC}
\end{equation}
where $\lambda$ is a constant. Using this, the mean field Hamiltonian is
\begin{equation}
\begin{split}
H^{Q}_{MF} = &\sum_{\bm{r}} \left\{\bm{c}^\dagger(\bm{r}) h^o \bm{c}(\bm{r}) - A_0(\bm{r}) \bm{c}^\dagger(\bm{r}) \bm{c}(\bm{r})\right.\\&+[\lambda e^{i\phi_x(\bm{r}) + i A_{xy}(\bm{r})}c_2^\dagger(\bm{r}+\hat{x}+\hat{y})c_4(\bm{r}+\hat{y})\\& + \lambda e^{-i\phi_x(\bm{r}) } c_1^\dagger(\bm{r})c_3(\bm{r}+\hat{x}) \\&+\lambda e^{i\phi_y(\bm{r}) + i A_{xy}(\bm{r})}c_2^\dagger(\bm{r}+\hat{x}+\hat{y})c_3(\bm{r}+\hat{x})\\ &- \left.\lambda e^{-i\phi_y(\bm{r}) }c_1^\dagger(\bm{r})c_4(\bm{r}+\hat{y}) + h.c.]\right\},
\end{split}\label{Aeq:MFMinimalCoup}
\end{equation}
and 
\begin{equation}
\begin{split}
\frac{\partial H^Q_{MF}}{\partial{A_{xy}(\bm{r})}} =& [ i \lambda e^{i\phi_x(\bm{r})}c_2^\dagger(\bm{r}+\hat{x}+\hat{y})c_4(\bm{r}+\hat{y}) \\&+ i \lambda_{1y} e^{i\phi_y(\bm{r})}c_2^\dagger(\bm{r}+\hat{x}+\hat{y})c_3(\bm{r}+\hat{x}) h.c.].
\end{split}\label{Aeq:CurDefMF}
\end{equation}
As a consistency check, we note that combining Eq. \ref{Aeq:CurrDef1} and \ref{Aeq:HSSelfConsistent2DC} also leads to Eq. \ref{Aeq:CurDefMF}. 

To proceed, we will introduce a new non-interacting Hamiltonian
\begin{equation}
\begin{split}
H^{Q}_{q} = &\sum_{\bm{r}} \left\{\bm{c}^\dagger(\bm{r}) h^o \bm{c}(\bm{r}) \right.\\&+[\lambda e^{i\phi_x(\bm{r}) + i q}c_2^\dagger(\bm{r}+\hat{x}+\hat{y})c_4(\bm{r}+\hat{y})\\& + \lambda e^{-i\phi_x(\bm{r}) } c_1^\dagger(\bm{r})c_3(\bm{r}+\hat{x}) \\&+\lambda e^{i\phi_y(\bm{r}) + i q}c_2^\dagger(\bm{r}+\hat{x}+\hat{y})c_3(\bm{r}+\hat{x})\\ &- \left.\lambda e^{-i\phi_y(\bm{r}) }c_1^\dagger(\bm{r})c_4(\bm{r}+\hat{y}) + h.c.]\right\},
\end{split}\label{Aeq:MFMinimalCoup2}
\end{equation}
which is equivalent to Eq. \ref{Aeq:MFMinimalCoup} at $A_0(\bm{r}) = 0$ and $A_{xy}(\bm{r}) = q$ (const). In particular, at $A_{xy}(\bm{r}) = q = 0,$ Eq. \ref{Aeq:MFMinimalCoup} and \ref{Aeq:MFMinimalCoup2} are equal to each other, and by extension have the same ground states and excited states. Additionally, we also have that
\begin{equation}
\sum_{\bm{r}}\frac{\partial H^Q_{MF}}{\partial{A_{xy}(\bm{r})}} = \frac{\partial H^Q_{q}}{\partial q},
\end{equation}
where the derivatives are evaluated at $A_{xy}(\bm{r}) = q= 0$. Combining this with Eq. \ref{Aeq:KuboSpectralQuad} we find that the change in the quadrupole moment in the mean field limit is
\begin{equation}
\begin{split}
\frac{\partial}{\partial \theta} Q_{xy} = \lim_{\epsilon \rightarrow 0}\frac{i}{\epsilon L_x L_y}\sum_{n\neq 0}&\left[\frac{\bra{0} \frac{\partial H^Q_{q}}{\partial\theta}\ket{n}\bra{n} \frac{\partial H^Q_{q}}{\partial q}\ket{0}}{\epsilon + E_0 - E_n}\right. \\ & - \left. \frac{\bra{0}  \frac{\partial H^Q_{q}}{\partial q}\ket{n}\bra{n} \frac{\partial H^Q_{q}}{\partial\theta} \ket{0}}{\epsilon + E_n - E_0}\right],
\label{Aeq:KuboSpectralQuad2}
\end{split}
\end{equation}
where $\ket{n}$ are the energy eigenstates of $H^Q_{q}$ at $q= 0$ (which are equal to energy eigenstates of $H^Q_{MF}$ at $A_{xy}(\bm{r}) = 0$). After some algebra, we can express the total change in the quadrupole moment in terms of the eigenfunctions of $H^Q_{q}$ as
\begin{equation}\begin{split}
\Delta Q_{xy} = \Delta\left[(-i) \sum_{\alpha \in occ} \int \frac{d^2\vec{k}}{4\pi^2} \bra{\vec{k},\alpha}\partial_q \ket{\vec{k},\alpha}\right],
\end{split}\label{Aeq:BerryDef2}
\end{equation}
where$\ket{\vec{k},\alpha}$ is a single particle eigenfunction of $H^Q_{q}$ with momentum $\vec{k}$ and band index $\alpha$, and the sum is over the occupied bands. We have also integrated over $\theta$ to derive Eq. \ref{Aeq:BerryDef2}. This formula can be used to calculate the change in the quadrupole moment during a given adiabatic process, such as the one in Sec. \ref{ssec:ContDipolePumping}.

We can also use this procedure to find the change in the quadrupole moment for the mean field limit of Eq. \ref{eq:LatticeHamM2D}. To do this, we will introduce the Hamiltonian 
\begin{equation}
\begin{split}
H^{ M }_{q'} = &\sum_{\bm{r}}\left[ \bm{c}^\dagger(\bm{r}) h^{oM} \bm{c}(\bm{r}) - A_0(\bm{r}) \bm{c}^\dagger(\bm{r}) \bm{c}(\bm{r})\right.\\&+[\lambda_x e^{i \phi_x + i q'}c_2^\dagger(\bm{r}+\hat{x}+\hat{y})c_4(\bm{r}+\hat{y})\\& +\left. \lambda_x e^{-i \phi_x } c_1^\dagger(\bm{r})c_3(\bm{r}+\hat{x}) h.c.]\right],
\end{split}\label{Aeq:MFMinimalCoup2M}
\end{equation} where $h^{oM} $ is defined as in Eq. \ref{eq:LatticeHamM2D}. Here, we have included a parameter $q'$, which we will use to find the quadrupole moment. Following the same steps as before, we find that the change in the quadrupole moment is given by
\begin{equation}\begin{split}
\Delta Q_{xy} = \Delta \left[(-i) \sum_{\alpha \in occ} \int \frac{d^2\vec{k}}{4\pi^2} \bra{\vec{k},\alpha}\partial_{q'} \ket{\vec{k},\alpha}\right],
\end{split}\label{Aeq:BerryDef2M}
\end{equation} where $\ket{\vec{k},\alpha}$ are the energy eigenstates of the Hamiltonian in Eq. \ref{Aeq:MFMinimalCoup2M}.

\section{Role of the Phase Fields in the Mean Field Limit}\label{app:PhaseFields}
In decomposing the ring exchange interaction into the sum of terms that are quadratic in the fermion creation/annihilation operators, we found it was necessary to introduce phase fields $\phi_x$ and $\phi_y$. Under phase shifts that depend linearly on position, $c({\bf{r}}) \rightarrow c({\bf{r}}) e^{i \beta_1 x + i \beta_2 y}$, the phase fields transform as $\phi_x \rightarrow \phi_x + \beta_1$ and  $\phi_y \rightarrow \phi_y + \beta_2$. Similarly, under a gauge transform $\Lambda$, the phase fields transform as $\phi_i \rightarrow \phi_i + \Delta_i \Lambda$ for $i = x,y$. These fields enter the mean field Hamiltonians $h_{MF}$, by shifting the lattice momentum $k_i \rightarrow k_i + \phi_i$. In this appendix, we shall show that integrating over the phase fields projects out states with non-vanishing many-body electric polarization. Here, we shall only consider the $\phi_x$ field. The analysis for the $\phi_y$ field can be done analogously. 

To show that the integration projects out states with non-vanishing many-body polarization, let us consider the $h_{MF}$. If we set $\phi_x = 0$, then the Hamiltonian $h_{MF}(k_x)$ is diagonalized by the single-particle states 
\begin{equation}
\gamma^\dagger_n(k_x) = u^{a\dagger}_n(k_x) c^\dagger_a(k_x),\label{Aeq:singlepartDef}
\end{equation} 
where $n$ is the band index. Here we are suppressing all the dependence on all momenta except $k_x$ for simplicity. In terms of these single-particle eigenstates, the ground state of the Hamiltonian with $\phi_x = 0$ is given by 
\begin{equation}
\begin{split}
\ket{0,0} &= \prod_{\{n,k_x\}\in \text{occ}} u^{a\dagger}_n(k_x) c^\dagger_a(k_x)\ket{vac}\\
&= \prod_{\{n,k_x\}\in \text{occ}} u^{a\dagger}_n(k_x) \sum_{x}\frac{e^{i k_x x}}{\sqrt{L_x}} c^\dagger_a(x)\ket{vac}.
\end{split}\label{Aeq:state1}
\end{equation}
For $\phi_x \neq 0$, the mean field Hamiltonian is given by $h_{MF}(k_x+\phi_x)$. Hence for $\phi_x\neq 0$, the state in Eq. \ref{Aeq:state1} becomes
\begin{equation}
\begin{split}
\ket{0,\phi_x} &= \prod_{\{n,k_x\}\in \text{occ}} u^{a\dagger}_n(k_x+\phi_x) c^\dagger_a(k_x)\ket{vac}\\
&= \prod_{\{n,k_x\}\in \text{occ}} u^{a\dagger}_n(k_x+\phi_x) \sum_{x}\frac{e^{i k_x x}}{\sqrt{N_x}} c^\dagger_a(x)\ket{vac},
\end{split}\label{Aeq:state2}
\end{equation}
where $N_x$ is the length of the lattice system in the $x$-direction. 

To proceed we will make two assumptions. First, we will assume that the single particle bands defined via Eq. \ref{Aeq:singlepartDef} are completely filled in the ground state (as they are in non-interacting band insulators). Second, we will assume that either $\phi_x$ can be written in the form $2\pi n/N_x$ for $n \in \mathbb{Z}$ (which means that the dipole gauge transformations must obey periodic boundary conditions), or that we are in the thermodynamic limit ($N_x \rightarrow \infty$). When these conditions are true, we can shift $k_x \rightarrow k_x - \phi_x$, and rewrite the state $\ket{0,\phi_x}$ as
\begin{equation}
\begin{split}
\ket{0,\phi_x} &= \prod_{\{n,k_x\}\in \text{occ}} u^{a\dagger}_n(k_x) c^\dagger_a(k_x-\phi_x)\ket{vac}\\
&= \prod_{\{n,k_x\}\in \text{occ}} u^{a\dagger}_n(k_x) \sum_{x}\frac{e^{i k_x x-i\phi_x x }}{\sqrt{L_x}} c^\dagger_a(x)\ket{vac}\\
& = e^{-i\phi_x x n(x)}\prod_{\{n,k_x\}\in \text{occ}} u^{a\dagger}_n(k_x) \sum_{x}\frac{e^{i k_x x}}{\sqrt{L_x}} c^\dagger_a(x)\ket{vac}\\
& = e^{-i\phi_x x n(x)}\ket{0,0},
\end{split}\label{Aeq:state3}
\end{equation}
where $n(x)=\sum_a c^\dagger_a(x)c_a(x)$. The operator $e^{-i\phi_x x n(x)}$ from Eq. \ref{Aeq:state3} is related to the expectation value of the polarization operator for periodic systems via
\begin{equation}
    P_x = -\frac{1}{\phi_x N_x}\text{Im}\log\langle e^{-i\phi_x x n(x)}\rangle.
\end{equation}
where $P_x$ is electronic contribution to the polarization of the system. 

We will now rewrite the $\phi_x = 0$ ground state from Eq. \ref{Aeq:state1} as a sum of eigenstates of the operator $e^{-i\phi_x x n(x)}$
\begin{equation}
\begin{split}
\ket{0,0} &= \sum_{n,X} a_{X,n} \ket{X,n}\\
\end{split}\label{Aeq:state4}
\end{equation}
where $\ket{X,n}$ satisfies $e^{-i\phi_x x n(x)}\ket{X,n} = e^{-i\phi_x X}\ket{X,n} $. From this we can conclude that the state $\ket{X,n}$ has \textit{total} polarization $X/N_x + P_{ion}$, where $P_{ion}$ is the contribution to the polarization from the ions. The ionic contribution must be included in order for the system is charge neutral, and for the polarization to be invariant with respect to a change of coordinates. For the models we are considering, a simple calculation shows that $P_{ion} = 0$ mod$(1)$. 

Combining Eq. \ref{Aeq:state3} and Eq. \ref{Aeq:state4}, the $\phi_x \neq 0$ state is given by
\begin{equation}
\begin{split}
\ket{0,\phi_x} &= e^{-i\phi_x x n(x)}\sum_{n,X} a_{X,n} \ket{X,n}\\
& = \sum_{n,X} a_{X,n} e^{i\phi_x X}\ket{X,n}.
\end{split}\label{Aeq:state5}
\end{equation}
From this we can see that integrating over $\phi_x$ will project out any states with $X/N_x \neq 0\text{ mod(1)}$. These are exactly the unpolarized many-body states.

\section{Continuum Analysis of $M\mathcal{T}$ model} \label{app:MContAnalysis}
In this appendix, we will analyze the $3$D $M\mathcal{T}$ symmetric model given in Eq. \ref{eq:LatticeHamM} in the continuum, near $V = 0$.  At $V = 0$, the system is quasi-1d, i.e., a decoupled $3$D array of $1$D wires oriented in the $z$-direction. If we pass to the continuum along the $z$-direction, each of these wires consists of 4 massless fermions (which correspond to the fluctuations of the lattice fermions with momentum near $k_z = \pi$), and 4 massive fermions (which correspond to the fluctuations of the lattice fermions with momentum near $k_z = 0$). At $V = 0$, the continuum Lagrangian for these fields is 
\begin{equation}
\begin{split}
&\mathcal{L} = \sum_{\bm{r}} \bm{\psi}^\dagger(\bm{r})G^{-1}_0(\omega+A_0(\bm{r}),p_z+A_z(\bm{r}))\bm{\psi}(\bm{r})\\ &\phantom{====} +\bm{\Psi}^\dagger(\bm{r})G^{-1}_\Lambda(\omega+A_0(\bm{r}),p_z+A_z(\bm{r}))\bm{\Psi}(\bm{r}) \\
&G^{-1}_0(\omega,k_z) =  \omega \text{I} + p_z \Gamma^0,\\
&G^{-1}_\Lambda(\omega,k_z) =  \omega \text{I} - p_z \Gamma^0 - M(\cos(\kappa)\Gamma^2 + \sin(\kappa)\Gamma^4),
\end{split}\label{Aeq:ContLag1}
\end{equation}
where $\bm{\psi} = (\psi_1,\psi_2,\psi_3,\psi_4)$ are the light fermion operators, $\bm{\Psi} = (\Psi_1,\Psi_2,\Psi_3,\Psi_4)$ are the heavy fermions operators, and $\kappa$ is defined such that $\tan(\kappa) = \gamma'/\gamma$. We will take $M$ to be the UV cutoff for this theory. 

We can now consider the interactions in the continuum. As before, we will use a Hubbard-Stratonovich transformation to decompose the lattice ring exchange interaction. Using Eq. \ref{eq:LatticeHS3D} we see that in the continuum the Hubbard-Stratonovich fields $\lambda_{1/2x}$ and $\lambda_{1/2y}$ couple to both the light fermions $\psi$ and the heavy fermions $\Psi$. Since they are gapped, the heavy fermions can be integrated out, leaving a Lagrangian in terms of the light fermions and the Hubbard-Stratonovich fields. After integrating out the heavy fermions, the Lagrangian can be written as
\begin{equation}
\mathcal{L}^{dCS} = \mathcal{L}_{\psi} + \mathcal{L}_{\lambda\lambda}, 
\end{equation}
where $\mathcal{L}_{\psi}$ contains all terms involving the light fermions $\psi$ and $\mathcal{L}_{\lambda\lambda}$ contains all couplings between the Hubbard-Stratonovich fields. If we ignore any terms that osculate like $(-1)^z$, $\mathcal{L}_{\psi}$ is given by
\begin{equation}
\begin{split}
\mathcal{L}_{\psi} = &\sum_{\bm{r}} \bm{\psi}^\dagger(\bm{r})G^{-1}_0(\omega+A_0(\bm{r}),p_z+A_z(\bm{r}))\bm{\psi}(\bm{r})\\& - [ \lambda_{2x}(\bm{r}) \psi_1^\dagger(\bm{r})\psi_3(\bm{r}+\hat{x}) \\
&\phantom{=} + \lambda_{1x}(\bm{r})\psi_2^\dagger(\bm{r}+\hat{x}+\hat{y})\psi_4(\bm{r}+\hat{y}) \\
&\phantom{=}+\lambda_{2y}(\bm{r})\psi_1^\dagger(\bm{r})\psi_4(\bm{r}+\hat{y})\\
&\phantom{=}+\lambda_{1y}(\bm{r})\psi_2^\dagger(\bm{r}+\hat{x}+\hat{y})\psi_3(\bm{r}+\hat{x}) + h.c.].
\end{split}\label{Aeq:ContHS}
\end{equation}
To determine $\mathcal{L}_{\lambda\lambda}$ we must integrate out the heavy fermions $\Psi$. At one loop order, $\mathcal{L}_{\lambda\lambda}$ is given by
\begin{equation}
\begin{split}
\mathcal{L}_{\lambda\lambda} = &\sum_{\bm{r}} -u_1 [\lambda_{1x}(\bm{r})\lambda^*_{1x}(\bm{r})+ \lambda_{2x}(\bm{r})\lambda^*_{2x}(\bm{r}) \\
&+ \lambda_{1y}(\bm{r})\lambda^*_{1y}(\bm{r})+ \lambda_{2y}(\bm{r})\lambda^*_{2y}(\bm{r})] \\
&+ [u^x_2 \lambda_{1x}(\bm{r})\lambda_{2x}(\bm{r}+\hat{y}) + u^y_2 \lambda_{1y}(\bm{r})\lambda_{2y}(\bm{r}+\hat{x}) \\
&\phantom{=} + \frac{2}{V}\lambda_{1x}(\bm{r})\lambda_{2x}(\bm{r})e^{-iA_{xy}(\bm{r})}\\ &\phantom{=} + \frac{2}{V}\lambda_{1y}(\bm{r})\lambda_{2y}(\bm{r})e^{-iA_{xy}(\bm{r})}
+ h.c.],
\end{split}\label{Aeq:AdditionalInteractions}
\end{equation}
where $u_1 = \frac{\log(4)-1}{16\pi},$  $u^x_2 = \frac{\cos^2(\kappa)}{16\pi}$, and $u^y_2 = \frac{\sin^2(\kappa)}{16\pi}$. 

As discussed in the main text, we shall employ the self-consistent mean field theory approximation. Similar to before, the Hubbard-Stratonovich fields acquire expectation values of the form
\begin{equation}
\begin{split}
\lambda_{1x}(\bm{r}) &= \lambda_x e^{i\phi_x(\bm{r}) + iA_{xy}(\bm{r})},\\
\lambda_{2x}(\bm{r}) &= \lambda_x e^{-i\phi_x(\bm{r}) },\\
\lambda_{1y}(\bm{r}) &= 0,\\
\lambda_{2y}(\bm{r}) &= 0.\\
\end{split}\label{Aeq:ContSelfConsist}
\end{equation}
Here, $\lambda_{1y}$ and $\lambda_{2y}$ have vanishing expectation values, while $\lambda_{1y}$ and $\lambda_{2y}$ do not, since $u^x_2>u^y_2$ in Eq. \ref{Aeq:AdditionalInteractions}. The value of $\lambda_x$ is determined by the effective potential
\begin{equation}
\begin{split}
H_{\lambda\lambda} = &\frac{4}{V_x'} \lambda_x^2 + \frac{1}{2\pi}\lambda_x^2\Big(\log(\frac{\lambda_x^2}{M^2})-1\Big),
\end{split}
\end{equation} 
where $V_x' = (V^{-1} + \frac{u_1+u^x_2}{2} )^{-1}$. The effective potential is minimized by $\lambda_x = M e^{-\frac{4\pi}{V_x'}}$. In agreement with the numeric results, we find that $\lambda_x$ vanishes when $V \rightarrow 0,$ and increases monotonically with increasing $V$. Additionally, due to the $u^x_2$ term in Eq. \ref{Aeq:AdditionalInteractions}, at low energies 
\begin{equation}
\begin{split}
&\phi_x(\bm{r}+\hat{y})-\phi_x(\bm{r}) = \Delta_y\phi_x(\bm{r})  = A_{xy}(\bm{r}).
\end{split}\label{Aeq:ContPhaseConstraint}
\end{equation}
As noted before, under a gauge transformation $\Lambda$, the phase fields $\phi_x$ transform as $\phi_x(\bm{r}) \rightarrow \phi_x(\bm{r}) + \Delta_x \Lambda(\bm{r})$. Because of this, all terms in Eq. \ref{Aeq:ContPhaseConstraint} have the same gauge transformation, as desired. 

To find the effective response theory for the continuum model in the mean field limit, the effective response action can be found by evaluating the current-current correlation functions. Here we are interested in the correlation functions between the currents  associated with $A_0(\bm{r})$, $A_z(\bm{r})$, $A_{xy}(\bm{r}),$ and $\phi_x(\bm{r})$ . These currents are found by taking the functional derivative of the mean field fermionic action with respect to one of the aforementioned fields. 
In the bulk, we find that in the low frequency and momentum limit, the effective response theory response is
\begin{equation}
\begin{split}
\mathcal{L}_{eff, \text{bulk}} = \frac{1}{4\pi}& (A_{xy} -  \partial_y \phi_x ) \partial_t A_z\\& - (A_{xy} -  \partial_y \phi_x) \partial_z A_0,
\end{split}
\end{equation}
where we have passed to the continuum in the $x$ and $y$ directions as well. Using Eq. \ref{Aeq:ContPhaseConstraint}, we find that the bulk action vanishes. However, at the boundaries normal to the $x$ and $y$-directions, as well as the hinges separating these surfaces, we find that the response Lagrangians are 
\begin{equation}
\begin{split}
L_{eff,\pm x} = 0\\
L_{eff,\pm y} =\pm & \frac{1}{2\pi}\big[ A_0 \partial_x A_z + \phi_x \partial_z A_0 - \phi_x\partial_t A_y \big],\\
L_{eff,\pm x , \pm y} =- & \frac{1}{4\pi} A_0A_z. \\
L_{eff,\pm x, \mp y} = + & \frac{1}{4\pi} A_0A_z.
\end{split}\label{Aeq:EffectiveActionBoundM}
\end{equation}
This is consistent with what we found using the mean field analysis in Sec. \ref{ssec:MFM}.
\end{document}